\documentclass[aps,onecolumn,11pt,floatfix,altaffilletter,
               tightenlines,showpacs,showkeys,notitlepage,nofootinbib]{revtex4-1}

\usepackage[colorlinks=true,citecolor=blue,linkcolor=blue]{hyperref}
\usepackage[normalem]{ulem}
\usepackage{amsmath,amssymb}
\usepackage{verbatim}
\usepackage{enumitem}
\usepackage{relsize}
\usepackage{mathtools}
\usepackage{epsfig}
\usepackage{graphicx}               % Standard graphics package
\usepackage{url}
\usepackage{color}
\usepackage{multirow}
\usepackage{placeins}
\usepackage[dvipsnames]{xcolor}
\usepackage{epstopdf}
\usepackage{nicefrac}
\usepackage{siunitx}
\usepackage{csquotes}
\usepackage{pifont}
\usepackage{soul}

\usepackage[capitalize]{cleveref}
\usepackage{lipsum}
 \usepackage{gensymb}
 
\allowdisplaybreaks

\bibpunct{[}{]}{,}{n}{}{,}

\newcommand{\cmark}{\ding{51}}%
\newcommand{\xmark}{\ding{55}}%


\begin{document}

%=============================================================================
\title{
Model Independent Bounds on the Non-Oscillatory Explanations\\
of the MiniBooNE Excess}

\author{Vedran Brdar}   \email{vbrdar@mpi-hd.mpg.de}
\author{Oliver Fischer} \email{oliver.fischer@mpi-hd.mpg.de }
\author{Alexei Yu.~Smirnov} \email{smirnov@mpi-hd.mpg.de}
\affiliation{Max-Planck-Institut f\"ur Kernphysik,
       69117~Heidelberg, Germany}
%=============================================================================

\begin{abstract}

We consider the non-oscillatory explanations of the low
energy excess of events detected by MiniBooNE.
We present a systematic search 
for phenomenological scenarios  based on new physics which can produce the excess. 
We define scenarios as series  of transitions and processes 
which connect interactions of 
accelerated protons in target with  single shower events in the MiniBooNE 
detector.   
The key elements of the scenarios are production and decay of 
new light  $\mathcal{O}(\text{keV}-100\,\text{MeV})$ particles
(fermions or/and  bosons). We find  about $20$ scenarios 
with minimal possible number of new particles  and interaction points. 
In practice, they are all reduced to few generic
scenarios and in this way we develop the effective
theory of the MiniBooNE excess.
We consider tests of the scenarios with near or close detectors in  neutrino
experiments T2K ND280, NO$\nu$A, MINER$\nu$A as well as in  NOMAD and PS191. 
The scenarios immediately connect
the MiniBooNE excess and expected numbers of new physics
events in these detectors.  We compute the expected numbers  
of events as functions of lifetimes and masses of new particles and 
confront them with the corresponding experimental bounds.   
We indicate scenarios  that are excluded or strongly disfavored by one or several experiments. Given our general approach, this work can also be regarded as the effective theory of new physics at accelerator based neutrino experiments, being relevant for future projects such as DUNE.

\end{abstract}

\maketitle

\section{Introduction}
%%%%%%%%%%%%%%%%%%%%%%%%%%%%%%%%%%%%%%%%%%%%%%%%%%%%%%%%%%%%%%%%%%%%%%%%%%%%%%%%%%%%%%
\noindent
The jury is still out on whether new physics 
effects are necessary for explanation of   
the low energy excess of the $e-$like events observed by 
MiniBooNE~\cite{Aguilar-Arevalo:2018gpe,Aguilar-Arevalo:2020nvw}. 
In this work we assume that the answer to this question 
is affirmative. 
%% by focusing on testing various beyond the Standard Model realizations. 
The popular explanation based on oscillations driven by mixing with a new 
eV-scale neutrino is very strongly disfavored, if not excluded\footnote{An alternative
oscillation scenario was discussed in ref.~\cite{Asaadi:2017bhx} where 
short baseline oscillations are due to very strong medium potential  
generated by new resonance scattering of neutrinos on the local overdense relic neutrino background.  
}.  
Not only the global neutrino oscillation fit  \cite{Dentler:2018sju}
but also properties of the excess 
(energy and angular distributions) are behind the last statement.

In this connection various non-oscillatory explanations of the excess were proposed. 
Most of them  make use of possible mis-identification of 
the MiniBooNE events which can be due to electrons, photons, collinear   
$e^+ e^-$  as well as  $\gamma \gamma$ pairs. The explanations are based on production and decay 
of new heavy neutrinos $N$ or/and bosons $B$ with masses $\mathcal{O}(\text{keV}-100\,\text{MeV})$. 
They include:

\begin{itemize}

\item 
The $N-$production in the  MiniBooNE detector via the $\nu_\mu-$upscattering 
and then the radiative $N-$decay \cite{Gninenko:2009ks}; 

\item
Production of $N$ in the decay pipe via mixing in $\nu_\mu$ 
and further radiative decay 
along the beamline, and mainly, in the detector \cite{Fischer:2019fbw};
%%between the decay region and the detector as well as in 

\item 
The $N-$production in the detector via $\nu_\mu-$upscattering and  
decay with appearance
of the $e^+ e^-$ pair. Two versions have been proposed:
the 3-body decay:  $N  \rightarrow \nu e^+ e^-$  
\cite{Ballett:2018ynz,Abdullahi:2020nyr,Ballett:2019pyw}, and 
the 2-body decay: $N  \rightarrow \nu B$  
followed by the decay of an on-shell boson  $B \rightarrow e^+ e^-$. 
Here $B$ can be a new gauge boson $Z'$ \cite{Bertuzzo:2018itn} or a
scalar $B=S$ \cite{Datta:2020auq,Dutta:2020scq,Abdallah:2020vgg,
Abdallah:2020vgg}.  
In these models, $B$ has a decay length which is much smaller 
than the size of the detector,  
$\lambda_B \ll d^{MB}$, so that the event looks like a 
local decay of $N$. There is 
an important kimematical difference from  
the 3-body decay of $N$ \cite{Ballett:2018ynz}, 
since here the invariant mass of the pair 
$e^+ e^-$ is determined by the mass of $B$ which is smaller than the mass of $N$. 

\item  
The $N-$production via mixing in the decay pipe followed by the decay 
$N \rightarrow \nu_e \phi$ along the baseline with emission 
of $\nu_e$. The latter, in turn,  produces an electron via the CCQE scattering 
in the detector \cite{deGouvea:2019qre,Dentler:2019dhz,Bai:2015ztj} (see also \cite{PalomaresRuiz:2005vf}).

\item
Production of the light scalar $B$ in the $\nu_\mu-$upscattering:  
$\nu_\mu A \rightarrow N B A'$,
which then decays as $B \rightarrow e^+ e^-$ \cite{Abdallah:2020biq}. 
(In the model  \cite{Abdallah:2020biq} $B$ is produced via coupling with gauge 
boson mediator of the upscattering process.) The new neutrino $N$ does not 
contribute to the MiniBooNE signal in contrast to the previous mechanisms.  

\end{itemize}

%{\color{red} \st{It should be mentioned that all these explanations do not provide a perfect fit to the 
%MiniBooNE excess; they are disfavored by some other data, 
%and most  of them do not reproduce the LSND excess in contrast to oscillations.}} 

{\color{black}
It should be mentioned that a number of explanations do not
reproduce the LSND excess in contrast to oscillations
(see \cref{table:scenario}).}

Recent measurements of the bunch timing \cite{Aguilar-Arevalo:2020nvw} 
do not show  deviation 
(shift or widening) of the time distribution of the MiniBooNE events from the one 
due to usual light neutrinos \cite{Aguilar-Arevalo:2020nvw}. This essentially 
excludes mechanisms of decay of 
heavy neutrinos in the second item above and restricts parameters of the mechanism in 
the first item.

Do other possibilities of this type exist or is everything already covered?   
In this connection, we perform a systematic search of all possible  
phenomenological scenarios  that can explain  the MiniBooNE excess. 
We identify the simplest scenarios with a minimal number of new particles and 
new interaction points. 
%%The scenarios can be classified by the number of points of new physics interactions. 
Clearly, an increase of these points would introduce additional smallness
since there are various restrictions 
on new interactions.

The goal of this paper is to perform model independent tests 
of explanations of the MiniBooNE excess.  
For this, we introduce scenarios, that is, sets of transitions and processes which connect proton interactions 
on target with the appearance of single shower events in MiniBooNE.  
To test the explanations we  use data from accelerator neutrino experiments with near or relatively 
close detectors.  
The scenarios allow us to directly connect numbers of events in these detectors with the MiniBooNE 
excess.  Various model-dependent features cancel in this consideration. 
We describe these scenarios by a small number of parameters. 
%%Furthermore, we will not consider all observational consequences of the scenarios,  
%%but focus on  effects in near detectors of neutrino experiments which have  setups
%%similar to the MiniBooNE setup. 
%%The expected numbers of events  in the near detector experiments 
%%are confronted with the experimental bounds. 
Notice that the scenarios can be further (and in some cases even more strongly) 
restricted by other observations. 

The paper is organized as follows.  
In \cref{sec:options} we present a systematic search for the simplest phenomenological 
scenarios which explain the MiniBooNE excess. 
In relevant aspects  they are reduced to few  
qualitatively different possibilities. 
In \cref{sec3} we present general formulas for the number of events in the 
detectors as functions of parameters of the experimental setups and parameters of the scenarios. 
The latter mainly include the lifetimes and masses of new particles. 
In \cref{sec:limits}, we present  parameters of the  employed experiments and 
derive experimental upper bounds on the number of events due to new physics. 
In \cref{sec:results} we compute the expected number of events due to  new physics in different scenarios 
and confront them with experimental bounds. Discussion and conclusions follow in \cref{sec:sum}.

%%%%%%%%%%%%%%%%%%%%%%%%%%%%%%%%%%%%%%%%%%%%%%%%%%%%%%%%%%%%%%%%%%%%%%%%%%
\section{Scenarios for the MiniBooNE excess}
\label{sec:options}
%%%%%ssss2%%%%%%%%%%%%%%%%%%%%%%%%%%%%%%%%%%%%%%%%%%%%%%%%%%%%%%%%%%%%%%%%%%%

%%In general, non-oscillatory new physics that explains the MiniBooNE excess 
%%of electron-like events can involve new particles, new dynamics (processes for known particles), 
%%or both. Here, 

\subsection{General bounds on explanations of the excess}
%%%%%%%%%%%%%%%%%%%%%%%%%%%%%%%%%%%%%%%%%%%%%%%%%%%%%%%%%%%%%%%%%%%%%%%%%%%%

\noindent
MiniBooNE (MB) observed the excesses of  $1sh-$ events of 
$560.6 \pm 119.6$ and $77.4 \pm 28.5$ in the neutrino and antineutrino mode (horn polarities), 
respectively~\cite{Aguilar-Arevalo:2018gpe}. The collected data 
corresponds to  $18.75\times10^{20}$ POT  ($11.27 \times 10^{20}$ POT) in neutrino (antineutrino) mode.  
 We will use the sum  of the $\nu$ and $\bar{\nu}$  excesses: 
\begin{equation}
N^{MB}_{1sh,exp} = 638.0 \pm 132.8. 
\label{eq:numbmb}
\end{equation}
We assume that this excess is due to new physics rather than underestimated or missed
background or oscillations related
to existence of the eV-scale sterile neutrino.

The source of events is the 8 GeV proton beam from the Booster that hit the Beryllium target,
producing secondary particles. The 
818 ton liquid scintillation detector  observes 
via the Cherenkov radiation  the single shower, $1sh$,  events: 
\begin{equation}
p + A ~[target] \rightarrow [ X ] \rightarrow  1sh~ events~ [detector].
\label{eq:setup}
\end{equation}
The recoil nucleon can produce scintillation, but this additional
source of light was not considered in the MB reconstruction of events\footnote{Being included 
in the analysis, the information on the recoil
could help excluding various possibilities and distinguish between the decay  
and  upscattering explanations.}. 
The MiniBooNE detector is not capable to identify particle(s) which
induce these EM showers. 

Appearance of the $1sh$ events is time-correlated with  the $pA-$collisions in the target.
Therefore it should be a mediator(s) system $X$  which connects the ends: 
the $pA-$ interaction in the target 
and the EM shower in the detector. Furthermore, the arrival time distribution of events 
%%the $1sh$ events appear within time window of bunches arrival at the target
%%with delay which 
was found to be consistent with the arrival time of the usual neutrinos. 
%% propagation with velocity of
%%light [[add insertion]].
We will not discuss the LSND result: the  requirement of joint 
explanation imposes additional restrictions on scenarios.

What is the ``black box" $X$ in Eq. (\ref{eq:setup})? It can be production and propagation
of new particles, or some new dynamics related to known particles
like Lorentz violation \cite{Katori:2006mz}, non-standard decoherence \cite{Farzan:2008zv}, {\it etc}.
We will assume that (i) the mediator system is some
new particle (or system of particles) $X_s$ that is produced in the source, (ii) $X_s$ 
evolves, in general, via a chain of processes:  $X_s \rightarrow X_{det}$, and (iii) then $X_{det}$  
interacts or decays in the detector  producing the $1sh$ events: 
\begin{align}
p + A ~[target] \rightarrow X_{s} \,[~ \rightarrow ~]\,  X_{det} \rightarrow 1sh ~~ [detector].
\label{eq:setup1}
\end{align}

There are certain observations that  allow us to eliminate 
many possibilities and make the  
first step toward connecting ``the ends'':\\

1. The proton beam energy, $E \sim 8$ GeV,  restricts the mass scale 
of new particles to be at most around a few  GeV. 
Since charged particles at this mass scale are excluded, 
the new particles should be electrically neutral.\\

2. The numbers of excess events as compared to the  $\nu_\mu-$
 and $\nu_e-$CC events equal
\begin{equation}
\frac{N^{MB}_{1sh}}{N^{MB}_{\mu}} \simeq 10^{-2} \,, 
~~~~\frac{N^{MB}_{1sh}}{N^{MB}_{e}} = 0.53\,.
\label{eq:nuexcnumue}
\end{equation}
Therefore the processes, which lead to the excess, 
should not be very rare. In fact,  the yield should be
comparable with the yield of usual neutrinos unless we assume that
$X$ has strong interaction. \\

3. The excess is absent in the beam-dump run~\cite{Aguilar-Arevalo:2018wea}: 
In this run according to number of POT about 30 events should be produced,  
%%given that the neutral 
%%pion flux is almost independent on the beam mode; 
but no excess was observed.\\

4. The ratio of excesses in $\nu$ and  $\bar{\nu}$ modes (horn polarities) 
corresponds to what is expected for usual neutrinos. \\

%%In general, these neutral particles can be fermions, vectors and scalars
%%The electrically neutral fermions can mix with LH neutrino. 
%%Connecting the ends.

The implications of these results follow.  \\

{\bf From the source side:} In general, $X_s$ can be produced 

\begin{itemize}

\item
on target in the $pA-$collisions immediately,

\item 
in decays (interactions) of known particles
produced in the $pA-$collisions, such as $\pi$, $K$, heavy mesons,

\item 
by usual neutrinos $\nu_\mu$ in  detector or/and surrounding matter 
along the baseline.

\end{itemize}

The beam-dump mode results and the $\nu - \bar{\nu}$  results 
exclude the first possibility.
The number of excess events in $\nu$ and  $\bar{\nu}$ 
modes  corresponds to what is  expected for usual neutrinos
which implies the same differences of $X_s$ production  and 
$X_{det}$ interaction as for netrinos. 
Neutral particle decays as sources of $X_s$ are excluded 
since they  are not affected by the magnetic field and beam-dump \cite{Jordan:2018qiy}.
Thus, we arrive at the conclusion that $X_s$ should be produced in
the charged $\pi-$ and  $K-$decays immediately or  by usual neutrinos from 
these decays.

Notice that apart from three possibilities described above 
one can consider production of $X_s$ in upscattering of muons from $\pi-$  
and $K-$ decays in a shield and dirt.\\

{\bf From the detector side:} the $1sh$ MiniBooNE events can be produced
by $e$, $\gamma$, collimated $e^+ e^-$ pair and collimated $\gamma \gamma$ pair, that is, 
by state $\xi$
\begin{equation}
\xi = e, \, \, \,  \gamma,\, \, \,   e^+ e^-, \, \, \,    \gamma \gamma.
\label{eq:emsyst}
\end{equation}
We will not consider more complicated systems, since their
production will  bring additional suppression.
Fluxes of particles $\xi$ from the outside are suppressed
by absorption in walls of the detector, rejection by anticoincidence system and
fiducial volume cut. Furthermore, radial distribution
of events shows that the excess increases toward the center
\cite{Aguilar-Arevalo:2020nvw}. Therefore $X_{det}$  in (\ref{eq:setup1}) should be some neutral particle that
enters MiniBooNE and produces $\xi$ in interaction or decay inside the detector.

The particle(s) $X_{det}$ as well as $X_{s}$  can be fermion $N$
or boson $B$, and the latter can be scalar  or vector bosons.
%%For fermions, we consider spin 1/2  
%%although spin $3/2$ like gravitino can also be considered.  
%%\footnote{One may also consider particles 
%%of higher spin numbers and composite systems. However, MiniBooNE is not the most appropriate experiment 
%%for such next-to-minimal realizations. For instance compositeness can be tested at LHC; 
%%in addition note that particles of higher spin suffer from non-renormalizability.}.
For definiteness we will mainly explore  spin 1/2 fermion\footnote{Spin $3/2$ particles, 
like the gravitino, can also be considered.} 
and boson cases: $X = N, B$. 
%%   also because
%%all explanations in literature are based on existence of
%%new fermions. Many results
%%can be immediately generalized to the bosons case.
%%We will consider separately the cases with bosons which show some qualitatively new features. \\

If $X_s$ is a new heavy neutrino $X_s = N$,  it can be produced  via mixing in
$\nu_\mu$. Therefore, the relevant channels of production are the same as
for $\nu_\mu$  with substitution $\nu_\mu \rightarrow N$.
If $X_s = B$, the decays are the same as the standard decay modes
of $K$ and $\pi$ with additional $B$ emission (bremsstrahlung)
$K \rightarrow \mu \nu B$, $\pi \rightarrow \mu \nu B$,
or standard modes in which
one of pions is substituted by $B$: $K \rightarrow \pi B$,
$K \rightarrow \pi \pi B$. Details of these decays, values of
couplings, bounds {\it etc.} are not important for our analysis.

The electromagnetic systems $\xi$ (\ref{eq:emsyst}) can be produced in decays of $N$
or in $N-$interactions.
Due to fermionic nature the $N-$decays can proceed with emission of the usual neutrinos
or a new neutral fermion $N'$:
%\begin{equation}
$$
N \rightarrow \nu + \xi, \,  \,  \, \,  N \rightarrow N' + \xi.
$$
%\label{eq:dec}
%\end{equation}
The simplest possibilities include the radiative decay ($\xi = \gamma$):
%\begin{equation}
$$
N \rightarrow \nu + \gamma,
$$
%\label{eq:decg}
%\end{equation}
the 3-body decay ($\xi = e^+  e^-$)
%\begin{equation}
$$
N \rightarrow \nu + e^+ + e^-, 
$$
%\label{eq:decee}
%\end{equation}
and decay via production of on-shell boson (double decay):
%\begin{align}
$$
N \rightarrow \nu + B, \, \, \,  \,  B \rightarrow  e^+ + e^-  \, \, \,
{\rm or} \, \, \,  B \rightarrow  \gamma + \gamma.
$$
%\label{eq:decee2}
%\end{align}
Here, $B$ can be $\pi^0$ or some new scalar or vector boson.
%%There is no way to produce single $e$ in $N$ decay.

Alternatively, $\xi$ can be produced in $N-$interactions with electrons or nucleons (A):   
%\begin{equation}
$$N + e \rightarrow e  + N',~~~~~
N + A \rightarrow e  + A'\,, 
$$
%\label{eq:scN}
%\end{equation}
%%which can proceed via CC or/and NC interactions.
where $N'$ can coincide with the usual neutrinos $\nu_\mu$ or
$\nu_e$.\\

In the case of new boson,  $X_{det} = B$, the state 
$\xi$ can be produced in the 2-body decays:
%\begin{equation}
$$
B \rightarrow  e^+ + e^-,  ~~~~B \rightarrow  \gamma + \gamma, ~~~~ 
B \rightarrow  B' + \gamma, 
$$
%%
%\label{eq:decbgb}
%\end{equation}
or the 3-body decay: 
%\begin{equation}
$$
B \rightarrow  B' +  e^+ + e^- .
$$
%\label{eq:decbgb2}
%\end{equation}
Also, $\xi$ can appear in  $B-$interactions with nuclei 
and electrons: 
%\begin{align}
$$
B  + A \rightarrow A +  e^+ + e^-, ~~~ B  + A \rightarrow A +  \gamma\,, 
~~~B  + e \rightarrow B + e\,. 
$$
%\label{eq:decbgb3}
%\end{align}

%%%%%%%%%%%%%%%%%%%%%%%%%%%%%%%%%%%%%%%%%%%%%%%%%%%%%%%%%%%%%%%%%%%%%%%%%%%%5
\subsection{Combinatorics of connections.  Scenarios.} 
\label{subsec:2b}
%%%%%%%%%%%%%%%%%%%%%%%%%%%%%%%%%%%%%%%%%%%%%%%%%%%%%%%%%%%%%%%%%%%%%%%%%%%

Let us consider all possible connections of the  source and detector parts, 
%%we arrive at the following simplest possibilities.
{\it i.e.},  transition  $X_{s} \rightarrow  X_{det}$.
In the simplest case, $X_{s}$ and  $X_{det}$ coincide:  
$X_{s} = X_{det}$. The next possibility is that  $X_{det}$ is  produced in
decays of $X_{s}$ or in interactions of $X_{s}$  with the medium
on the way to a detector or inside the detector.
Several particles can be involved 
via a chain of processes connecting the ends:  
$X_{s} \rightarrow  X_1 \rightarrow X_2 ~~... \rightarrow X_{det}$. 
At this point, we will employ criteria of minimality:  the 
simplest links with minimal number of chains or interaction points will be identified. 
Notice that, in general, any new vertex or additional new particle typically 
brings  an additional suppression and  it is difficult to 
produce the required number of events in MiniBooNE. \\

Let us consider transitions  with two and more 
interaction points which include production and decay of
a new fermion $N$ or boson $B$\footnote{Notice that the 
simplest scenario would be with single non-standard interactions vertex,  when 
$X_s = X_{det} = \nu_\mu$. Now, $\nu_\mu$, from standard 
$\pi$ and $K$ decays, produce electrons in the detector 
via the charged current non-standard interaction 
(CC NSI) $\nu_\mu + A \rightarrow e + A'$ (This implies that $\nu_\mu$ is not orthogonal to $\nu_e$)
or via neutral current (NC) NSI on electrons. Such a possibility is restricted very strongly. 
}.\\

Heavy neutrino $N$ can be produced

\begin{itemize}

\item
in decays of usual mesons $\pi$ and $K$
in a  decay pipe (for $N$ it is due to mixing with
usual neutrinos). We call this element of the scenario $M$  (Mixing). 

\item
by the $\nu_\mu-$interactions with matter
outside the pipe, that is, by the $\nu_\mu-$upscattering, $U_N$. 

\end{itemize}

In the mixing case the $N-$flux
is formed in the decay pipe,
while in the $U_N-$case,  $N$ are produced outside the pipe.\\

In turn, $N$ can decay

\begin{itemize}

\item
immediately into $\xi$ (we denote this process by $D_\xi$); 

\item
into a state with $\nu_e$, $D_\nu$, which then produces $\xi = e$ 
interacting in the detector ($U_e$);

\item
into new neutral particles  $N \rightarrow B$ which then
decay into $\xi$ ($D_B D_\xi$).

\end{itemize}

Instead of decay,  $N$ can upscatter on nucleons and electrons 
in a detector and outside the detector in dirt to produce $\xi$ ($U_\xi$).
But this would  involve another smallness
due to additional non-standard interaction.  
Indeed, the probability of $N$ interactions equals $P_N = \sigma_N n l$, where 
$\sigma_N$ is the cross section,  $n$ is the number density of scatterers  and $l$ is the length of trajectory 
along which $N$ interacts.    
For new 4-fermion interactions characterized by coupling $G_N$ and $\sigma_N \propto G_N^2 E_N$, where $E_N$ 
is the energy of $N$, we obtain 
\begin{equation}
P_N \approx 5 \cdot 10^{-11} 
\left(\frac{l}{ 10{\rm m }} \right)
\left(\frac{n}{3 n_A}       \right)
\left(\frac{E_N}{1~ {\rm GeV }} \right)
\left(\frac{G_N}{G_F} \right)^2\,,
\label{eq:probint}
\end{equation}
where $n_A$ is the Avogadro number.  Let us compare this probability with 
the probability of $N-$decay. 
If $N$ is produced at the distance $l$ from a detector and the size of 
a detector is $d$,  then the probability of its decay in the detector equals 
\begin{equation}
P_{dec} =  e^{- l/\lambda_N}
\left(1 -  e^{- d/\lambda_N}\right). 
\label{eq:final2}
\end{equation}
Here  $\lambda_N$ is the decay length of $N$:
\begin{equation}
\lambda_N (E_N,m_N) = \frac{E_N}{m_N} \, c \tau^0_N, ~~~~
\label{eq:declength}
\end{equation}
where $c$ is velocity of light, $\tau^0_N$ is the lifetime of $N$ in the rest frame 
and $m_N$ is the mass of $N$.

For fixed $l$ and $d$ the maximum of $P_{dec}$ is achieved at 
\begin{align}
\lambda_N  =  d \, [\text{log}(1 + d/ l)]^{-1} \approx l, 
\end{align}
where the second equality is for $d \ll l$.  
The probability at $\lambda_N = l$ and typical  values of $d$ and $l$ 
equals 
\begin{equation}
P_{decay}^{max} = \frac{d}{e\, l} \sim 10^{-2},  
%\label{eq:maxp2}
\end{equation}
($e \approx 2.7$). 
Therefore, the $N-$decay can be substituted by upscattering of 
$N$,  if $P_N > 10^{-2}$, which implies, according to  
(\ref{eq:probint}),  that  $G_N > 10^4 \, G_F$. The latter is difficult to realize. \\

Connecting  two $N-$production mechanisms (mixing, upscattering) and three decay
possibilities listed above we arrive at the following 6 scenarios for $X = N$.
The number of possibilities  multiplicates due to various $\xi$ (\ref{eq:emsyst}).

\emph{1)}  $M_N D_\xi$, Mixing - Decay scenario:  $N$ is produced in the 
$K-$ and $\pi-$decay via mixing in $\nu_\mu$  and it decays as
$N \rightarrow N' + \xi$. Here  $\xi$ is any state in Eq.  (\ref{eq:emsyst})
except the electron, and $N'$ can be a standard neutrino $\nu$.
Only decays inside a  detector give an observable signal.\\

\emph{2)}  $M_N D_\nu U_e$,   Mixing - Decay into $\nu_e$  scenario:
$N$ produced via mixing decays with emission of $\nu_e$:
 $N \rightarrow \nu_e + B$. Then $\nu_e$ upscatters in  detector, 
producing electron.\\ 

\emph{3)} $M_N D_B D_\xi$,  Mixing-double decay scenario: $N$ produced 
via mixing decays invisibly
into another new particle $B$, which, in turn,  decays into
(or with emission of) $\xi$. \\
%%This double decay scenario is characterized
%%by $U_{\mu 4}$,  $m_N$, $\tau_N$, $m_B$, $\tau_B$.\\
%%(Only fast $N$ decay has been considered in literature (Bertuzzo).\\

\emph{4)} $U_N D_\xi$, Upscattering - decay  scenario: $N$ is produced in  the
$\nu_\mu$ interactions  with particles
of medium between a source and a detector as well as 
inside the detector. Then $N-$decay in detector
produces $\xi$.  If interactions of $N$ with  medium can be neglected,
the  $N-$flux will be  accumulated along the way to a detector.\\
%%The parameters of this scenario are $m_N$, $\tau_N$ and $\sigma_N$.\\

\emph{5)} $U_N D_\nu U_e$,  Upscattering - decay into $\nu_e$ scenario: 
$N$ produced by the $\nu_\mu-$upscattering
decays with emission of $\nu_e$, which then
scatters in  detector via CCQE  producing the $e-$shower.\\

\emph{6)} $U_ND_BD_\xi$, Upscattering - double decay scenario: $N$ produced by 
the $\nu_\mu-$upscattering
undergoes double decay: $N \rightarrow B \rightarrow \xi$.\\

Scenarios 1, 2, 4, 5 contain two vertices with new particles,
scenarios 3 and 6 are of higher (third)
order in  new physics interactions.

Two more scenarios can be identified in which $\xi-$state 
is produced by upscattering of $N$. They have additional suppression in comparison to
$\xi$ production in  decays.  The first scenario is $M_N U_\xi$, {\it i.e.},   
the Mixing - $N-$upscattering. Here $N$ produced via mixing 
in $\nu_\mu$ upscatters in a detector with production of electron:  
$N + A \rightarrow e + A'$. This implies the  lepton number violation 
since $N$ is mixed in $\nu_\mu$ but produces $e$ in interactions.  
The second scenario is  $U_N U_\xi$,  which is  
double upscattering.   $N$ is produced in 
upscattering of $\nu_\mu$ and then upscatters with production of $\xi$ ($e$). \\

The six scenarios described above are not completely independent from
the geometrical point of view and even coincide in
certain limits of values of parameters.
Thus, for short lifetime of $B$ we have
\begin{align}
U_N D_\xi \approx U_N D_B D_\xi\,,
\end{align}
with the only difference that in the double decay case
the invariant mass of particles in the final state is fixed
by the mass of $N$. \\

For $X=B$ we have similar  mechanisms of  
production and decay. As far as propagation features are concerned,  the scenarios with $B$ 
coincide with scenarios for $N$, but  
differ from the model building side. Also in this case instead of mixing in $\nu_\mu$,
$B$ are produced in $\pi-$ and  $K-$ decays and therefore $M$ should be 
interpreted as $B$ production in the Meson decays.  
For bosons  we have the following scenarios:

$(i)$ $M_B D_\xi$ - production of $B$ in a decay pipe in meson decays and
further decay $B \rightarrow \xi$, $B \rightarrow B' \xi $; 

$(ii)$ $M_B D_\nu D_e$  -  $B-$ decays with emission of $\nu_e$,
$B \rightarrow \nu_e \bar \nu_e$ or $B \rightarrow \nu_e N'$; 

$(iii)$ $M_B D_{B'} D_\xi$ - double decay, which 
is a non-minimal and  complicated version of (i). 

Three other mechanisms differ from (i - iii) by
$B$ production mechanism, namely,  instead of decays in a pipe,
$B$ is produced via the $\nu_\mu-$ upscattering in
a detector and the surrounding medium. These three scenarios include  

$(iv)$ $U_B D_\xi$ - with $B$ decays as in (i), see Ref. \cite{Abdallah:2020biq}; 

$(v)$ $U_B D_\nu D_e$ - $B-$decay into $\nu_e$, which in turn, 
produces $e$ in CCQE in a detector; 

$(vi)$ $U_B D_{B'} D_\xi$ - double decay which is non-minimal version of (iv). 

Throughout the paper we focus on scenarios with $X=N$.

{\color{black}
In \cref{table:scenario} we associate the proposed non-oscillatory
models for the MiniBooNE
anomaly to aforementioned scenarios.
There, we also indicate whether a given proposal
can also reasonably well explain the LSND anomaly
(3rd column).
}

\begin{table}[]

{\color{black}{

{\Large{
\begin{tabular}{|c|c|c|}
\hline
Model & Scenario  & LSND \\ \hline
 \cite{Fischer:2019fbw} & $M_N D_\gamma$   & \xmark \\ \hline
 \cite{Gninenko:2009ks} & $U_N D_\gamma$   & \xmark \\ \hline
 \cite{Ballett:2018ynz,Abdullahi:2020nyr,Ballett:2019pyw} & $U_N D_{ee}$   & \xmark \\ \hline
\cite{Bertuzzo:2018itn,Datta:2020auq,Dutta:2020scq} & $U_N D_B D_{ee}$   & \xmark \\ \hline
\cite{Abdallah:2020vgg} & $U_N D_B D_{ee}$   & \cmark \\ \hline
 \cite{deGouvea:2019qre,Dentler:2019dhz,Bai:2015ztj,PalomaresRuiz:2005vf} & $U_N D_\nu U_e$   & \cmark \\ \hline
 \cite{Abdallah:2020biq} & $ U_B D_{ee}$    & \xmark \\ \hline
\end{tabular}
}}
\caption{Mapping of the proposed models
that aim at explaining the MiniBooNE
anomaly (references in the 1st column)
on to scenarios introduced in this paper
(2nd column). In the 3rd column check marks (crosses) indicate whether a given proposal can (can not) fit the LSND data.}
\label{table:scenario}

}}
\end{table}

%%%%%%%%%%%%%%%%%%%%%%%%%%%%%%%%%%%%%%%%%%%%%%%%%%%%%%%%%%%%%%%%%%%%%%%%%%%%
\subsection{Bounds on parameters of  scenarios from timing}
\label{subsec:timing}
%%%%%%%%%%%%%%%%%%%%%%%%%%%%%%%%%%%%%%%%%%%%%%%%%%%%%%%%%%%%%%%%%%%%%%%%%%%%%%%

The key parameters of the scenarios are masses and lifetimes of new particles. Therefore, 
the bounds from timing of the MB events are crucial for our consideration. 
The bounds differ for scenarios with $N$ production in a 
decay pipe via mixing and in a detector via upscattering.
In the first case, $N$ propagates from a  production point in a pipe to
a  detector, {\it  i.e.} the distance equals the baseline,  $l$.
A delay of the events produced by $N$
with respect to the signal from usual neutrinos equals 
\begin{align}
\Delta t 
= \frac{l}{c} \left[ \frac{1}{\sqrt{1 - (m_N/E_N)^2}} -1 \right] \approx \frac{l}{c} \frac{m_N^2}{2E_N^2}, 
\label{eq:deldel}
\end{align}
%%where $c$ is velocity of light,  
%%$y \equiv m_N/E_N$ 
and the last equality in (\ref{eq:deldel}) is for $m_N/E_N \ll 1$. 
Numerically, we have
\begin{align}
\Delta t = 8 \, \text{ns} \, \left(\frac{l}{500\, \, {\rm m}}\right)
\left(\frac{m_N}{0.1\, \,  {\rm GeV}}\right)^2
\left(\frac{1\, \,  {\rm GeV}}{E_N}\right)^2\,.
\label{eq:deltat}
\end{align}
Using the typical excess energy $E_N = 0.3$ GeV and $\Delta t = 1$ ns
we find the from (\ref{eq:deltat}) the upper bound on the mass: $m_N < 10$ MeV.
In the case of $N-$ and $B-$decays  this bound leads to  
very forward excess of events in MiniBooNE which contradicts data.
Indeed, the observed angular spectrum of the MiniBooNE excess requires 
$m_N$ to be above 200 MeV \cite{Jordan:2018qiy}. 
%%(  case the dominant 
%%production of heavy neutrinos are kaon decays at the source.)
Such a possibility can still be considered if there is 
a two component interpretation of the angular distribution of the excess
which, in fact, is favored by recent data. One component, e.g., due to underestimated background is nearly
isotropic  and another one due to new physics contribution peaks in the forward direction.
Keeping this in mind we will  consider such scenarios.\\

Another possibility is that $N$ and $B$ are produced via
the $\nu_\mu-$ (or another light particle) upscattering.
In the upscattering case, the typical decay length is  smaller
than a detector size:  $\lambda_N < d$.  
Therefore, we should take  $\lambda_N$  as a conservative estimate of 
the distance of $N-$decay. 
Substituting $l$ by  $\lambda_N =  c \tau^0 E_N/m_N$ in the  expression (\ref{eq:deldel})  
we can write the upper bound on lifetime of $N$ which ensures a delay smaller than a given $\Delta t$:
\begin{equation}
{c\tau^0} <  c \Delta t \frac{m_N}{E_N} \left[ \frac{1}{\sqrt{1 - (m_N/E_N)^2}} -1 \right]^{-1}\, . 
\end{equation}
For $ m_N/E_N  \ll 1$ this gives
\begin{equation}
c\tau^0 <  2 c \Delta t \frac{E_N}{m_N}.
\end{equation}
Taking $\Delta t = 1$ ns and $E_N = 0.8$ GeV we obtain the following
upper bounds on $c\tau^0$  for values  
$m_N = (0.15,~ 0.25,  ~ 0.35)~ {\rm GeV}$ respectively
\begin{align}
c\tau^0 < (3.2, ~1.92, ~1.37)~ \text{m}\,.
\end{align}
%%[[In the plots we can simply cut the lines above these values of $c\tau^0 (m_N)$.]]

%One comment is in order. 
$N$ production via the $\nu_\mu-$upscattering
usually implies $N$ mixing in $\nu_\mu$. Therefore,
in general, one has to sum the contributions from  $N$ produced via the mixing and
upscattering mechanisms. However, these two mechanisms are effectively operative in different
ranges of $c\tau^0 $. In  the upscattering case,
$N$ should decay within detector volume ($c\tau^0 \leq 1$ m) unless it decays into another new particle $B$,
while in the case of $N-$production in a decay pipe via mixing 
$N$ should reach a detector,
\emph{i.e.} survive  about  several
hundred meters, implying that $c\tau^0 \gtrsim 100$ m.
Therefore for a given value of $c\tau^0$ only one mechanism dominates.

%%%%%%%%%%%%%%%%%%%%%%%%%%%%%%%%%%%%%%%%%%%%%%%%%%%%%%%%%%%%%%%%%%%%%%%%%%%%%%
\subsection{Signature factors and efficiencies}
\label{subsec:sig}
%%%%%%%%%%%%%%%%%%%%%%%%%%%%%%%%%%%%%%%%%%%%%%%%

%%The state $\xi$ shows up in the detectors as event of the type $s^i$. 
A detector $i$ observes events of various types 
$s^i$,  which depend on features of the  detector. 
We will call $s^i$  signatures.
In particular, MiniBooNE observes 1 and 2 showers events, while
ND T2K with better particle ID can observe -- $\gamma$ showers, $e-$ showers (tracks),
and $2-$showers events:
\begin{equation}
s^{MB} = \{ 1sh, \, 2sh \}, \,  \, \, ~~~
s^{ND} = \{ \gamma -sh, \,  e-sh,  \, \, 2sh \}. 
\label{eq:events}
\end{equation}
Because of mis-identification, 
the observed events do not correspond uniquely to certain 
original states
$\xi$. To  quantify this, we introduce the signature factors 
$f^i_{\xi - s^i}$ which
give the fraction of cases in which a given state $\xi$
shows up as $s^i$ event in the
$i-$detector. Equivalently,  $f^i_{\xi - s^i}$ can be considered as the probability that a state $\xi$
will show up as $s^i$ event.

$f^i_{\xi - s^i}$ depends on the parameters of the state $\xi$
- energies of particles, masses,  as well as  on properties of detectors.
For MiniBooNE, a single  electron will be detected
%%(if detected which is given by $\epsilon$) 
as 1sh event, namely $f^{MB}_{e -  1sh} = 1$. Similarly,
for $\gamma$: $f^{MB}_{\gamma - 1sh} = 1$.
Also $e^+e^-$ state can show up as 1 shower event but 
$f^{MB}_{ee -  1sh} < 1$ and the fraction depends on the
kinematical variables of $e^+$ and $e^-$. 
%%We will discuss the signature 
%%factors in more details for other experiments  in
%%section devoted to specific experiments and observables.\\

%%[[repetition. Move complete description here]]

The numbers of events depend also on experimental reconstruction efficiency 
for a given signature $\epsilon^i_{s}(E_N,m_N)$. It is an empirical 
function which depends on properties of the signature, 
such as energies and angles. For simplicity, we take it to be a constant 
value for a given experiment and signature. 

We can introduce the signature factor in different way (taking one step back), considering 
final process  (decay or scattering) in which the state $\xi$ is produced.  
Then one can introduce $f_{s^i}$ as fraction of $N-$decays or $\nu-$scatterings  
in which the $s^i$ event is produced.

%%%%%%%%%%%%%%%%%%%%%%%%%%%%%%%%%%%%%%%%%%%%%%%%%%%%%%%%%%%%%%%%%%%%%%
\section{Numbers of new physics events in the generic scenarios}
\label{sec3}
%%%%%%ssss2a%%%%%%%%%%%%%%%%%%%%%%%%%%%%%%%%%%%%%%%%%%%%%%%%%%%%%%%%%%%%%%%%

%%%%%%%%%%%%%%%%%%%%%%%%%%%%%%%%%%%%%%%%%%%%%%%%%%%%%%%%%%%%%%%%%%%%%%
\subsection{General expression for number of events}
%%%%%%ssss2a%%%%%%%%%%%%%%%%%%%%%%%%%%%%%%%%%%%%%%%%%%%%%%%%%%%%%%%%%%%%%%%%

For the scenarios described in \cref{sec:options}
we will compute the number of expected events of type $s^i$ in $i-$ detector 
$N^i_{s, exp}$, in the following way 
\begin{equation}
N^{i}_{\xi, exp}  =  N^{MB}_{1sh, exp} ~
\frac{N^i_{\xi - s^i}}{N^{MB}_{1sh}}~,
\label{eq:NDevents}
\end{equation}
where $N^{MB}_{1sh, exp}$ is given in (\ref{eq:numbmb}),  
%%is the sum of the $\nu$  and $\bar{\nu}$ excesses of events observed by MB,  
$N^{i}_{\xi - s^i}$ and $N^{MB}_{\xi- 1sh}$ are the theoretical numbers of events 
in a detector $i$ and MiniBooNE correspondingly. 
That is, we normalize the numbers of events of type $\xi - s^i$  
in a given  detector $i$ to the MB excess of 1sh events, $N^{MB}_{1sh}$. 
In this way we ensure that  a given scenario explains the MB excess. 
Furthermore, various factors cancel in the ratio  of predictions 
such as mixing parameter, coupling constants, normalization of cross sections,  {\it etc}. 
%%but other quantities are different in the experiments and
%%should be taken carefully.

The signal in $i-$detector predicted
in terms of the MiniBooNE excess (\ref{eq:NDevents}) is determined by difference
(ratio) of  theoretical values of signals in the $i-$ and MiniBooNE detectors.
(Recall that we are considering experiments with 
qualitatively similar setups.) 
In what follows we will derive general expressions for the numbers of events. 
Apart from the external parameters such as numbers of POT,
$\epsilon$, detector mass $M$, the difference stems from geometry - values of  
the length of decay pipe $l_p^i$, the distance between the end of the pipe and the detector $b^i$, 
so that $l^i = l_{p}^i + b^i$ is
the total baseline, the effective length of a detector $d^i$, 
the energy spectra, and masses of particles involved, in particular
$m_N$, $m_B$. The difference depends on characteristics
of detectors and first of all,
particle ID, efficiencies of event selection etc.,  which is encoded in the signature factors.
Other characteristics cancel. 

For simplicity, superscripts $i$ that indicate  experiment/detector
will be omitted.  We will recover them when needed.

Scenarios for MiniBooNE excess are the chains of interactions
and propagations of new as well as standard model particles. The interactions include
upscattering and decays.  In each interaction one  leading particle, $Y_k$, is absorbed and
another one, $Y_{k + 1}$,  is  produced which eventually gives an observed signal $\xi$ in a detector.
We assume that the leading particles move along the line which  connects the source and detector,
thus neglecting all the  scattering and emission  angles but the angles in the detector.
The latter will be included into significance factors and efficiencies.
%%So, the problem is reduced to 1D problem.
At the same time we will take into account the change of energy of the leading particle
in all interactions. In a given interaction vertex $k$  with coordinate $x_k$ 
a leading particle with energy $E_k$ is absorbed and leading particle with energy $E_{k + 1}$ is emitted.

The general expression for number of events can be written
as a product of  several factors $I_k$ associated to vertices $k$ of interactions.
The initial flux is the flux of $\pi-$ and $K-$ mesons produced at a target
$d \phi^0_{\pi} (E_\pi)/dE_\pi$ and $d \phi^0_{K} (E_\pi)/dE_K$.  
So, the first vertex is $\pi-$ (or $K-$) decay in a decay pipe: $I_1 = D_1$.
There are two possibilities:

1. New particles $N$ or $B$ are produced in these decays.

2.  $\nu_\mu$ is produced and as initial state we can consider
the  $\nu_\mu-$ flux at the exit of the decay pipe $d \phi^0_\nu (E_\pi)/dE_\nu$.
Since $\nu_\mu$ is stable  the first vertex should
be upscattering: $I_1 = U_1$.

In the  1D approximation (straight propagation of  the leading particles) the flux integrated over time
should be multiplied by the area of a detector $A$.
For a vertex with decay the following factors are associated:
\begin{equation}
D_k (E_k)  =
\int dE_k \frac{d \Gamma_k (E_k, E_{k + 1})}{\Gamma^{tot}_k ~dE_{k + 1}}
\int \frac{dx}{\lambda_k}
S_k (E_k, x_k - x_{k - 1}).
\label{eq:dvertex}
\end{equation}
Here
\begin{equation}
S_k (E_k, x_k - x_{k - 1}) \equiv  e^{- (x_k - x_{k - 1})/\lambda_k}, ~~~~
\left(\lambda_k \equiv \frac{E_k}{m_k} c\tau_k^0 \right) 
\label{eq:dvertex1}
\end{equation}
is the survival probability: since particle $Y_k$ (which enters vertex $k$) decays,
it should survive between  $x_{k}$ and the production point
$x_{k - 1}$.
Notice that we can not perform  integration over $E_{k + 1}$ in (\ref{eq:dvertex}),
since other factors in the product of  $I_i$  on
the RHS from a given $I_k$ can depend on $E_{k + 1}$.

For vertex with upscattering of stable particle $Y_k$ the factor reads as
\begin{equation}
U_k (E_k)  =
\int dE_k \frac{d \sigma_k (E_k, E_{k + 1})}{dE_{k + 1}} \int dx_k n_k(x_k),
\label{eq:uvertex}
\end{equation}
where $n_k(x)$ is the density of a layer in which $Y_k$ interacts.
In the case of constant density
the spatial integral can be written as
$$
n_k l_k \int \frac{dx_k}{l_k},
$$
where we introduced $l_k$, the length of layer of the $k$ particle production
to make integrals dimensionless.

If upscattered particle is unstable, a survival probability should be added under spatial integral
in Eq. (\ref{eq:uvertex}):
\begin{equation}
U_k' (E_k)  =
\int dE_k \frac{d \sigma _k (E_k, E_{k + 1})}{dE_{k + 1}}
\int dx_k n_k(x_k) S_k (E_k, x_k - x_{k - 1}) .
\label{eq:duvertex}
\end{equation}

Thus, the general expression for the number of events in a scenario with $n$ vertices can be written as
\begin{equation}
N_{\xi - s}  =
A\int dE_\pi \frac{d \phi^0_\pi (E_\pi)}{dE_\pi} \times
\Pi_{k = 1}^{n - 1} I_k (E_k)  \times I_n (E_n) f_{\xi - s}(E_\xi) \epsilon ,
\label{eq:ngenvertex}
\end{equation}
where $I_k = \{D_k, U_k, U_k' \}$ are introduced in (\ref{eq:dvertex}), 
(\ref{eq:uvertex}) and (\ref{eq:duvertex}). 
This expression can be factorized into the part that  depends
on kinematic variables (energies),  and the propagation part which depends on the coordinates.
In particular, the propagation or decay  part equals
\begin{equation}
P_{dec}  =
\Pi_i  \int \frac{dx_i}{l_i} [S_i(E_i, x_i - x_{i - 1})]^{g_i}~
\Pi_j \int \frac{dx_j}{\lambda_j} S_j(E_j, x_j - x_{j - 1}).
\label{eq:probpart}
\end{equation}
Here the first product of integrals over $i$ corresponds to upscattering vertices with $g = 0$ for stable
and $g = 1$ for unstable upscattered particle $i$.  The second product over $j$ corresponds
to vertices with decays.
In this expression the order and limits of integrations depend on specific scenario.

Spins of the propagating (leading) particles are not important for general
expression (\ref{eq:ngenvertex}). They, however,  are important for characteristics of interactions,
decay rates and cross sections.

%%%%%%%%%%%%%%%%%%%%%%%%%%%%%%%%%%%%%%%%%%%%%%%%%%%%%%%%%%%%%%%%%%%%%%%%%
\subsection{Mixing-Decay,  $M_ND_\xi-$ scenario}
\label{subsec:MD}
%%%%ssmd%%%%%%%%%%%%%%%%%%%%%%%%%%%%%%%%%%%%%%%%%%%%%%%%%%%%%%%%%%%%%%%%%%%%%%%

\begin{figure*}[h!]
\centering
\includegraphics[scale=0.4]{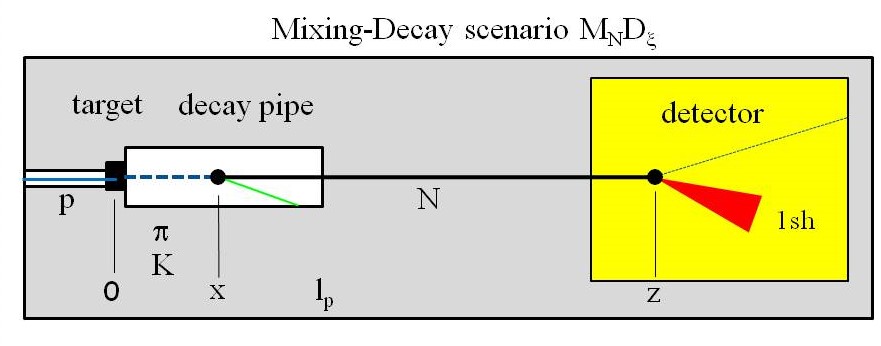}
\caption{
\emph{Schematic depiction of the Mixing-Decay scenario}. 
Black blobs show the interaction points, the red triangle denotes the EM shower,
$l_p$ is the length of decay pipe.
}
\label{fig:MD}
\end{figure*}

Recall that in this scenario (schematically shown in \cref{fig:MD}), the heavy neutrinos, $N$, are produced 
in the $\pi-$ and $K-$ decays via mixing in $\nu_\mu$ in a decay pipe.  
Then $N$ decay ($N \rightarrow \xi + \nu$) along the baseline,
from the production point in a pipe  to a detector.   
Mostly, N decays in a detector that produce the observable events. 
This mechanism gains with respect to upscattering mechanisms since no interactions with matter in a detector
is needed. But it loses because $N$ decays everywhere. 
(One expects lateral phenomena: some signal from $N$-decay outside a detector.) 
As we discussed, the optimal decay length, which maximizes signal, 
is comparable to the baseline $\lambda \sim l$.

%%%%%%%%%%%%%%%%%%%%%ffff1%%%%%%%%%%%%%%%%%%%%%%%%%%%%%%%%%%%%%%%%%%%
%%%%%%%%%%%%%%%%%%%%%%%%%%%%%%%%%%%%%%%%%%%%%%%%%%%%%%%%%%%%%%%%%%%%%%%%%%%%%%%%%%%%%

Due tue to the arrival time restrictions, $m_N < 10 $ MeV, (see sect. IIC),  $N$
is mainly produced in $\pi-$ decays. Therefore the initial flux is 
the $\pi-$ flux at the target 
%\begin{equation}
$d\phi_\pi^0(E_\pi)/dE_\pi$. 
%\label{eq:fkdt1}
%\end{equation}
%%where $d\phi_\pi^0 (E_\pi)/dE_\pi$ is the flux of pions at the target. 
This is the  two vertices scenario with decays in both vertices. 
Then, according to general  formulas (\ref{eq:ngenvertex})
the number of events in detector is given by 
\begin{equation}
N_{\xi - s} =   \epsilon A \int dE_N f_{\xi - s}(E_N) \int dE_\pi
\frac{\phi_\pi^0 (E_\pi)}{dE_\pi} \frac{d \Gamma_{\pi N}(E_\pi, E_N)}{\Gamma^{tot}_\pi dE_N}
P_{dec}(E_\pi, E_N),
\label{eq:final1int}
\end{equation}
where the mixing parameter $|U_{\mu 4}|^2$ is included in the decay rate 
$d \Gamma_{\pi N}/dE_N$. 
The decay factor (\ref{eq:probpart}) equals  the integrals 
\begin{equation}
P_{dec}  =
\int_0^{l_p} \frac{dx}{\lambda_\pi} S_\pi (x)
\int_{l_p +b}^{l_p +b + d} \frac{dz}{\lambda_N} S_N(z - x)\,,
\label{eq:probpart1}
\end{equation}
with the limits of integrations immediately seen in Fig. \ref{fig:MD}.  
In (\ref{eq:probpart1})
$$
S_\pi (x) = e^{-x/\lambda_\pi}, ~~~S_N(z - x) = e^{-(z - x)/\lambda_N} \,.
$$
Explicit computation gives 
\begin{equation}
P_{dec}(E_\pi, E_N)  =
e^{- b/\lambda_N}
\left(1 -  e^{- d/\lambda_N}\right)
\left(1 - \frac{\lambda_\pi}{\lambda_N} \right)^{-1}
\left[e^{- l_p/\lambda_N} - e^{- l_p/\lambda_\pi} \right]\,.
\label{eq:finaldecf}
\end{equation}

Since $\lambda_\pi \ll \lambda_N$ and $\lambda_\pi < l_p$,
the dependence of $P_{dec}$ on $E_\pi$ is weak and $P_{dec}$ can be moved 
out of integration over $E_\pi$, with $E_\pi$ substituted by an effective pion energy
$\bar{E}_\pi$. Then, by introducing the $N-$flux at the target
which would be in the case of stable $N$, 
\begin{align}
\frac{\phi_N^0 (E_N)}{dE_N} \equiv  
%%|U_{\mu N}|^2 
\int dE_\pi \frac{d\phi_\pi^0(E_\pi)}{dE_\pi} 
\frac{d \Gamma_{\pi N} (E_\pi, E_N)}{\Gamma^{tot}_\pi dE_N}, 
\end{align}
the Eq. (\ref{eq:final1int}) can be reduced to 
\begin{equation}
N_{\xi - s} =   \epsilon A 
\int dE_N \frac{d\phi_N^0 (E_N)}{dE_N}f_{\xi - s} (E_N) P_{dec}(\bar{\lambda}_\pi)\,.
%%e^{- l/\lambda_N} \left(1 - e^{- d/\lambda_N} \right).   
\label{eq:final2b}
\end{equation}
Here $\bar{\lambda}_\pi = c\tau_\pi^0 \bar{E}_\pi/m_\pi$. 
If also $d \ll \lambda_N$,  the probability of decay in a detector
is much smaller than 1 and the decay factor
%%(recall $\lambda_{ND} > \lambda_{MB}$)
becomes 
\begin{equation}
P_{dec} \approx   \frac{d}{\lambda_N}  
~e^{- l /\lambda_N}. 
%%~~~ i = ND, MB. 
\label{eq:pdec}
\end{equation}

Qualitatively, the dependence of the predicted numbers of events 
 (\ref{eq:final1int}) on $c\tau_0$ 
can be understood considering the ratio of the decay factors 
(\ref{eq:pdec}) for a given experiment $i$ and 
MiniBooNE taken at certain effective energies 
in experiments, $E^i$ and $E^{MB}$: 
\begin{equation}
r_d \equiv \frac{P_{dec}^i}{P_{dec}^{MB}} 
= \left(\frac{d^{i}}{d^{MB}} \right)
\left(\frac{{E}_N^{MB}}{{E}_N^{i}}\right)
e^{(L^{MB} - L^i)/c \tau^0} ,
\label{eq:rapp1}
\end{equation}
where
\begin{equation}
L^{i} \equiv l^{i} \frac{m_N}{{E}_N^{i}}. 
\label{eq:Lf}
\end{equation}
According to (\ref{eq:Lf}), the dependence of $N^{i}_{s}$ on $c \tau^0$ is determined by
baseline lengths rather than  sizes of detectors.  
Among all the detectors we consider, 
$l$ is the longest and $E_N$ is the smallest for MiniBooNE,
therefore  $L^{MB} > L^i$. 
%%where ${E}_N^{i}$ are averaged energies of $N$.
Numerically,
\begin{equation}
L^{MB} = 6.7~ {\rm m}~ \left(\frac{m_N}{10 ~{\rm MeV}}\right).  \, \, \,
%%L_{ND} = 14~ {\rm m} ~\left(\frac{m_N}{0.15 {\rm GeV}}\right).
\label{eq:Lmbnd}
\end{equation}
 
For $c\tau^0 \gg (L^{MB} - L^i)$, the ratio $r_d$,
and consequently $N^{i}_{\xi - s}$, do  not depend on
$c\tau^0$ as well as $m_N$. In this limit decays of $N$ before the detector can be neglected. 
With decrease
of $c\tau^0$, first the MiniBooNE detection is affected by 
the $N-$decays and then
$i$ detector does. As a result, at
\begin{align}
c\tau^0 <  c\tau^0_{up} \equiv L^{MB} - L^i = 
m_N \left(\frac{l^{MB}}{E_N^{MB}} - \frac{l^{i}}{E_N^{i}} \right)\, 
\end{align}
the ratio turns up and shows exponential growth
(in agreement with figures in \cref{sec:results}).
With increase of $m_N$, the upturn  shifts to larger $c\tau_0$.
The dependence of the number of events on 
$m_N$ is determined in addition 
by the $m_N-$dependence of the  $N-$fluxes, cross sections  and signature factors. 

In the asymptotics $c\tau^0 \gg \Delta L$ the theoretical number of events
can be estimated using (\ref{eq:final2b}) and (\ref{eq:pdec}) as  
\begin{align}
N_{\xi - s} =   \epsilon A d  \frac{m_N}{c \tau^0}
\int dE_N  \frac{1}{E_N} f_{\xi - s}(E_N) \frac{d\phi_N^0 (E_N)}{dE_N} . 
\label{eq:final3}
\end{align}
Then, assuming that $f_{\xi - s}(E_N) = const$,  
the expected number of events  (\ref{eq:final1int}) can be written as 
\begin{equation}
N_{s, 1sh exp}^{i} = N_{exp}^{MB}
\left(\frac{V^{i}}{V^{MB}}\right)
\left(\frac{{E}_N^{MB}}{{E}_N^{i}}\right)
\left(\frac{f_{\xi - s}^{i}}{f_{1sh}^{MB}}\right)
\left(\frac{\epsilon_{\xi - s}^{i}}{\epsilon_{1sh}^{MB}}\right)
\left(\frac{\phi_N^{i}}{\phi_N^{MB}}\right), 
\label{eq:ndass}
\end{equation}
where $V^{i} = A^{i}d^{i}$ is the volume of a detector $i$, and $\phi_N^i \propto \phi_\nu^i$
is the integral flux of $N$ at a detector. 

%%%%%%%%%%%%%%%%%%%%%%%%%%%%%%%%%%%%%%%%%%%%%%%%%%%%%%%%%%%%
\subsection{Upscattering - decay, $U_N D_\xi-$ scenario}
\label{subsec:UD}
%%%%%%%%%ffff%%%%%%%%%%%%%%%%%%%%%%%%%%%%%%%%%%%%%%%%%%%%%%%%%%%

\begin{figure*}[h!]
\centering
\includegraphics[scale=0.4]{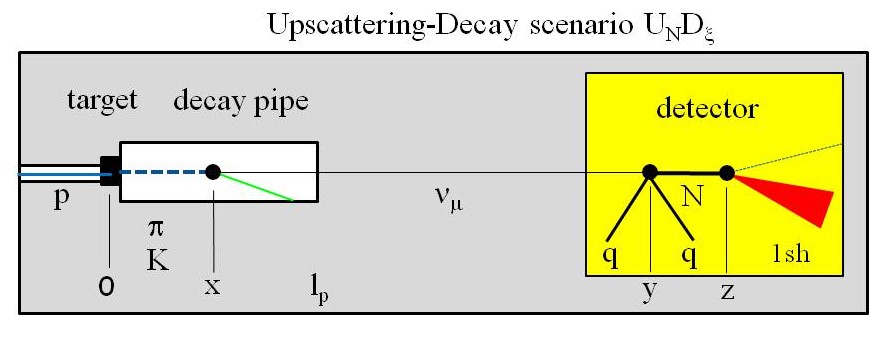}
\caption{The same as in Fig. \ref{fig:MD} but for the 
\emph{Upscattering-Decay scenario}.
%%Black blobs show the interaction points, red triangle denotes the EM shower,
%%$l_p$ is the length of decay pipe.
}
\label{fig:UD}
\end{figure*}

In this scenario (schematically shown in \cref{fig:UD}) $N$ is produced 
by the $\nu_\mu$ upscattering on material 
along a baseline and then it decays as $N \rightarrow \nu + \xi$. 
The $N-$decays inside a detector give an observable signal, while 
$N$ itself can be produced both in the detector and  
in surrounding material. If  $\lambda_N \gg d$, 
a large part of the $N-$flux can be formed outside a detector. 
The initial flux is the $\nu_\mu-$flux at the exit from the decay pipe, 
$d \phi_\nu^0 (E_\nu)/dE_\nu$. 

%%ffff2%%%%%%%%%%%%%%%%%%%%%%%%%%%%%%%%%%%%%%%%%%%%%%%%%%%%%
%%%%%%%%%%%%%%%%%%%%%%%%%%%%%%%%%%%%%%%%%%%%%%%%%%%%%%%%%%%%%%%%%%%%%%

Let us first consider both production and sequential 
decay of $N$ inside a detector.  Following the general formulas in sect. IIIA  
%%and integrating over the energies of  $\nu_\mu$ and $N$  
we obtain the number of $s-$events
\begin{equation}
N^{in}_{\xi - s} =  \epsilon V_d n_d
\int dE_N f_{\xi - s}(E_N)\frac{d \phi_N^\sigma(E_\nu)}{dE_N} 
~P_{dec}^{in}, 
\label{eq:dfact2in}
\end{equation}
where $V_d \equiv A d$ and 
\begin{equation}
\frac{d \phi_N^\sigma (E_N)}{dE_N} \equiv 
\int dE_\nu \frac{d \phi_\nu^0 (E_\nu)}{dE_\nu} 
\, \frac{d\sigma(E_\nu, E_N)}{dE_N}.
\label{eq:dfact3in}
\end{equation}
Notice that $n d \phi_N^\sigma(E_N)/dE_N$ is the density  
of $N-$flux produced in the detector.  In the prefactor of (\ref{eq:dfact2in})
the product $A d n = V_d n = M_d$  gives the mass of a detector.

According to Fig. \ref{fig:UD}, the  decay factor equals 
\begin{equation}
P_{dec}^{in} = 
\int_l^{l+d} \frac{dy}{d}  \int_y^{l+d} \frac{dz}{\lambda_N} S_N(z - y),
\label{eq:dfact3}
\end{equation}
which gives explicitly
\begin{equation}
P_{dec}^{in} =  1 - \frac{\lambda_N}{d} 
\left(1 - e^{-d/\lambda_N}\right).
\label{eq:dfact2}
\end{equation}
In the asymptotics,  $\lambda_N \gg d$, this factor converges to 
\begin{equation}
P_{dec}^{in}  \approx \frac{d}{2\lambda_N} \,, 
\label{eq:dfact3}
\end{equation}
and in the opposite case,  $\lambda_N \ll d$, we have  $P_{dec} \rightarrow 1$.

Let us find the contribution to the number of events in a detector 
from  $N$ produced  in surrounding material (dirt). We denote by $\Delta$ the distance 
between a  detector and dirt (usually the air in a detector pit).   For simplicity we 
consider uniform surrounding medium 
%%between the end of the decay tunnel and detector (length $b$) 
with density $n_b$ and length $b$. 
Similarly to (\ref{eq:dfact2in}) 
%the number of $N-$decays inside  a detector, which gives 
the number of observable events, equals
\begin{equation}
N^{out}_{\xi - s} = \epsilon  N_b  
\int dE_N \frac{d \phi_N^\sigma (E_N)}{dE_N} f_{\xi - s}
P_{dec}^{out}(E_N), 
\label{eq:dfact}
\end{equation}
where  $N_b = n_b A b$ is the number of scatterers in medium 
The  decay factor $P_{dec}^{out}$ differs from $P_{dec}^{in}$ by limits of integration:
\begin{equation}
P_{dec}^{out} =
\int_{l_p}^{l_p + b} \frac{dy}{b}  \int_{l_p + b + \Delta}^{l_p + b + \Delta + d} 
\frac{dz}{\lambda_N} S_N(z - y),  
\label{eq:dfact3}
\end{equation}
which gives 
\begin{equation}
P_{dec}^{out}(E_N) = \frac{\lambda_N}{b} e^{-\Delta/\lambda_N} \left(1 - e^{-b/\lambda_N} \right)  
\left(1 - e^{-d/\lambda_N} \right)\, . 
\label{eq:outdfac}
\end{equation} 
Here $e^{-\Delta/\lambda_N} $ is the survival probability of $N$ between 
the end of dirt and the detector.  
If a detector and a pit have non-rectangular form, the parameters 
$\Delta$ and $d$ depend on the distance to the center (axis) 
of the setup $h$,  and one needs to integrate over $h$.

In the limit $b \gg \lambda_N$ we obtain
\begin{equation}
N^{out}_{\xi - s}   =  A n_b \epsilon  
\int dE_N  \lambda_N \frac{d \phi_N^\sigma (E_N)}{dE_N} f_{\xi - s}
e^{-\Delta/\lambda_N} \left(1 - e^{-d/\lambda_N} \right). 
\label{eq:dfactout}
\end{equation}
In this limit, the $N-$flux is collected along the distance 
of the order $\lambda_N$ in front of a detector. 

The total number of events due to $N$ produced in a detector 
and surrounding materials  can be written as 
\begin{equation}
N^{tot}_{\xi - s} =  N^{in}_{\xi - s} + N^{out}_{\xi - s} = 
A d n_d \epsilon  \int dE_N \frac{d \phi_N^\sigma (E_N)}{dE_N} f_{\xi - s}
\left(P_{dec}^{in} + \frac{b n_b}{d n_d} P_{dec}^{out}\right), 
\label{eq:totn1}
\end{equation}
or explicitly, 
\begin{equation}
N^{tot}_{\xi - s} =  A d n_d \epsilon  
\int dE_N \frac{d \phi_N^\sigma (E_N)}{dE_N} f_{\xi - s}
\left\{
1 +  \frac{\lambda_N}{d} \left(1 - e^{-d/\lambda_N} \right)
\left[\frac{n_b}{n_d} e^{-\Delta/\lambda_N} \left( 1 - e^{-b/\lambda_N} \right) - 1 \right]
\right\}.
\label{eq:totnee}
\end{equation}
In the limit $b \gg \lambda_N$ the number of events equals
\begin{equation}
N^{tot}_{\xi - s} =  A d n_d \epsilon  
\int dE_N \frac{d \phi_N^\sigma (E_N)}{dE_N} f_{\xi - s}
\left[
1 +  \frac{\lambda_N}{d} \left(1 - e^{-d/\lambda_N} \right)
\left(e^{-\Delta/\lambda_N} \frac{n_b}{n_d}  - 1 \right)
\right].
\label{eq:totnee2}
\end{equation}
For $\lambda_N >  d$ and  $\Delta < \lambda_N$ the contribution from  
dirt can be several times larger than 
the one from a detector.

Let us consider the dependence of numbers of events (\ref{eq:totnee2}) on $c \tau^0$.   
It  is largely determined by the ratios of decay factors
for the detector $i$ and MiniBooNE taken at certain effective 
energies $E^{MB}_N$  and $E^{i}_N$. 
%%The energies depend on $m_N$.  
%%Two contributions discussed above have different 
%%dependences on $c \tau^0$ and we will consider them in oder. 
For the contribution due to $N$ production inside a detector $i$, the  
dependence of the number of events on $c \tau^0$ is determined 
by the ratio of decay factors $P_{dec}^{in}$ (\ref{eq:dfact2}) which can be written as  
\begin{equation}
r_{dec} = \frac{1 - \frac{c\tau^0}{D^i} \left(1 - e^{-D^i/ c\tau^0}\right)}
{1 - \frac{c\tau^0}{D^{MB}} \left(1 - e^{-D^{MB}/c\tau^0}\right)}, 
\label{eq:decrat2}
\end{equation}
%%According to (\ref{eq:dfact2}) the predictions 
where 
\begin{align}
D^i \equiv  d^i \frac{m_N}{E^i}\,,
\end{align}
are the ``reduced" sizes of detectors 
($d/\lambda = D/c\tau^0$). 
Among the experiments we consider,  
MiniBooNE has the largest reduced size, $D^{MB} > D^i$. 
Numerically, for MiniBooNE ($d_{MB} = 8$ m and $E^{MB}_N = 0.8$ GeV) we obtain    
\begin{equation}
D_{MB} = 1.5 {\rm m}  \left(\frac{m_N}{0.15~ {\rm GeV}}\right). 
\label{xi-minb}
\end{equation}
Taking this into account we find from (\ref{eq:decrat2})\\

$(i)$ for $c \tau^0 < D^{i}$ m both decay probabilities 
(for MiniBooNE and $i$ detector) are close to 1, so that $r_{dec} \approx 1$. 
Consequently, the ratio of number of events 
does not depend on $c\tau^0$ as well as on $m_N$. 
The dependence of expected number of events on  $m_N$ follows from fluxes and cross sections.  

$(ii)$ In the interval $D^i < c\tau^0 < D^{MB}$, 
$N$ still has space  to decay in MiniBooNE and $P^{MB}_{dec} \sim 1$, while   
the $N-$decay length becomes larger than $i$ detector length and therefore 
$P^{i}$  decreases. As a result, the number of $i$ detector 
events should decrease. 

$(iii)$ For $c\tau^0 > D_{MB}$  the particles  
$N$ decay  only partially  in both detectors,  
and the ratio of decay factors  converges to  
\begin{equation}
r_{dec}^{\infty}   = \frac{P^{i}_{dec}}{P^{MB}_{dec}} = 
\frac{D^{i}}{D^{MB}} = 
\frac{d^{i} E^{MB}_N}{d^{MB} E^{i}_N}.  
%%\approx 0.07. 
\label{eq:rdecmusc}
\end{equation}
Again, dependences of $r_{dec}$ and prediction of the number of events 
on $c\tau^0$ as well as on $m_N$ disappear. 

In the limit $c\tau^0 \rightarrow 0$ the decay factors 
$P_{dec} \approx 1$  and the number of events can be estimated as 
\begin{equation}
N_{\xi - s}^{ND} = N_{1sh, exp}^{MB}
\left(\frac{M^{i}}{M^{MB}}\right)
\left(\frac{f_{\xi - s}^{ND}}{f_{1e}^{MB}}\right)
\left(\frac{\epsilon_{\xi - s}^i}{\epsilon_{1sh}^{MB}}\right)
\left(\frac{\sigma^i}{\sigma^{MB}}\right)
\left(\frac{\phi_\nu^i}{\phi_\nu^{MB}}\right),   
\label{eq:ndass0}
\end{equation}
%%where the ratio of the $\nu_\mu$ fluxes  was estimated 
as $\phi_\nu^{i} \propto (POT)^{i}$ \cite{Abe:2020vot}.

For $N$ production in the dirt and then decay in a detector we have 
\begin{equation}
r_{dec} = 
\frac{\frac{\lambda_N^i}{d^i} \left(1 - e^{-d^i/\lambda_N^i} \right)
\frac{n_b^i}{n_d^i} e^{-\Delta^i/\lambda_N^i} \left( 1 - e^{-b^i/\lambda_N^i} \right)}
{1 +  \frac{\lambda_N^{MB}}{d^{MB}} \left(1 - e^{-d^{MB}/\lambda_N^{MB}} \right)
\left[\frac{n_b^{MB}}{n_d^{MB}} e^{-\Delta^{MB}/\lambda_N^{MB}} 
\left( 1 - e^{-b^{MB}/\lambda_N^{MB}} \right) - 1 \right]}\,.
\label{eq:rdec4}
\end{equation}
Now, the decay factor (\ref{eq:outdfac}) is proportional to $\lambda_N$ and   
in the limit $c \tau^0 \rightarrow 0$ the ratio (\ref{eq:rdec4})  equals 
\begin{align}
r_{dec}^0 = \frac{\lambda_N^i}{d^i} \frac{n_b^i}{n_d^i}\,,
\end{align}
so that the  contribution from dirt  vanishes. 
In the opposite limit, $c \tau^0 \rightarrow \infty$,  we have  
$$
r_{dec}^{\infty} = 
\left(\frac{n_d^{MB}}{n_d^{i}}\right)
\left(\frac{n_b^{i}}{n_b^{MB}}\right)
\left(\frac{b^{i}}{b^{MB}}\right)
\left(\frac{\lambda_N^{MB}}{\lambda_N^{i}}\right).
$$
That is, the dirt contribution converges to a constant.

%%%%%%%%%%%%%%%%%%%%%%%%%%%%%%%%%%%%%%%%%%%%%%%%%%%%%%%%%%%%%%%%%%%%%%%%%%%
\subsection{Upscattering - Double Decay scenario, $U_N D_B D_\xi-$scenario.}
%%%%%%%%%%%%%%%%%%%%%%%%%%%%%%%%%%%%%%%%%%%%%%%%%%%%%%%%%%%%%%%%%%%%%%%%%%%%%

\begin{figure*}[h!]
\centering
\includegraphics[scale=0.4]{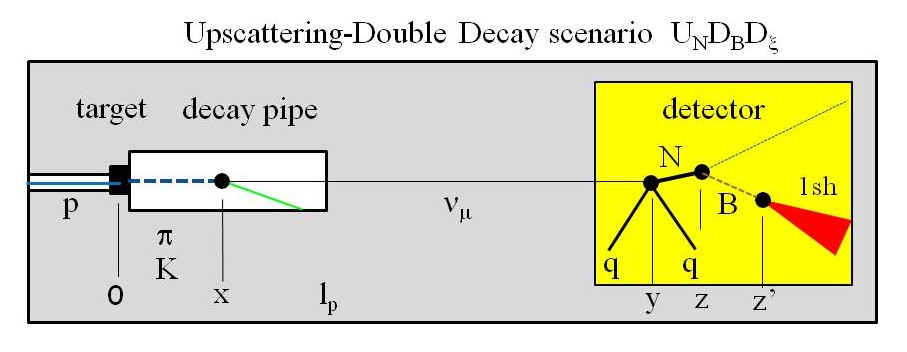}
\caption{
\emph{Upscattering -  Double Decay scenario}.
Black blobs show the interaction points, the red triangle denotes the EM shower,
$l_p$ is the length of decay pipe.
}
\label{fig:UDD}
\end{figure*}

This scenario (schematically shown in \cref{fig:UDD})
has three vertices with one $\nu_\mu-$upscattering and two sequential decays. 
The initial state and initial part are the same as in the previous scenario. 
When $B$ decays promptly,  this scenario  
is similar to the $U_ND_\xi$ described in  \cref{subsec:UD}.  
In this case the only but rather relevant difference is that the  
invariant mass of  $\xi$  is fixed by the mass of a boson, $B$: $W_\xi = m_B$.  
The latter  can be substantially smaller than the mass of $N$ which affects the signature factor.  
In \cref{sec:results} we will show results for short $B$ lifetime.

%%%%%%%%%ffff3%%%%%%%%%%%%%%%%%%%%%%%%%%%%%%%%%%%%%%%%%%%%%%%%%%%%%%%%%%%%%%%
%%%%%%%%%%%%%%%%%%%%%%%%%%%%%%%%%%%%%%%%%%%%%%%%%%%%%%%%%%%%%%%%%%%%%%%%%%%%%%%%%%%%%%%%%%%%

In what follows we will consider the new contribution from $N-$ production outside a detector. 
The number of expected events can be written as 
\begin{equation}
N^{out}_{\xi - s} = \epsilon  N_b
\int dE_B f_{\xi - s}(E_B)
\int dE_N \frac{d \phi_N^\sigma (E_N)}{dE_N}  \frac{d\Gamma_N (E_N, E_B)}{\Gamma_N^{tot} dE_B}   
P_{dec}^{out}(E_N, E_B),
\label{eq:nout}
\end{equation}
where $d\phi_N^\sigma /d E_N$ was defined in (\ref{eq:dfact3in}), 
and additional integration was introduced over $dE_B$.  
The distribution $d\Gamma_N (E_B, E_\xi)/\Gamma_N^{tot} dE_\xi$ is included in $f_{\xi - s}(E_B)$. 
The decay factor is given by 
\begin{equation}
P_{dec}^{out}(E_N, E_B) = \int_{l_p}^{l_p +b} \frac{dy}{b} 
\int_{l}^{l +d} \frac{dz'}{\lambda_B} 
\int_{y}^{z'} \frac{dz}{\lambda_N} S_N (z-y) S_B (z' - z). 
\label{eq:outdfac1}
\end{equation}
Here $S_N  = e^{-(z-y)/\lambda_N}$, $S_N  = e^{-(z' - z)/\lambda_B}$ and the 
limits of integrations can be immediately read off from 
Fig. \ref{fig:UDD},  but with the $\nu_\mu-$upscattering in a dirt. 
Explicit integration gives 
\begin{equation}
P_{dec} =
\frac{\lambda_N^2}{(\lambda_N - \lambda_B) b}
\left[1 - e^{- b/\lambda_N)} \right]
\left(1 - e^{- d/\lambda_N} \right) + 
\frac{\lambda_B^2}{(\lambda_B - \lambda_N) b}
\left[1 - e^{- b/\lambda_B)} \right]
\left(1 - e^{- d/\lambda_B} \right). 
\label{eq:forn2}
\end{equation}
The decay factor is symmetric with respect to interchange 
$\lambda_N  \leftrightarrow \lambda_B$. In the limit $\lambda_B \rightarrow 0$ 
(fast $B-$decay) it coincides with $P^{out}_{dec}$ in Eq. (\ref{eq:outdfac}). 
The result is symmetric with respect to $N$ and $B$. 

If $\lambda_B = \lambda_N = \lambda$,  we obtain 
\begin{equation}
P_{dec} =
\frac{2\lambda}{b}
\left(1 - e^{- b/\lambda)} \right)
\left(1 - e^{- d/\lambda} \right). 
\end{equation}

%%%%%%%%%%%%%%%%%%%%%%%%%%%%%%%%%%%%%%%%%%%%%%%%%%%%%%%%%%%%%%
\subsection{Mixing - Decay into $\nu_e$, $M_N D_\nu U_e-$scenario.} 
\label{subsec:jk}
%%%%%%%%%%%%%%%%%%%%%%%%%%%%%%%%%%%%%%%%%%%%%%%%%%%%%%%%%%%%%%

\begin{figure*}[ht!]
\centering
\includegraphics[scale=0.4]{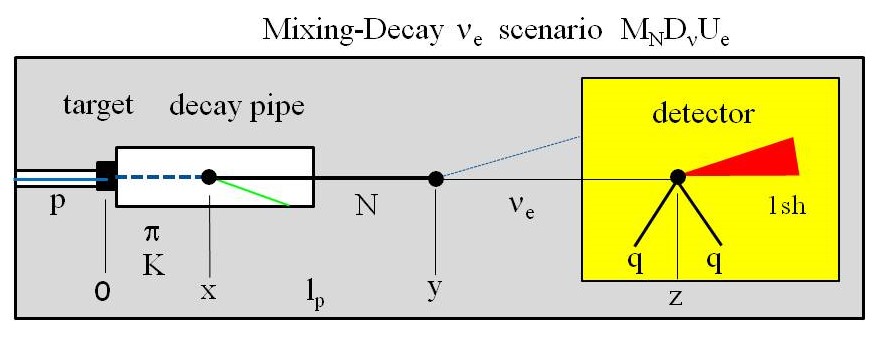}
\caption{The same as in Fig. \ref{fig:MD} but for the 
\emph{Mixing - Decay into $\nu_e$ scenario}.
%%Black blobs show the interaction points, red triangle is the EM shower,
%%$l_p$ is the length of decay pipe.
}
\label{fig:Kopp-a}
\end{figure*}

This scenario (schematically shown in \cref{fig:Kopp-a}) essentially provides an additional source of $\nu_e$
at low energies. Therefore, there is no restriction
from angular dependence  of the observed MiniBooNE events, but   
$N$ should be light enough  to satisfy the timing bound.  
Therefore it is dominantly produced 
in the $\pi-$decay. 

Relatively light $N$  produced via mixing with $\nu_\mu$ 
decays into  $\nu_e$ and a new light scalar or vector 
boson along the beamline: $N \rightarrow \nu_e + B$. 
In turn, the bosons $B$ may decay
into $\nu_e \bar\nu_e$ pair, thus enhancing 
the $\nu_e-$flux at low energies.
Here there are  more interaction points in comparison  to previous 
scenarios (although in one point the interactions are standard). 

%%%%%%%%%%%%%%%%%%%%%%%%%%%%%%ffff4%%%%%%%%%%%%%%%%%%%%%%%%%%%%%%%%%%%
%%%%%%%%%%%%%%%%%%%%%%%%%%%%%%%%%%%%%%%%%%%%%%%%%%%%%%%%%%%%%%%%%%%%%%

Since $N$ can decay already in the decay tunnel, 
the consideration should start from $\pi-$decay as 
in the $M_N D_\xi$ scenario 
of  \cref{subsec:MD}. 
In contrast to $M_ND_\xi$,  decay of $N$ in a pipe 
do contribute to the observable signal, since 
$\nu_e$ are stable and  can travel to a detector. 
This requires different consideration from  
$M_N D_\xi$.

The initial flux is the pion flux produced in a
proton target  $\phi_\pi^0$. Then using general formulas of sect.  IIIA  
we can write expression for the number of expected events 
\begin{eqnarray}
N_{e - 1sh} =  \epsilon A d n_d 
%%|U_{\mu N}|^2 
\int dE_\nu \sigma^{CC} (E_\nu) f_{e - 1sh}(E_\nu)
\int dE_N   \frac{d\Gamma_N(E_N, E_\nu)}{\Gamma_N^{tot}dE_\nu}
\nonumber\\
\times \int dE_\pi \frac{d \phi_\pi^0(E_\pi) }{d E_\pi}
\frac{d \Gamma_{\pi N} (E_\pi, E_N)}{\Gamma_\pi^{tot} dE_N} 
P_{dec}(\lambda_\pi, \lambda_N).  
\label{eq:nuenumb2}
\end{eqnarray}
%%We do not introduce integration over the electron momentum,
%%considering that all electrons, that appear, produce
%%the observable signal. (Otherwise, one should use the differential
%%neutrino cross section $d\sigma (E_\nu, E_e)/dE_e$
%%and perform integration over $E_e$ from certain
%%experimental threshold energy, say $E_e = 100$ MeV).
%%
The decay factor equals
%\begin{equation}
$$
P_{dec} = 
\int_l^{l+d} \frac{dz}{d} \int_0^{l_p} \frac{dx}{\lambda_\pi} \int_x^z \frac{dy}{\lambda_N}
e^{- x/\lambda_\pi}
~e^{- (y - x)/\lambda_N},  
$$
%\label{eq:pdecint}
%\end{equation}
and explicitly, 
\begin{equation}
P_{dec} = \left[1 -  e^{-l_p/\lambda_\pi} - 
g (\lambda_\pi, \lambda_N) \frac{\lambda_N}{d} e^{-b/\lambda_N}
\left( 1 - e^{-d/\lambda_N}\right) \right],
\label{eq:decfact3}
\end{equation}
where 
\begin{equation}
g(\lambda_\pi, \lambda_N) = \left(1 - 
\frac{\lambda_\pi}{\lambda_N} \right)^{-1}
\left[e^{- l_p/\lambda_N} - e^{- l_p/\lambda_\pi} \right].
\label{eq:gbbb}
\end{equation}

If $d \ll \lambda_N$, the equation (\ref{eq:decfact3}) reduces to
\begin{equation}
P_{dec} \approx 
\left(1 -  e^{-l_p/\lambda_\pi} - g(\lambda_\pi, \lambda_N) e^{-b/\lambda_N} \right).
\label{eq:nuenumbif}
\end{equation}
Let us consider two limits of this result: 

1) $\lambda_N \rightarrow 0$ (very fast $N-$ decay): we have from (\ref{eq:nuenumbif})
%\begin{equation}
$$
P_{dec} \approx \left(1 -  e^{-l_p/\lambda_\pi}\right) \,,
$$
%\label{eq:nuenumb3}
%\end{equation}
which is nothing but the decay probability of pions in a pipe. It gives 
the $\nu_\mu-$flux at a detector.\\

2) $\lambda_N \rightarrow \infty$ (very slow $N-$decay):  
in the lowest order in $l/\lambda_N$ we find from (\ref{eq:nuenumbif})
%\begin{equation}
$$
P_{dec} \approx \left(1 -  e^{-l_p/\lambda_\pi}\right)
\frac{l^{eff}}{\lambda_N},
$$
%\label{eq:nuenumbz}
%\end{equation}
where $l^{eff}$ is the effective baseline: 
\begin{equation}
l^{eff} \equiv b  + l_p \left( 1 - e^{-l_p/\lambda_\pi} \right)^{-1} 
- \lambda_\pi.
\label{eq:iieff}
\end{equation}
In the limits $\lambda_\pi \rightarrow 0$ 
and $\lambda_\pi \rightarrow \infty$  this equation gives   
$l^{eff} =  b  + l_p$ and $l^{eff} = b$ 
correspondingly. For a typical situation with $\lambda_\pi = l_p$ we find
from (\ref{eq:iieff})
\begin{equation}
l^{eff} =  b  + l_p (e - 1)^{-1} \approx b + 0.58\, l_p\,.
\label{eq:iieffop}
\end{equation}

If $c \tau^0 \rightarrow 0$, the ratio of decay factors converges to $r_{dec}^0 = 1$, while
for $c \tau^0 \rightarrow \infty$
\begin{align}
r_{dec}^{\infty} = 
\frac{E^i}{E^{MB}}
\frac{z^i (b^i + \lambda^i_\pi) + l_p^i}{z^{MB} (b^{MB} + \lambda^{MB}_\pi) + l_p^{MB} }\,,
\label{eq:xx}
\end{align}
where $z^i \equiv (1 - e^{l_p^i/\lambda_\pi^i})$. 
Consequently, in both limits the number of events does not depend on 
$c \tau^0$.

%%%%%%%%%%%%%%%%%%%%%%%%%%%%%%%%%%%%%%%%%%%%%%%%%%%%%%%%%%%%%%%%%%%%%%%%%
\subsection{Mixing-Double Decay scenario,  $M_N D_B D_\xi$}
\label{subsec:5th}
%%%%ssmd%%%%%%%%%%%%%%%%%%%%%%%%%%%%%%%%%%%%%%%%%%%%%%%%%%%%%%%%%%%%%%%%%%%%%%%

\begin{figure*}[h!]
\centering
\includegraphics[width=0.57\textwidth]{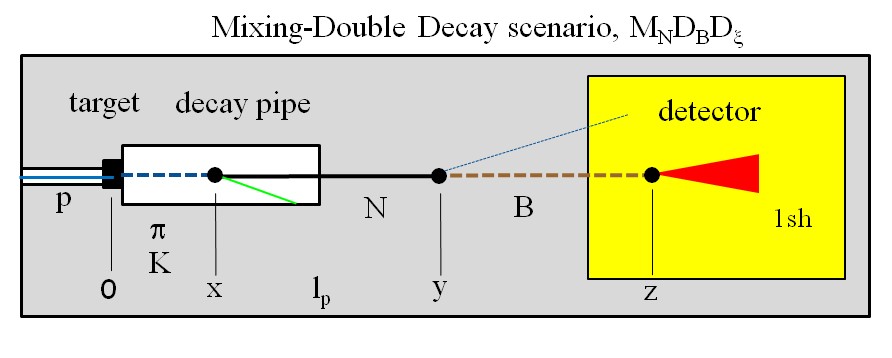}
\caption{
The same as in Fig.1 but for the \emph{ Mixing-Double Decay scenario}. }
\label{fig:scenE}
\end{figure*}

According to this scenario,  $N$ is produced
in the $\pi-$ and $K-$decays via mixing in $\nu_\mu$ within a decay pipe.
Then $N$ decays  along the baseline with emission of boson $B$, 
$N \rightarrow \nu + B$,
and the latter decays  $B \rightarrow \xi $  or $B \rightarrow \xi + B'$.
The $B-$ decay should occur  in a detector(see Fig. \ref{fig:scenE}).
This scenario  reproduces various features of the previously described scenarios:
in particular, for fast decaying $B$, $\lambda_B \ll d$,  
it is reduced to the $M_N D_\xi-$ scenario.

%%%%%ffff5%%%%%%%%%%%%%%%%%%%%%%%%%%%%%%%%%%%%%%%%%%%%%
%%%%%%%%%%%%%%%%%%%%%%%%%%%%%%%%%%%%%%%%%%%%%%%%%%%%%%%%
The initial flux is the flux of pions (also K-mesons) produced in the target.
All three processes involved are decays. According to Fig. \ref{fig:scenE}
the limits of integrations  are the following: 
The coordinate  of $\pi-$ ($K-$) decay 
is in the interval  $x = [0 -  l_p]$; the coordinate of $N-$ decay (and production of $B$) $y = [x -  z]$; 
the point of $B$ decay should be within the detector: $z = [l - (l + d)]$.
With this, according to the general formulas in Sect. IIIA  
the expression for the number of events can be written as
\begin{equation}
N_{\xi - s} =   \epsilon^i_s A
%%|U_{\mu 4}|^2 
\int dE_\pi \frac{\phi_\pi^0 (E_\pi)}{dE_\pi}
\int dE_N
\frac{d \Gamma_{\pi N} (E_\pi, E_N)}{\Gamma^{tot}_\pi~ dE_N}
\int dE_B
\frac{d \Gamma_N (E_N, E_B)}{\Gamma^{tot}_N ~dE_B}
f_{\xi - s}(E_B) P_{dec}(E_\pi, E_N, E_B).
\label{eq:numbint1}
\end{equation}
Here the mixing parameter $|U_{\mu 4}|^2$ is included in $d\Gamma_\pi/dE_N$. 
The decay factor (\ref{eq:probpart}) equals
\begin{equation}
P_{dec}(E_\pi, E_N, E_B)  = \int^{l_p}_0 \frac{dx}{\lambda_\pi}
\int^{l + d}_l \frac{dz}{\lambda_B} \int^{z}_x \frac{dy}{\lambda_N}
e^{-x/\lambda_\pi} e^{-(y - x)/\lambda_N} e^{-(z - y)/\lambda_B}.
\label{eq:decf1}
\end{equation}
%%The propagation (decay) factor is given by integration of product of survival
%%probabilities between the decay points.
%%The limits integration and the
%%order of integration depend on observational features.
Explicit integration over coordinates gives
\begin{equation}
P_{dec} =   \frac{\lambda_N}{\lambda_N - \lambda_B} P_{MD}(\lambda_N) +
\frac{\lambda_B}{\lambda_B - \lambda_N} P_{MD}(\lambda_B),
\label{eq:sumnb1}
\end{equation}
where $\lambda_B = (E_B/m_B) c \tau_B^0$ and
\begin{equation}
P_{MD}(\lambda) =
\frac{1}{(1 - \lambda_\pi/\lambda)}
e^{-l/\lambda}
\left[1 - e^{- l_p (1/\lambda_\pi - 1/\lambda)} \right]
\left(1 - e^{- d/\lambda} \right),
\label{eq:forn2}
\end{equation}
which coincides with the decay factor  in the $M_ND_\xi-$scenario Eq. (\ref{eq:finaldecf}).
Notice that the expression in (\ref{eq:sumnb1}) is symmetric with respect to
$\lambda_N \leftrightarrow \lambda_B$.

The scenario is determined by 4 parameters $c\tau_N^0$, $m_N$, $c\tau_B^0$, $m_B$.
In the limit $\lambda_B \rightarrow 0$ (very fast $B-$decay),
$P_{dec} \rightarrow  P_{dec}^N (\lambda_B = 0)$ and the latter coincides 
with expression (\ref{eq:finaldecf}) 
for $M_N D_\xi$ scenario.  In the limit  $\lambda_N \rightarrow 0$ (very fast $N$ decay)
$P_{dec} \rightarrow  P_{dec}^B (\lambda_N = 0)$. That is, we obtain the same
expression (\ref{eq:finaldecf}) with just substitution $\lambda_N  \rightarrow \lambda_B$.

Let us consider the case $\lambda_N = \lambda_B$ which 
is reduced to 2 parameters case and
one expects the largest deviation from the result of the $M_N D_\xi$ scenario.
In the limit $\lambda_B \rightarrow \lambda_N$ we can expand
\begin{equation}
P_{MD}(\lambda_B) = P_{MD}(\lambda_N) + \left .\frac{dP_{MD}}{d\lambda_B}\right|_{\lambda_B =
\lambda_N}(\lambda_B - \lambda_N).
\label{eq:expansion}
\end{equation}
Inserting this expression into (\ref{eq:sumnba}) we find
\begin{equation}
P_{dec}(\lambda) =   P_{MD}(\lambda_N) + \lambda \frac{dP_{MD}}{d\lambda},
\label{eq:sumnba}
\end{equation}
which gives
\begin{equation}
P_{dec} = P_{MD}(\lambda)
\left[
1 - \frac{\lambda_\pi}{\lambda - \lambda_\pi}
+ \frac{l}{\lambda}
+ \frac{l_p}{\lambda} \frac{1}{e^{l_p (1/\lambda_\pi - 1/\lambda)} -1}
- \frac{d}{\lambda} \frac{1}{e^{d/\lambda}  - 1}
\right]. 
\label{eq:pprop1}
\end{equation}
For small size detector, $d \ll \lambda$, we find 
\begin{equation}
P_{dec} \approx  P_{MD}(\lambda)
\left[
- \frac{\lambda_\pi}{\lambda - \lambda_\pi}
+ \frac{l}{\lambda}
+ \frac{l_p}{\lambda} \frac{1}{e^{l_p (1/\lambda_\pi - 1/\lambda)} -1}
\right].
\label{eq:pprop2}
\end{equation}
If $\lambda_\pi \ll \lambda$,  it can be rewritten as
\begin{equation}
P_{dec} \approx  P_{MD}(\lambda) \frac{L(\lambda, l, l_p)}{\lambda}, 
\label{eq:pprop3}
\end{equation}
where
%\begin{equation}
$$
L(l, l_p) = \left(l + \frac{l_p}{e^{l_p/\lambda_\pi} - 1}  - \lambda_\pi
\right).
$$
%\label{eq:defL}
%\end{equation}
The more precise expression  weakly depends on $\lambda$, and in the first approximation $L = l$.

Using similar approximations in $P_{MD}(\lambda)$ we obtain explicitly
\begin{equation}
P_{dec} \approx  h \frac{L}{\lambda} \frac{d}{\lambda} e^{- l/\lambda}
\label{eq:pprop3}
\end{equation}
and $h \approx 1 - e^{l_p / \lambda_\pi}  \approx 1$.
The ratio of the decay factors (\ref{eq:pprop3}) for a given detector $i$ and MiniBooNE
can be written as
\begin{equation}
\frac{P_{dec}^i}{P_{dec}^{MB}} =
\left(\frac{d^i}{d^{MB}}\right)
\left(\frac{L^i}{L^{MB}}\right)
\left(\frac{E^{MB}}{E^i}\right)^2
\exp (l^{MB}/\lambda^{MB} - l^{i}/\lambda^{i}).
\label{eq:ratprob}
\end{equation}
As in the $M_N D_\xi$ scenario, the dependence of number of events 
on $c\tau^0$ shows up via the exponential upturn determined 
by the MiniBooNE parameters 
$l^{MB}$ and $\lambda^{MB}$ and constant asymptotics for large $c\tau^0$.  
The difference in comparison to the $M_N D_\xi$ scenario 
is the appearance of the additional factor
\begin{equation}
\frac{L^i}{L^{MB}} \frac{E^{MB}}{E^i}.
\label{eq:dopfac}
\end{equation}
%in comparison to $M_N D_\xi$. 
For ND280 this factor equals 0.4.

So, in all these special cases $P_{dec}$ are  reduced to the two-parameters 
expression (\ref{eq:finaldecf}).

%%%%%%%%%%%%%%%%%%%%%%%%%%%%%%%%%%%%%%%%%%%%%%%%%%%%%%%%%%%%%%%%%%%%%%%%%%%
\subsection{Upscattering - Decay into $\nu_e$ scenario, $U_N D_\nu U_e$}
%%%%%%%%%%%%%%%%%%%%%%%%%%%%%%%%%%%%%%%%%%%%%%%%%%%%%%%%%%%%%%%%%%%%%%%%%%

\begin{figure*}[h!]
\centering
\includegraphics[width=0.57\textwidth]{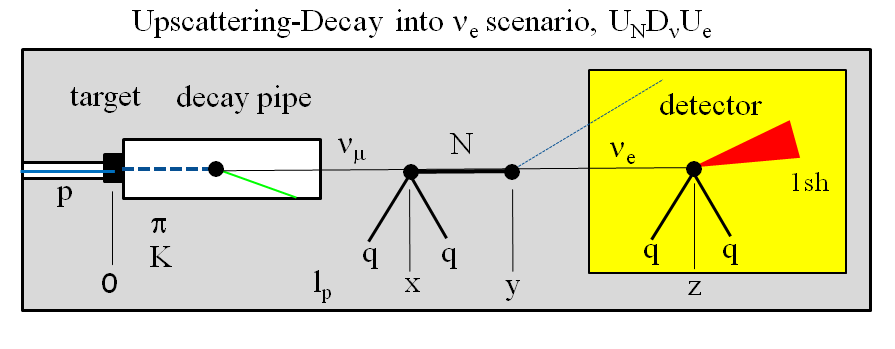}
\caption{
The same as in Fig.1 but for {\it Upscattering - Decay into $\nu_e$ scenario}. }
\label{fig:scenF}
\end{figure*}

Here $N$  is produced via the $\nu_\mu-$ upscattering (point $x$) 
outside the decay pipe(see Fig. \ref{fig:scenF}).
It  decays  into  $\nu_e$ (the point $y$)  and new light scalar or vector
boson $N \rightarrow \nu_e + B$. Then $\nu_e$ via the CC interactions
produces electron in a detector (point $z$).

It is similar  to the $M_N D_\nu U_e-$scenario, where the $N$ production via mixing is substituted by
$\nu_\mu-$upscattering. That can bring a smallness as we discussed in Sect. II.
In contrast to  $M_N D_\nu U_e-$ scenario, here there is no  production of  $N$ in a  decay pipe.
There are two standard model vertices with production of
$\nu_\mu$ and upscatering of $\nu_e$. The non-standard interactions appear in 
production and decay of $N$.

%%%%%ffff6%%%%%%%%%%%%%%%%%%%%%%%%%%%%%%%%%%%%%%%%%%%%%
%%%%%%%%%%%%%%%%%%%%%%%%%%%%%%%%%%%%%%%%%%%%%%%%%%%%%%%%

Since $N-$production via the $\nu_\mu-$upscattering occurs outside a decay pipe,
we can use the $\nu_\mu$ flux at the exit
from the  pipe $d\phi^0(E_{\nu_\mu}, l_p)/dE_{\nu_\mu}$ as the initial flux.
Therefore according to the  general consideration in Sect. IIA,
the number of events can be written as
\begin{eqnarray}
N_{e - s} = \epsilon A
\int dE_{\nu_\mu} \frac{d\phi^0_{\nu_\mu} (E_{\nu_\mu})}{dE_{\nu_\mu}}
\int dE_N \frac{d\sigma (E_{\nu_\mu}, E_N)}{d E_N} n_N l_N
\nonumber\\
\times
\int dE_{\nu_e} \frac{d\Gamma_N (E_N, E_{\nu_e})}
{\Gamma^{tot}_N dE_{\nu_e}} \sigma(E_{\nu_e}) n_d d~
f_{e-s}(E_{\nu_e})  P_{dec}.
\label{eq:n-udn}
\end{eqnarray}
Here $n_N$ and  $l_N$ are the density and the length
of  a layer in which $N$ production occurs.
In Eq. (\ref{eq:n-udn}) we used the integrated $\nu_e$
cross section and effective (integrated) signature factor, without
introducing dependences on the electron energy.
Since only one unstable particle ($N$) is involved,
the decay factor $P_{dec}$ depends on a single
survival probability $S_N(y - x)$.
%%\begin{equation}
%%P_{dec} = \int \frac{dx}{l_N}
%%\int \frac{dy}{\lambda_N}
%%\int \frac{dz}{d} e^{-(y - x)/\lambda_N}.
%%\label{eq:prop-udn}
%%\end{equation}

There are two contributions to the total number of events
related to the $N$ production in a dirt (outside a detector) and
in a detector. For simplicity we consider a dirt as uniform medium with density $n_b$.
For the first contribution we use  $l_N = b$, $n_N = n_b$, 
and consequently, the propagation factor equals
\begin{equation}
P_{dec}^{out} = 
\int_{l_p}^l \frac{dx}{b}
\int_l^{l + d} \frac{dz}{d} 
\int_x^z \frac{dy}{\lambda_N}
e^{-(y - x)/\lambda_N}.
\label{eq:prop-udn}
\end{equation}
Explicit integration gives
\begin{equation}
P_{dec}^{out} = 1 - \frac{\lambda_N^2}{bd}
\left(1 - e^{-b/\lambda_N}\right)
\left(1 - e^{-d/\lambda_N}\right).
\label{eq:prop-udn-out}
\end{equation}

For the $N-$production in a detector: $l_N = d$, $n_N = n_d$,
and the limits of integration  differ from those in
(\ref{eq:prop-udn}):
%\begin{equation}
$$
P_{dec}^{in} = \frac{1}{d^2}
\int_{l}^{l + d} dx
\int_x^{l + d} \frac{dy}{\lambda_N}
\int_y^{l + d} dz ~e^{-(y - x)/\lambda_N}.
$$
%\label{eq:prop-udnin}
%\end{equation}
Integration gives
\begin{equation}
P_{dec}^{in} = \frac{1}{2} - \frac{\lambda_N}{d}
+ \frac{\lambda_N^2}{d^2} \left(1 - e^{-d/\lambda_N}\right).
\label{eq:prop-udninex}
\end{equation}
In the limit of very fast $N-$ decay, $\lambda_N \rightarrow 0$,
the propagation factors converge to $P_{dec}^{out} \rightarrow  1$ and  $P_{dec}^{in} \rightarrow 1/2$.
In the opposite limit, $\lambda_N \rightarrow \infty$,
both factors vanish: $P_{dec}^{out}, \,  P_{dec}^{in} \rightarrow 0$.

The sum of two contributions (from ``out" and ``in" production) is proportional to
%\begin{equation}
$$
n_b n_d b d \left(P_{dec}^{out} +
\frac{n_d}{n_b} \frac{d}{b} P_{dec}^{in}
\right).
$$
%\label{eq:prop-tot}
%\end{equation}

Decay factors similar to eq. (\ref{eq:prop-udn-out})  appear in the $U_N D_\xi$ scenario
(see eqs. (49), (52)).  The difference is that in $U_N D_\xi$ scenario $N$ decays immediately
into the observed particles $D_\xi$, and therefore the decay should occur in a detector.
In the present scenario $\xi = e$  is produced in two steps $D_\nu U_e$, and  therefore $N$  can decay 
both in detector and in dirt.

Let us consider $c \tau_0$ dependence of the number of events.
Since $n_b b  \gg n_d d$, 
in the first approximation the $N$ production in a detector can be neglected.
Then according to (\ref{eq:prop-udn-out}) there are two characteristic scales
in the setup: $b$ and $d$ which correspond to two regions of the $\nu_\mu-$upscatterings followed
by decays.

In the limit $\lambda_N \gg b, d$ the Eq. (\ref{eq:prop-udn-out}) gives
%\begin{equation}
$$
P_{dec}^{out} \approx  \frac{b + d}{\lambda_N} \approx \frac{b }{\lambda_N}.
$$
%\label{eq:limbig}
%\end{equation}
In the intermediate range $d \ll \lambda_N \ll b$ we obtain
$
P_{dec}^{out} \approx 1 -  \lambda_N/b,
$
and for very fast decay,  $\lambda_N \ll b, d$,
$ P_{dec}^{out} \approx 1 -  \lambda_N^2/b d$.
Ratio of the decay factors for a given experiment and MiniBooNE has the following dependence on
$c \tau_0$.
In the asymptotics $c \tau_0 \gg b^{MB} m_N/E^{MB}$ the ratio is constant:
$$
 r^i \equiv  \frac{P_{dec}^{out, i}}{P_{dec}^{out, MB}} \approx
\frac{b^i}{b^{MB}} \frac{E^{MB}}{E^i},
$$
and for experiments under consideration:  $r^i < 1$. With decrease of $c \tau_0$
the ratio increases mainly in the intermediate region
 $d^i m_N/E^i <   c \tau_0 <   b^{MB} m_N/E^{MB}$, and then converges to 1
at $ c \tau_0 <  d^i m_N/E^i$.
So, qualitatively the dependence is similar to the dependence 
for other upscattering scenarios 
with, however,  longer transition region between the two asymptotics. \\

We described this scenario for completeness. 
It will be difficult (if possible) to construct a viable model 
that matches this scenario. Indeed,  here there are 
two ($\nu_\mu-$ and $\nu_e-$) upscattering vertices  
which bring smallness to the number of events. Furthermore, 
the transitions can be treated as flavor violating non-standard interaction 
(NSI) that transform $\nu_\mu-$ to $\nu_e-$ and there are stringent bounds 
on this NSI. Therefore in what follows 
we will not present detailed phenomenological studies of this scenario.

%%%%%%%%%%%%%%%%%%%%%%%%%%%%%%%%%%%%%%%%%%%%%%%%%%%%
%%%%%%%%%%%%%%%%%%%%%%%%%%%%%%%%%%%%%%%%%%%%%%%%%%%%
\section{Signature factors, cross sections, experiments and bounds. }
\label{sec:limits}
%%%%%%%%%%%%%%%%%%%%%%%%%%%%%%%%%%%%%%%%%%%%%%%%%%%%
%%%%%%%%%%%%%%%%%%%%%%%%%%%%%%%%%%%%%%%%%%%%%%%%%%%%
\noindent
The key idea is that new physics scenarios  that explain the MiniBooNE excess should produce visible numbers 
of events in the near detectors of various neutrino experiments. 
That will allow us to put bounds on the scenarios. 
Here we describe the relevant 
features of different experiments as well as the theoretical and experimental results. 
We compute the upper limits on numbers of event  due to new physics.

%%%%%%%%%%%%%%%%%%%%%%%%%%%%%%%%%%%%%%%%%%%%%%%%%%%%
\subsection{Signal}
%%%%%%%%%%%%%%%%%%%%%%%%%%%%%%%%%%%%%%%%%%%%%%%%%%%%

The observable signal is given by a deposit of electromagnetic energy from a final state $\xi$.
Depending on the particle ID capabilities of a detector $i$, 
a given state $\xi$ can be (mis-)identified 
with a number of other  particle states.
Associated with this identification are detector and analysis efficiencies.
Below we describe our approach for quantifying this.
We also discuss the cross section input used for the upscattering scenarios.

%%%%%%%%%%%%%%%%%%%%%%%%%%%%%%%%%%%%%%%%%%%%%%%%%%%%
\subsubsection{Efficiency}
%%%%%%%%%%%%%%%%%%%%%%%%%%%%%%%%%%%%%%%%%%%%%%%%%%%%

The experiments $i$ quote signature efficiencies for the signatures $s^i$ which are, 
in general, a product of a detector efficiency $\epsilon^i_{\xi}$, 
a particle (mis-)identification efficiency $f^i_{\xi - s^i}$ 
and signal selection efficiency $\epsilon^i_{s^i}$.

The detector efficiency $\epsilon^i_{\xi}$ quantifies the probability 
that a final state $\xi$ is  registered in any way. 
%%one way or another. 
In what follows we assume that $\epsilon^i_{\xi} = 1$.  
The misidentification efficiency or the signature factor 
$f^i_{\xi - s^i}$ is  the fraction of cases when  final state $\xi$ 
produces a signature $s^i$. 
The signal selection efficiency $\epsilon^i_{s^i}$ quantifies the so-called quality cuts 
(which include kinematic cuts) 
of the events that are needed to enhance the signal-over-background ratio. 
These efficiencies depend strongly on the considered signatures and we take their values from experiments.

%\textcolor{red}{REMOVE: Here we give two examples of employed kinematic cuts that are relevant 
%for enhancing signal efficiency: following \cite{Abe:2019kgx} $\cos \theta>0.99$ is imposed 
%for the search of two tracks at T2K ND and for 
%the analysis conducted in \cite{Park:2015eqa} we have 
%$E \theta^2 < 0.0032$ GeV. 
%Here $E$ is the shower energy and $\theta$ is the angle between the direction of emitted 
%charged particle(s) that yield a shower and incoming active neutrino. 
%For estimating the T2K ND cut $\cos \theta>0.99$ we perform our own Monte Carlo 
%simulation of final state angular distributions. 
%We account for the cut $E \theta^2 < 0.0032$ GeV in MINER$\nu$A with an estimated 
%selection efficiency that is inferred from SM processes in Fig. 4 of \cite{Park:2015eqa}.}

%%%%%%%%%%%%%%%%%%%%%%%%%%%%%%%%%%%%%%%%%%%%%%%%%%%%
\subsubsection{Signature factor}
%%%%%%%%%%%%%%%%%%%%%%%%%%%%%%%%%%%%%%%%%%%%%%%%%%%%

In general, the signature factor includes an integration 
over the phase space of kinematical variables, and (mis-) identification 
factors  $I_e^i$, which depend on the type of detector.   

Some detectors can distinguish events induced by a single photon, an $e^+e^-$ pair, 
from those induced by a single electron. 
This is usually accomplished via measuring the energy loss, $dE/dx$, over the whole trajectory, 
or only in its initial part (like in MINER$\nu$A).
Detectors that have a magnetic field, like NOMAD or T2K ND280 also use the bending of tracks for particle ID.

We can introduce the  signature factors a in different way considering 
final interactions (scattering or decay) which produce the state $\xi$. 
Then $f$ can be defined as 
fraction of the final interactions in which the event $s^i$ appears. Formally that 
implies summation over $\xi$. 

Let us consider first scattering. 
For electrons that are produced by the CCQE $\nu_e-$scattering  on nucleons ($\xi = e$) we can write  
\begin{equation}
f_{e - s^i}(E_\nu) = \int_{E_e^{th}} dE_e  I_{s^i}(E_e)  
\frac{1}{\sigma^{tot}}\frac{d\sigma(E_\nu, E_e)}{ d E_e}, 
\label{eq:fetosi}
\end{equation}
where $I_{s^i}(E_e)$ is the probability that the electron with energy $E_e$ will show up as the $s^i$ event. 
In experiments capable to disentangle showers induced by $\gamma$ and $e$, the factor $I_{e-1sh}(E_e) \sim 1$ 
which then leads to $f_{e - 1sh} \approx 1$.  \\

Let us consider final states $\xi$ that originate from $N$ or $B-$decays.  
For $\xi = \gamma$
\begin{equation}
f^i_{\gamma - 1sh}(E_N) =  
\int  d E_\gamma \frac{1}{\Gamma_N(E_N)}\frac{d \Gamma_N(E_N, E_\gamma)}{  d E_\gamma} I_{\gamma - 1sh} (E_\gamma)\,. 
\label{eq:rate1shgf}
\end{equation}
Again, if $I_{\gamma - 1sh} (E_\gamma) \approx 1$, the definition (\ref{eq:rate1shgf}) gives 
$f^i_{\gamma - 1sh} \approx 1$.  

In general, the signature factor for $s^i-$event can be written as 
\begin{equation}
f^i_{\xi - s^i}(E_N, m_N) =  
\frac{1}{\Gamma_N(E_N,m_N)} \int^{\Pi_{s^i}} d \Pi_\xi~   
\frac{d \Gamma_N(E_N,m_N,\Pi_\xi)}{d \Pi_\xi}   I_{\xi - s^i} (\Pi_\xi)  \,,
\label{eq:rate1shgf1}
\end{equation}
where $\Pi_{s^i}$  is the final state phase space
in which the produced state $\xi$ shows  
up as a $s^i$ event in the experiment $i$.

%with the final state phase space $\Pi_{s^i}$ 
%in which the produced state $\xi$ shows  
%up as a $s^i$ event in the experiment $i$.  

For the final state being $\nu \gamma$ ($\xi = \gamma$), the relevant phase space 
is above the energy threshold, which is for instance $E_\gamma > 100$ MeV 
in MiniBooNE (used to suppress cosmic ray backgrounds).
%%% QUESTION
% \textcolor{red}{[[in all experiments? is not consistent with 55 MeV]].} 
%Then 
In experiments without the $\gamma - e$ identification,  
and for high energies of  $N$: $\Pi^i_\gamma$ is nearly the entire phase space. 
Thus,  $f^i_{\gamma - 1sh}(E_N, m_N) \approx 1$.

The $e^+e^-$ pair ($\xi = e^+ e^-$) can produce two shower ($2e-$showers) events as well as 
single shower events, if one of the components is missing or if two components are nearly collinear. 
For several detectors the unique relevant criterion for differentiation between 
the single and double shower events is the invariant mass of pair, $W_{ee}$. If $W_{ee} < W_c$,  
where $W_c$ is a certain critical value, the pair shows up as a single shower event, 
while for $W_{ee} > W_c$ -- as the two shower event.  This means that 
$I_{ee - 1sh} (W_{ee}) = 1$ when $W_{ee} < W_c$, and 
$I_{ee - 1sh} (W_{ee}) = 0$ when $W_{ee} > W_c$.  
%%Here we neglect $\gamma - e$ misidentification. 

%%Therefore, $f_{ee - 1sh} = 1$ if $W_{ee}$ < W_c$, and 
%%$f_{ee - 1sh} = 0$ if  $W_{ee}$ > W_c$.   

When the $e^+e^-$ pair is created from the 3-body decay $N\to \nu e^+ e^-$, $W_{ee}$ is not fixed 
and one needs to use the function $I_{ee - 1sh} (W_{ee})$. 
The step-like $I_{ee - 1sh} (W_{ee})$ determines the limits of integration. 
The fraction of decays  with $W_{ee} < W_c$,  which  appear  as single shower event equals:
\begin{equation}
f^i_{{ee-}1sh}(x,m_N) = \frac{1}{\Gamma(N\to \nu e^+e^-)} 
\int_0^{W_c} d W_{ee} \frac{d\Gamma(N\to \nu e^+e^-)}{dW_{ee}} = 
\frac{W_c^8 + 2 W_c^2 m_N^6 - 2 W_c^6 m_N^2}{m_N^8}\,.
\label{eq:invariantmass}
\end{equation}
We take  $W_c = 30$ MeV for MiniBooNE \cite{BillLouis}, $W_c = 5$ MeV for the T2K near detector 
ND280 (cf.\ ref.~\cite{Abe:2019cer}), and we estimate $W_c = 30$ MeV for PS191. 
For other detectors we do not use an invariant mass threshold for our analysis, 
{\it i.e.} we assume that $e^+ e^-$ pairs and photons give the same signature.
Notice that $f$ defined in this way does not depend on $E_N$, 
which simplifies computations. 
 
If the $e^+ e^-$ pair appears from the 2-body decay of a new boson, $B \rightarrow e^+ e^-$, 
the invariant mass $W_{ee}$ is fixed: $W_{ee} = m_B$. Therefore, the signature 
factor is determined uniquely by the mass of B:  
For $m_B < W_c$ we have $f^i_{ee - 1sh} = 1$,  while for $m_B > W_c$:  $f^i_{ee - 1sh} = 0$.  
This is realised, e.g., in scenarios with the decay chain $N \to \nu B, ~B  \to  e^+e^-$, 
where an on-shell dark photon $B$ is produced.
%%Therefore  $f^i_{1sh}(E_N, m_N) = 1$ for $m_{\gamma'} \leq 10$ MeV.
For the 2-shower signature we have relation $f^i_{ee - 2sh} = 1 - f^i_{ee - 1sh}$.

%%In final states with a single photon or a dark photon 
%%of mass below about 10 MeV, $f^i_{\gamma - 2sh} = 0$.

%%%%%%%%%%%%%%%%%%%%%%%%%%%%%%%%%%%%%%%%%%%%%%%%%%%%%%%%%%%%%%%%%
\subsection{Cross sections and fluxes}
%%%%%%%%%%%%%%%%%%%%%%%%%%%%%%%%%%%%%%%%%%%%%%%%%%%%%%%%%%%%%%%
\label{subsec:crosssection}
In the presence of new physics, the cross sections of heavy or light neutrino interactions 
depend on specific model of interactions,  
{\it i.e.} on the mass of mediator, Lorentz structure of coupling, {\it etc.}
%%model dependent, both in magnitude and in differential shape. 
Since we compute the ratios of numbers of events, the model-dependence of the cross sections mostly cancels.

{\color{black}
However, there are still some uncertainties that depend on
the nature of the new particle mediating upscattering.
We find  that for ND280 such uncertainty
is at most at the level of $\sim 20\%$. This value was
obtained by comparing the predicted number of events
for the vector and scalar mediators at $c\tau^0 \rightarrow 0$.
The reason for this small uncertainty at ND280 (as well as at PS191)
is that the flux of neutrinos peaks at $\sim 1$ GeV
as is the case for MiniBooNE.
Therefore the cross sections should
be taken at similar energies and the effect of change from vector
to scalar mediator also cancels in the ratio \cref{eq:NDevents}.

Situation is different with MINER$\nu$A where the spin of
mediator matters. The typical energy of neutrino flux
at MINER$\nu$A ($E^{MV}$) is few GeV higher than that
at MiniBooNE ($E^{MB}$).
Depending on whether mediator is a new light vector or a scalar,
the cross section grows or decreases in this energy range
between $E^{MB}$ and $E^{MV}$.
Namely, in the scalar mediator case
of models [11, 12] the cross section
decreases with energy substantially, so
that MINER$\nu$A is not able to probe such a  scenario,
contrary to the case of vector mediator
[26]. Keeping in mind  this result for MINER$\nu$A, we will mainly
consider the  vector mediators cases.
Constraints from all experiments apart
form MINER$\nu$A are roughly independent
of the nature of mediator,
as argued above. NO$\nu$A also has  the flux
peaking at larger energy than the MiniBooNE flux
(given the same beam as for MINER$\nu$A).
Therefore, we use results from NO$\nu$A
to probe only a scenario where upscattering
goes via SM charged current process
($M_N D_\nu U_e$).

}

Furthermore, to cover all the possibilities we consider both partially coherent and incoherent interactions. 
For the partially coherent case, we take the mass of mediator 
in the upscattering process to be 30 MeV in accord with the benchmark point of \cite{Bertuzzo:2018itn}.
For the incoherent case, we calculate the cross section for the mediator mass 
of 1.25 GeV (using the cookbook presented in \cite{Kim:2020ipj})
which  corresponds to the benchmark point in \cite{Ballett:2018ynz}.
For the quasi-elastic scattering of $\nu_e$ we use the $\nu_\mu$ upscattering cross section 
from ref.~\cite{Formaggio:2013kya} 
%(figures 11 and 12). 
as a proxy. Differences of the cross sections due to difference of the electon and muon masses  
should be minor  because they are both small compared to the neutrino energies.

\subsection{Experiments and bounds}
%%%%%%%%%%%%%%%%%%%%%%%%%%%%%%%%%%%%%%%%%%%%%%%%%%%%%%%%%%%%%%%%%%%%

\subsubsection{MiniBooNE}
%%%%%%%%%%%%%%%%%%%%%%%%%%%%%%%%%%%%%%%%%%%%%%%%%%%%%%%%%%%%%%%%%%%%%%%

Some information on MB has already been presented in \cref{sec:options}. 
The total number of muon neutrinos 
that passed through the MiniBooNE detector in positive (negative) horn polarity mode 
is $8.12\times 10^{17}$ ($3.1\times 10^{17}$) \cite{AguilarArevalo:2008yp}. 
This corresponds to the muon neutrino flux per POT:
\begin{equation}
\phi^{MB} = 5.19 \cdot 10^{-10} {\rm cm}^{-2} (POT)^{-1}\,.
\label{eq:mb-flux}
\end{equation}
The relevant parameters of the experimental setup are:  the decay pipe length $l_p^{MB} = 50$ m, 
baseline $l^{MB} = 540$ m, average detector 
length $d^{MB}=8$ m and the target mass $m^{MB} = 800$ t.
The average electron reconstruction and selection efficiency is $\epsilon^{MB}_{1sh} \simeq 10\%$. {\color{black} Taking this average value of the  efficiency
instead of using the energy dependent efficiencies \cite{eff} introduces a negligible $2\%$ effect.}

Apart from single shower events  MiniBooNE observed also the  2 shower events 
and this can be a powerful probe of scenarios with  $\xi=ee$ and $\xi=\gamma\gamma$. We have, however, estimated 
that this gives weaker bounds on the scenarios than 
the 2 shower data from ND280.

\subsubsection{T2K ND280}
%%%%%%%%%%%%%%%%%%%%%%%%%%%%%%%%%%%%%%%%%%%%%%%%%%%%%%%%%%%%%%%%%%%%%%%%%%

The T2K ND280 (ND280 for brevity) 
is sourced by 30 GeV protons that interact with the  graphite target \cite{Abe:2011ks}.
The lengths involved  are $l_p \simeq 100$ m, $b=230$ m (dirt), and   $l^{ND}= 280$ m \cite{Kudenko:2008ia}.

ND280, placed at $2.5^\circ$ off axis, is a multicomponent detector which consists 
of the following main sub-detectors:

(i) the $\pi^0$ detector P0D.  
The P0D filled with water has a target mass $m^{ND}_{P0D} = 15.8$ t  
and a length $d^{ND}_{P0D} = 2$ m \cite{Assylbekov:2011sh};

(ii) the tracking detector containing the three Time Projection Chambers (TPC) 
filled in by Ar gas. 
Each TPC module has a mass of $0.3$ t and a length of $0.9$ m.

(iii)  two  Fine Grained Detectors (FGD) filled in by scintillatiors. 
The mass and the length of each  FGD are  $1.1$ t and  $0.365$ m, correspondingly \cite{Amaudruz:2012esa}.
The detectors are magnetized with a field strength of 0.2 T, which, 
together with energy loss tracking, allows for a very good particle identification capacity.
The distance between downstream edge of 
P0D and the upstream edge of FGD1 equals $\Delta^{ND} =  1$ m.

Strictly,  one has to consider interactions, decays and detection in all  these  detectors separately.
For simplicity we will neglect most of the detector substructures.
The neutrino flux is taken from  ref.~\cite{Abe:2012av}.
We use two data sets from two independent studies: a search for heavy neutrinos \cite{Abe:2019kgx} 
and an analysis of electron neutrino CCQE \cite{Abe:2020vot}. 
The latter gives bounds on numbers of $\gamma-$showers and $e-$showers. \\

\paragraph*{1. Resolved  $e^+  e^-$ pairs: 2showers.}  T2K searched the resolved  
$e^+$ and $e^-$ tracks  (showers) from hypothetical  heavy neutrino decays inside 
the Time Projection Chamber (TPC) ref.~\cite{Abe:2019kgx}.  
In this study $12.34\times 10^{20}\,(6.29\times 10^{20})$ POT 
in neutrino (anti-neutrino) mode were used. 
The selected events consist of two tracks of opposite charge originating from a vertex in  TPC, 
without other tracks being observed in the TPC itself or in the detector located directly upstream (including P0D).
This gives an effective detector length of 2.7 m.
%Particle ID is based on the energy loss and also on the curvature of the tracks in the magnetic field.
The invariant mass of 2-track system was restricted by $W_{ee} < 700$ MeV and the angle between two tracks $< 90^\circ$.  
The angle between system of the tracks and the beam axis for events passing selection criteria should be $\cos \theta > 0.99$. 
%\textcolor{blue}{
To implement this cut in computations of numbers of events we performed our own Monte Carlo 
simulation of final state angular distributions.
%}

For the indicated number of POT, the number of observed $e^+ e^-$ shower events 
in neutrino mode, which satisfy the selection criteria,  equals 
$N_{ee}^{ND_\nu,obs} =  62$.  The expected number of events from the standard sources 
(various neutrino interactions) is $N_{ee}^{ND,th} = 58 \pm 2.8$.  
In the antineutrino mode $N_{ee}^{ND_{\bar\nu},obs} = 16$ events have been observed,  
while  $N_{ee}^{ND,th} =  15.1\pm 1.6$ are expected.  
We sum the events from both modes.
We neglect the small error in the theory prediction ($2.8$), and  combine the statistical 
uncertainty, $\Delta N^{stat} = 8.8$,  with the systematic one in quadrature. 
For the latter we take 15\% relative uncertainty on the total number of observed events 
which gives $\Delta N^{syst} = 11.7$ 
(In what follows for experiments where systematic uncertainty is not explicitly quoted, we assume the uncertainty of 15\%).
With this, the following upper limits on a contribution from new physics are obtained
\begin{equation}
N_{2sh}^{ND}  <  20 ~(1 \sigma)\,,  ~~~ ~ 34~ (2 \sigma)\,, ~~~~  49 ~(2 \sigma)\,.
\label{eq:2sh-ndws}
\end{equation}
%
%\begin{equation}
%N_{2sh}^{ND}  <  30 ,  ~~~ \, \, \,    95\% ~{\rm C.L}.
%\label{eq:2sh-nd}
%\end{equation}

Due to particle ID capacity of ND280, the selected events can be produced by the $e^+ e^-$ pair only.   
We take the signature factor according to eq.~\eqref{eq:invariantmass} for the 3-body $N-$ decay,  and  $f_{ee - 2sh} = 1$ 
for the 2-body $B-$ decay if $m_B > 5$ MeV.  \\

\paragraph*{2. Unresolved (collinear) $e^+ e^-$: 1 shower events.}  
The $\nu_e$ CCQE interactions were detected as isolated $e-$shower events~\cite{Abe:2020vot}.
The photon background is the most important for these events. In this connection, 
T2K studied single photons converted into  $e^+ e^-$ pairs in the FGD1. 
The event selection criteria in the analysis include the following:
two tracks originate from the vertex in FGD1,
the energy losses in the tracks, $dE/dx$, are compatible with electrons.
The tracks correspond to particles of opposite sign.
The invariant mass is less than $W_{ee} < 55$ MeV (the latter was imposed to ensure that 
$e^+ e^-$ originate from photon conversion). As signature efficiency 
we adopt $\epsilon^{ND}_{\gamma} = 0.3$ from ref.~\cite{Abe:2020vot}.

A total numbers of events of this type $N_{\gamma}^{ND, obs} =  647$,  182, and  157 
were found in the analysis of the FHC data, the electron analysis of RHC data and  positron analysis of RHC data 
correspondingly.
The simulated numbers of events  that originate from SM processes 
(CCQE neutrino-nucleon scattering, resonant pion production, 
deep inelastic scattering, final state interactions of hadrons produced, {\it etc.}) 
turn out to be  larger: $N^{ND,th}_\gamma =  700.97$~(FHC),  193.73 (electron RHC) and  169.31 (positron RHC). 

We sum up the event numbers from FHC  and the positron RHC data\footnote{Including also the electron 
analysis would add information, but we have to take the correlation of 
the two analyses into account to which we have no access.}. 
The statistical error on the combined event numbers, $\Delta N^{stat} = 28.1$, 
and the $15\%$ systematic error, $\Delta N^{syst} = 118.8$,  are summed in quadrature.  
This gives the upper bounds on numbers of isolated $\gamma$'s from new physics 
\begin{equation}
N_{\gamma}^{ND} < 58 ~~(1\sigma), ~~~~
181 ~~(2\sigma), ~~~~
305 ~~(3\sigma). 
\label{eq:ngammaND}
\end{equation}
The deficit of observed signal events with respect to the prediction strengthen  the bound.   
Here, signature factor $f_{\gamma - 1sh}^{ND} = 1$. 

We will not use results of a dedicated search for  
the single photon events at T2K ND280 in ref.~\cite{Abe:2019cer} due to low statistics.\\

\paragraph*{3. Single $e-$shower.} In the same ND280 study  of the $\nu_e-$CCQE interactions 
ref.~\cite{Abe:2020vot} the total numbers of 697, 176 and  95 $e-$like events 
were found in the FHC, electron  RHC  and  positron  RHC analyses.
These  numbers  are smaller than the expected numbers  
from various standard neutrino interactions:  797,  175.92 and 99.99. 
As before, we combine the event numbers from the FHC mode and    
the positron RHC mode.
The statistical error, $\Delta N^{stat} = 28.3$,  and the $15\%$ relative
systematic error, $\Delta N^{syst} =  120.6$,  are added  in quadrature. 
% using 15\% relative uncertainty for the latter.
This leads to the upper bounds on numbers of $e-$ like events from new physics
\begin{equation}
N_{e}^{ND} < 17 ~~(1\sigma), ~~~~
139 ~~(2\sigma), ~~~~
261 ~~(3\sigma). 
\label{eq:nd-e}
\end{equation}
This analysis can be used to constrain scenarios with $\xi = e $. 
%%The $e - \gamma$ misidentification is important. 
The reconstruction (and selection)  efficiency for the $e-$like events equals   
$\epsilon^{ND}_{e-sh} =  0.3$ according to ref.~\cite{Abe:2020vot}. 
Notice that in future phases of experiment the T2K ND280 can substantially improve
these bounds.

%%%%%%%%%%%%%%%%%%%%%%%%%%%%%%%%%%%%%%%%%%%%%%%%%%%%%%%%%%%%%%%%%%%%%%%%%%%%
\subsubsection{MINER$\nu$A}
%%%%%%%%%%%%%%%%%%%%%%%%%%%%%%%%%%%%%%%%%%%%%%%%%%%%%%%%%%%%%%%%%%%%%%%%%%%%%%%%%%

The MINER$\nu$A experiment employs the Mine Injector beam line, where 120 GeV protons hit a graphine target. 
The produced neutrino flux has variable energy in the range (2 - 20) GeV. 
We use two  energy samples:  
ME (medium energy) with the peak %of $\phi \sigma$
 at $E_\nu^{MV} = 6$ GeV,  and LE (low energy) with the peak at $E_\nu^{MV} = 4$ GeV.
The flux of usual neutrinos is substantially larger than the MB flux:
\begin{equation}
\phi^{MV,ME} = 3 \cdot 10^{-8} {\rm cm}^{-2} (POT)^{-1}\,.
\label{eq:min-flux}
\end{equation}
%(ref.: 1906.00111 [hep-ex]). 
The ratio of fluxes per POT:  $\phi^{MV,ME}_\nu / \phi^{MB}_\nu = 15$.

The experimental setup  has the following sizes: 
$l^{MV}_p =  675$  m,   $l^{MV} = 935$ m,  $d^{MV}=3$ m; the  target mass equals $m^{MV} = 6.1$ tonnes. 
In computations we take the distance between the detector and the up-stream 
absorber (the dirt) to be $\Delta^{MV} = 10$ m.

The MINER$\nu$A detector consists of  scintillator strips, which provide 3D information on the tracks.
Good particle ID allows to distinguish the   
$1 e-$ from $1\gamma-$ and $e^+ e^-$ showers using the energy loss $dE/dx$ (along the track or in the first 4 strips).
%(Neutrino-electron scattering has the same signature (single e-like track)  
%as isolated  $\xi = \gamma$ or collimated $\xi = e^+ e^-$, apart from %the dE/dx.) 
Three different samples of data were explored:
the  CCQE $\nu$ interactions,  the  $\nu e -$ scatering data at LE and HE. \\

%%samples at different energies of $\nu e -$ scatering.}\\

\paragraph*{1. $e-$like events from the $\nu_e$ CCQE interactions.}   
A total number of $3204$ $e-$like events was observed, while 2931 events were expected \cite{Wolcott:2015hda}.
We sum the statistical uncertainty of the observed number of events,  
$\Delta^{stat} =  56.6$,  and $15\%$ systematic ucertainty, $N^{syst}= 480.7$,  quadratically  
which gives the  upper bounds on new physics contribution
\begin{equation}
N_{e}^{MV} < 757 ~~(1\sigma), ~~~~
1241 ~~(2\sigma), ~~~~
1725 ~~(3\sigma). 
\label{eq:mv-e}
\end{equation}
As the signature selection efficiency   
we use the energy-averaged selection efficiency for the electron showers from the  $\nu-e$ scattering analysis
in ref.~\cite{Park:2015eqa}:  $\epsilon^{MV}_{\gamma} = 70\%$.\\

\paragraph*{2. $\gamma-$like events from the $\nu-e$ scattering analysis.}   
The single EM shower events have been detected in interactions of the LE neutrino flux
produced by $3.43 \times 10^{20}$ POT, Ref.~\cite{Park:2015eqa}.  
The  dE/dx distribution of the events was constructed 
cf.\ fig. 3 of ref.~\cite{Park:2015eqa} which allows to disentangle events produced by electrons and 
gammas. 
%%In this analysis, electromagnetic shower candidates were selected and their dE/dx distribution 
%%was displayed prior to $\nu-e$ analysis cuts, 
%%From this distribution we sum the electromagnetic showers 
For  $dE/dx > 4.5$ (MeV/1.7cm)  171 photon-like events were observed 
which practically coincide with the number of expected 170 events.
The statistical error, $\Delta N^{stat}= 13.1$,  and  the systematic error, $\Delta N^{syst} = 17.1$, 
(using $10\%$ error according to  ref.~\cite{Park:2015eqa}) allow us to get  upper bounds on 
new physics contributions to single shower events
\begin{equation}
%%% TO BE REMOVED
%N_{e}^{MV},~~ 
N_{\gamma /ee}^{MV} < 23 ~~(1\sigma), ~~~~
45 ~~(2\sigma), ~~~~
66 ~~(3\sigma)\,.
\label{eq:mv-gamma/ee_LE}
\end{equation}
%
%%%OLD TEXT, REMOVE BECAUSE REDUNDANT
%\textcolor{red}{[[How do we use this???]]}
%[[repetition]] In this selected data set the collaboration observed 172 events while 170 events are predicted.
%
A similar analysis has been carried out with the ME data \cite{Valencia:2019mkf}, $1.16 \times 10^{21}$ POT.
Following the same procedure as above, 
1466 $\gamma$ events were observed and 1395 events were expected.
We add in quadrature the statistical error, $\Delta N^{stat}= 38.3$,  and the systematic error, 
$\Delta N^{syst} = 146.6$, which is the  $10\%$ error presented in  ref.~\cite{Valencia:2019mkf}.  
This gives the upper bounds on single shower events 
\begin{equation}
N_{\gamma /ee}^{MV} < 223 ~~(1\sigma), ~~~~
374 ~~(2\sigma), ~~~~
526 ~~(3\sigma)\,.
\label{eq:mv-gamma/ee}
\end{equation}
%%additional events to the $ee-$showers.
Since no photon PID cut on the data has been employed,
%%% QUESTION TO BE REMOVED?!
%\textcolor{red}{[[why gamma? we are dicussing here e-like]]} 
the results can be applied to  $\xi = \gamma$ and 
collimated electron-positron pairs, $\xi = e^+ e^-$. Our statistical analysis shows 
that constraints on the allowed number of additional photon-like events are 
the strongest when considering this ME dataset.

We set the probability that a $\xi$ is accepted as a single EM  shower to one: $f_{\xi - 1sh} = 1$.
%\textcolor{blue}{
We account for the cut $E \theta^2 < 0.0032$ GeV in MINER$\nu$A with an estimated
selection efficiency of $10\%$ that is inferred from SM processes in Fig. 4 of \cite{Park:2015eqa}.
Here $E$ is the shower energy and $\theta$ is the angle between the direction of emitted
charged particle(s) that yield a shower and incoming active neutrino.
We found that events surviving the cut on $E\theta^2$  would not induce observable hadronic activity in MINER$\nu$A.
%}

%%%%%%%%%%%%%%%%%%%%%%%%%%%%%%%%%%%%%%%%%%%%%%%%%%%%%%%%%%%%%%%%%%%%%%
\subsubsection{PS191}
%%%%%%%%%%%%%%%%%%%%%%%%%%%%%%%%%%%%%%%%%%%%%%%%%%%%%%%%%%%%%%%%%%%%%%%%%%%%%%%%

The PS191 experiment was sourced by the PS proton beam with  energy 
19.2 GeV interacting with a beryllium target and it collected $2 \cdot 10^{19}$ POT. 
The $\nu_\mu-$flux at the detector from pion decays was $\phi_{\nu_\mu}^\pi = 2.3 \cdot 10^{-4}$ cm$^{-2} \text{POT}^{-1}$. 
The setup has the parameters $l^{PS} = 128$ m,  $l^{PS}_p = 49.1$ m.
The detector was composed of a decay volume and a down-stream calorimeter. 
The decay volume of length $d^{PS}=12$ m was filled in with flash chambers for tracking 
and helium bags and therefore had negligible mass.
The calorimeter consisted of sandwiches made from flash chambers and 3 mm thick iron plates.
%For PS191 we estimate the distance between the detector and the up-stream absorber 
%%(the dirt) which we set to $b_{NOM} = 5$ m, based on a photograph in ref.~\cite{Bernardi:1987ek}.
Two studies have been performed.\\ 

\paragraph*{1. $2$ tracks in the decay volume.}
Events induced by heavy neutrino decays in the decay volume were searched for in ref.~\cite{Bernardi:1987ek}.
These events should have two tracks in the decay volume 
and an energy deposit in the calorimeter.  
The vertex of the two tracks can be reconstructed.
The criteria was that  the reconstructed vertex should be  more than 2 cm away from a flash chamber.
%Several SM processes can lead two tracks signal via $\nu_\mu$ interaction 
%with the track chambers, but none can give rise to a displaced vertex.
Not a single vertex was found;
this null result constrains the contribution from heavy neutrinos with decay  
into $\xi$ that leaves two charged tracks in the flash chambers.
The limit on events with 2 tracks reads \cite{Bernardi:1987ek}:
\begin{equation}
N_{2tr}^{PS,obs} < 2.3, ~~~~ 95\% {\rm C.L.} 
\label{ps:vertex} 
\end{equation} 
We apply  this limit for the final states $\xi=\gamma\gamma$ and $ e^+e^-$ with an invariant 
mass above the threshold $W_c^{PS} = 30$ MeV. 
This threshold was derived from  ref.~\cite{Bernardi:1987ek}, 
where heavy neutrinos with $m_N \approx 30$ MeV are still subject to constrains.
For the signature selection efficiency 
we use the signal selection efficiency $\epsilon^{PS}_{2tr} = 0.28$ 
taken from ref.~\cite{Bernardi:1987ek}.

\paragraph*{2. Single showers in the calorimeter.}
Good granularity of the calorimeter allows to distinguish  the photon showers  from the electron showers.
In  ref.~\cite{Bernardi:168324} the  electromagnetic showers  
with energies above 400 MeV were selected to suppress background from $\pi^0$ decay.
As a proxy for the signal selection efficiency
we use the reconstruction efficiency from ref.~\cite{Bernardi:1987ek}: $\epsilon^{PS}_{1sh}=0.7$.
%Showers can be from directly produced  \textcolor{red}{[[what does this mean direct?]]} 
Showers can be produced by $\nu_\mu$ interactions, in particular 
from final states including $\gamma$, $\pi^0$, $e$, and by hadrons. Hadron misidentification is atmost 1\%.
The sub-sample with an electron-likelihood selection cut yields an excess of the $e-$like events in the calorimeter 
\begin{equation}
N_{1sh}^{PS,obs} = 23\pm 8\,,
\label{ps:1shower} 
\end{equation} 
that was attributed to neutrino oscillations \cite{Bernardi:168324}.

%%%%%%%%%%%%%%%%%%%%%%%%%%%%%%%%%%%%%%%%%%%%%%%%%%%%%%%%%%%%%
\subsubsection{NO$\nu$A near detector}
%%%%%%%%%%%%%%%%%%%%%%%%%%%%%%%%%%%%%%%%%%%%%%%%%%%%%%%%%%%%%%%%%%

The NO$\nu$A experiment uses the NuMI neutrino beam 
sourced  by  interactions of 120 GeV protons  with a graphite target.
The parameters of setup are  
$l^{NOV} = 1000$ m,  $l^{NOV}_p = 675$ m, and  14.6 mrad off line detector.
The  detector is a tracking calorimeter composed of fine-grained cells of liquid scintillator with a total mass of 193~t.
Particle identification is based on the topological information from the tracking 
of particles and uses advanced pattern recognition algorithms.

\paragraph*{Single isolated e-shower.}
The event sample corresponds to $1.66 \cdot 10^{20}$ POT.
The analysis in ref.~\cite{Adamson:2016tbq} selects 
neutrino interaction candidates with total energy in the range 1.5 to 2.7 GeV 
and  maximal $\nu_e-$signal is expected around 2 GeV.
For the signature selection efficiency we adopt the signal selection efficiency: $\epsilon^{NOV}_{e} = 33\%$ \cite{Adamson:2016tbq}. 

The observed event distribution in the calorimetric energy shows 
good agreement between observed, $N_e^{NOVA,obs} = 2573$, and predicted,  
$N_e^{NOVA, th} =  2385$, numbers of  events.
%%% COMMENT: numbers obtained from digitizing fig. 1 of the reference.
Using the satistical uncertainty,  $\Delta N^{stat} = 50.7$,  and the $15\%$ systematic uncertainty, $\Delta N^{syst} =  385.9$, 
we find bounds on new physics contribution: 
\begin{equation}
N_{e}^{NOV} < 577 ~~(1\sigma), ~~~~
966 ~~(2\sigma), ~~~~
1355 ~~(3\sigma)\,.
\label{eq:nova-gamma/ee}
\end{equation}

%%%%%%%%%%%%%%%%%%%%%%%%%%%%%%%%%%%%%%%%%%%%%%%%%%%%%%%%%%%%%%
\subsubsection{NOMAD}
%%%%%%%%%%%%%%%%%%%%%%%%%%%%%%%%%%%%%%%%%%%%%%%%%%%%%%%%%%%%%%%

We also considered the NOMAD experiment with 450 GeV protons impinging 
on a beryllium target, a total POT of  $2.2 \times 10^{19}$,
a baseline of $620$ m, and a detector with length of $3.7$ m and target mass of $3.6$ t.
Among others, the collaboration performed a search for forward photons in ref.~\cite{Kullenberg:2011rd} 
to test the model from ref.~\cite{Gninenko:2009ks}.
We found that in general NOMAD has less testing power compared to the other detectors, 
hence we will not discuss it further in the following.

%%%%%%%%%%%%%%%%%%%%%%%%%%%%%%%%%%%%%%%%%%%%%%%%%%%%%%%%%%%%%%%%
\subsection{On discovery potential}
%%%%%%%%%%%%%%%%%%%%%%%%%%%%%%%%%%%%%%%%%%%%%%%%%%%%%%%%%%%%%%%%

%%%%%%%%%%%%%%%%%%%%%%%%%%%%%%%%%%%%%%%%%%%%%%%%%%%%%%%%%%%%%%
Experiments under consideration are all of the  same type:
accelerator experiments with near or relatively close detectors.
Therefore,
it is straightforward to compare their discovery potentials.
In various cases one can simply compare the ``strengths" of experiments
defined as the product of POT, efficiencies and masses of detectors:
$$
\kappa^i \equiv (POT)^i \times \epsilon^i \times M^i.
$$
Notice that for scenarios with decay, the active volume
of a detector is relevant,  and not the mass.

Apart from this product also other factors are important:
the energy of protons and composition of a target which determine
multiplicities of secondary particles, and consequently,  fluxes of neutrinos.
The length of baseline gives a spread of the neutrino or new particles beams, {\it etc.} 
Therefore,  instead of (POT), one can use immediately the neutrino fluxes
at  detectors:
$$
\kappa^i_\nu \equiv \phi^i_\nu \times \epsilon^i \times M^i,
$$
or the fluxes of heavy neutrinos.
The   MB strength is much higher than the ND one:
$\kappa^{MB} \simeq 2 \cdot 10^{23} $ tons,
while for ND280 $\kappa^{ND} = 4 \cdot 10^{21} $ tons.
Using the neutrino fluxes we obtain comparable strengths:
$\kappa^{MB}_\nu = 5.4 \cdot 10^{13}$ ton cm$^{-2}$,
$\kappa^{ND}_\nu = 2.1 \cdot 10^{13}$ ton cm$^{-2}$,
although the MB strength is still 2.5 times larger.

Further contribution  to the discovery potential comes
from  particle ID. Experiments with  better ID gain since
a smaller subset of events can be selected,  and
therefore stronger bounds on new physics contributions
can be obtained. This can be accounted by the ratio of the strength
over the upper bound on the observed number of events:
$\kappa^{i}_\nu / N^i$. Thus, MiniBooNE has observed 638 1-shower events
while ND280 upper bound is about 150. That is, ND280 gains factor of 3, 
and its discovery potential becomes even slightly higher than the one of MiniBooNE.
Further improvements can be related to specific scenario and
geometry of experiment. Thus,  ND280 can  gain in the decay
scenarios because of smaller baseline. This is precisely the  origin
of upturns (see below)  where the bound becomes stronger.
To a large extent this enhancement is artificial and related to
geometric suppression of number of the MB events. In upscattering
scenarios, sizes of detectors become important.
Similarly, one can  consider discovery potential
of other experiments and searches.\\

For convenience,  we summarize  relevant parameters of the experiments under discussion  in  
the \cref{tab:numbers}. We provide the salient information on analyses of data,   
signatures and the upper bounds on the number of new physics events in \cref{tab:exp-overview}. These bounds 
(see the fourth row) will be confronted with theoretical predictions in \cref{sec:results}.

%%%%%%%%%%%%%%%%%%%%%%%%%%%%%%%%%%%%%%%%%%%%%%%%%%%%%%%%%%%%%%%%%%%%%%%%%
\begin{table}[h!]
\begin{tabular}{|c|c|c|c|c|c|c|}
\cline{1-7}
{\textbf{experiment}} & {\textbf{MiniBooNE}} & {\textbf{T2K}} & {\textbf{NOMAD}} & {\textbf{PS191}} & 
{\textbf{MINER$\nu$A}} & {\textbf{NO$\nu$A}}   \\ \cline{1-7}
area ($\text{m}^2$) & $36\pi$ & $3.47$ & $6.76$  & $18$ & $1.71$ & $12.39\,$   \\ \cline{1-7}
%$\epsilon$ & 0.1 & 0.3 & 0.08 & 0.7 & 0.73 & 0.65  \\ \cline{1-7}
$d$ (m) & $2/3\cdot 12$  &  $d_1=1\,,\,d_2=0.9$   & $3.7$  & $3.55$   & 3   & 8  \\ \cline{1-7}
$l_p$ (m) & 50  & 94  & 290  & 49.1  & 675   & 675   \\ \cline{1-7}
POT ($\nu$+$\bar{\nu}$ mode) & $3 \times 10^{21}$ & $1.821 \times 10^{21}$ & 
$2.2 \times 10^{19}$  & $0.86 \times 10^{19}$  & $3.43 \times 10^{20}$  & $1.66\times \cdot 10^{20}$   \\ \cline{1-7}
$M$ (tonnes) & 818  & $m_{P0D}=15.8$\,,\,$m=1.1$ &  112  & 20  & 6.1 & 300  \\ \cline{1-7}
$\nu$ energy range (GeV) & $[0.1-5]$  & $[0.1-10]$ &  $[5-200]$  & $[0.1,5]$  & $[0.1-20]$ & $[0.1-20]$ \\ 
\cline{1-7}
\end{tabular}
\caption{Parameters that enter in the analysis. 
For T2K280, we list two numbers for detector mass and its length. This is
because we include the possibility of the upscattering in the P0D 
with 1 m distance from TPC-FGD system.
}
\label{tab:numbers}
\end{table}
%%%%%%%%%%%%%%%%%%%%%%%%%%%%%%%%%%%%%%%%%%%%%%%%%%%%%%%%%%%%%%%%%%%%%%

%%%%%%%%%%%%%%%%%%%%%%%%%%%%%%%%%%%%%%%%%%%%%%%%%%%%%%%%%%%%%%%%%%%%%%%%%%%%%%%%%%%%
\begin{table}
\centering
\begin{tabular}{ccccc}
Experiment & Analysis & Signature &  Upper limit $1\sigma /3\sigma$ & Reference \\
\hline\hline
T2K ND280 & Heavy neutrino decays & $e^+e^-$ & $20/49$ & \cite{Abe:2019kgx} \\
	& CCQE electrons & $e^-$ ($e^+$) & $17/261$ & \cite{Abe:2020vot} \\
	& CCQE electrons & single $\gamma$ & $58/305$ & \cite{Abe:2020vot} \\
\hline
NO$\nu$A & CCQE electrons & $e^-$ & $577/1355$ & \cite{Adamson:2016tbq} \\
\hline
%MINER$\nu$A & diffractive $\pi^0$ production  & $\gamma$ & $211/632$ & \cite{Wolcott:2016hws} \\
%{Based on the assumption that the diffractive $\pi^0$ production cross section is true 
%%but fluctuates 2 sigma down and the remaining two sigma are given by the signal.}
MINER$\nu$A	& CCQE electrons & $e^-$ ($e^+$) & $757/1725$ & \cite{Wolcott:2015hda}\\
	& Neutrino electron scattering & EM shower, or $\gamma,ee$ & $23/66$  & \cite{Park:2015eqa} \\
		& Neutrino electron scattering & EM shower, or $\gamma,ee$ & $223/526$ & \cite{Valencia:2019mkf} \\
\hline
NOMAD & Single photon search & single $\gamma$ & $18/50$ & \cite{Kullenberg:2011rd} \\
\hline
PS191 & Heavy neutrino decays & displaced vertex & $1.84 / 6.61$ & \cite{Bernardi:1987ek}\\
	&	Neutrino oscillation & electron-like events & $23 \pm 8$  & \cite{Bernardi:168324} \\
\hline
\end{tabular}
\caption{Summary of considered experimental searches, signatures  and the upper bounds that will be used 
to constrain scenarios explaining MiniBooNE.}
\label{tab:exp-overview}
\end{table}

%%%%%%%%%%%%%%%%%%%%%%%%%%%%%%%%%%%%%%%%%%%%%%%%%%%%%%%%%%%%%%%%%%%%%%%%%%%%%%%%%%
%%% NEW SECTION ON UNCERTAINTIES
%%%%%%%%%%%%%%%%%%%%%%%%%%%%%%%%%%%%%%%%%%%%%%%%%%%%%%%%%%%%%%%%%%%%%%%%%%%%%%%%%%
\subsection{Uncertainties}
{\color{black}
Let us briefly discuss uncertainties in the predictions
of the number of events $N^i$  according to \cref{eq:NDevents}. The
uncertainty from the cross section evaluations were discussed in \cref{subsec:crosssection}.\\

$1)$ The uncertainty of the number of  MiniBooNE excess event
are given by the collaboration: $\delta_{N^{MB}_{1sh, exp}}\approx 0.2$.\\

$2)$ The predictions use input parameters,
such as POT, baseline and detector properties (mass, density,
geometry, etc.). They  are  extremely well controlled
experimentally, and therefore we neglect the associated uncertainties.
A relevant source of uncertainty is the $\phi^0_\pi$ flux
estimated as $\mathcal{O}(10 \%)$.\\

$3)$ For the cases of upscattering in dirt we parametrize
the distance between upstream dirt
and the detector with a single quantity $\Delta$,
ignoring the 3D geometry of a setup. 
In principle, upscattering in the upstream dirt has to be taken into account via a Monte Carlo simulation, which covers the full geometry of the detector hall and the experiment, as well as the model-specific scattering cross section. This is clearly beyond the scope of our work. We estimated
a relative error due to this simplification
by varying $\Delta$ between 1 and 30 meters.
We find that variation affects the predictions of $N^i$
only for $c \tau^0 \gtrsim 1$ m and $\delta_\Delta = 0.4$.
This is comparable to the uncertainty in the density
of dirt: its variation
between 2 and 4 g/cm$^3$ leads to $\delta_{n_b} = 0.3$.\\

Since the above mentioned uncertainties are multiplicative factors
in our prediction, we combine them in quadrature.
Assuming a relative uncertainty for all fluxes to be 0.2, we have for the
mixing decay scenarios $M_N D_\xi$ and $M_N D_\nu U_e$:

\begin{equation}
\delta_M^i = \delta_N \oplus \delta_{\phi_0}^{MB} \oplus \delta_{\phi_0}^{i} \simeq 0.35\,.
\end{equation}

For the upscattering scenarios $U_N D_\xi$ and $U_N D_BD_\xi$ 
we have to add uncertainties due to parameters of
a dirt (distance to a detector, density) and cross
sections. For the latter we use $\delta_\sigma^{MB-i} = 0.2$
according to  \cref{subsec:crosssection}. As a result,

\begin{equation}
\delta_U^i = \delta_M \oplus \delta_{n_b}^{MB} \oplus \delta_{n_b}^{i} \oplus  \delta_\sigma^{MB-i}  \oplus \delta_\Delta \simeq \left\{ \begin{array}{cc} 
0.55 & c\tau^0<1\,\text{m} \\ 
0.68 & c\tau^0 \geq 1\,\text{m} 
\end{array}\right.\,.
\label{eq:deltaU}
\end{equation}

%For the uncertainty of the cross sections of MiniBooNE and experiment
%$i$ we use  $\delta_\sigma^{MB−i} = 0.2$
%according to the reasoning presented in sec.~%\ref{subsec:crosssection}.\\

$4)$ The further sources of uncertainties are
the detection efficiency and the selection efficiency.
Experimental collaborations evaluate both of these
efficiencies through Monte Carlo simulations for specific
models. This is, however, for the moment done only for SM processes; in other words, new physics interactions have not yet been properly implemented in generators such as GENIE.
Our approach is therefore, for a particular process, to adopt
the efficiency from a similar (identical signature, for instance 
single shower in the final state) SM process quoted by experimental 
collaborations. We expect this to yield a very good proxy for the new physics
process efficiencies. We list all employed efficiencies in \cref{sec:limits} and 
if experimental collaboration will move in the direction of considering such specific
processes in the years to come, our results could be straightforwardly rescaled with 
the new values that we do not expect to significantly depart from those employed in this paper.  

}

%%%%%%%%%%%%%%%%%%%%%%%%%%%%%%%%%%%%%%%%%%%%%%%%%%%%%%%%%%%%%%%%%%%%%%%%%
\section{Tests of scenarios}
\label{sec:results}
%%%%%%%%%%%%%%%%%%%%%%%%%%%%%%%%%%%%%%%%%%%%%%%%%%%%%%%%%%%%%%%%%%%%%%%%%%%%%%%%%%

\noindent
The bounds obtained in \cref{sec:limits} apply to the final states
of different scenarios. Therefore, two different scenarios with
the same final EM state have the same tests.
The difference is in implications, that is, in the level of restrictions of 
scenarios. Furthermore, due to misidentification, any signature $s^i$ provides 
bounds on all possible
final states $\xi$, and consequently, scenarios.
%%However, for certain scenarios particular signatures
%%and particular experiments give the best bounds and the problem
%%is to identify these experiments and signatures.
We call {\it the direct test} when the EM component of final state,
$\xi$,  coincides with signature: e.g.  $e - e$-shower, {\it etc.}
{\it The indirect tests} require misidentification.
The most stringent bounds (the best tests) are provided by the direct tests,
since misidentification brings certain smallness.

Several different experiments measure the same type of events (signatures) but
the best bound is given by experiment which has the highest strength.
The latter  allow us to identify the relevant experimental results for
each scenario.

Recall that, according to eq.~(\ref{eq:NDevents}), the predictions 
of numbers of events for all detectors
are normalized to the MiniBooNE excess, {\it i.e.},  to 
the number of $1$-shower events, $N^{MB}_{\xi - 1sh}$, and the latter  is
proportional to $f_{\xi - 1sh}$.

%%%%%%%%%%%%%%%%%%%%%%%%%%%%%%%%%%%%%%%%%%%%%%%%%%%%%%%%%%%%%%%%%%%
\subsection{Mixing - Decay scenario, $M_N D_\xi$}
%%%%%%%%%%%%%%%%%%%%%%%%%%%%%%%%%%%%%%%%%%%%%%%%%%%%%%%%%%%%%%%%%%%

This is the simplest scenario with only two
new physics interaction points: the production point of $N$ via mixing and
the $N-$decay point (see Fig. \ref{fig:MD}). $N$ with mass $m_N \leq 10$ MeV is produced 
in the $\pi$-decays in decay pipe and it decays along the beamline.

\begin{figure*}[h!]
\centering
\includegraphics[width=0.45\textwidth]{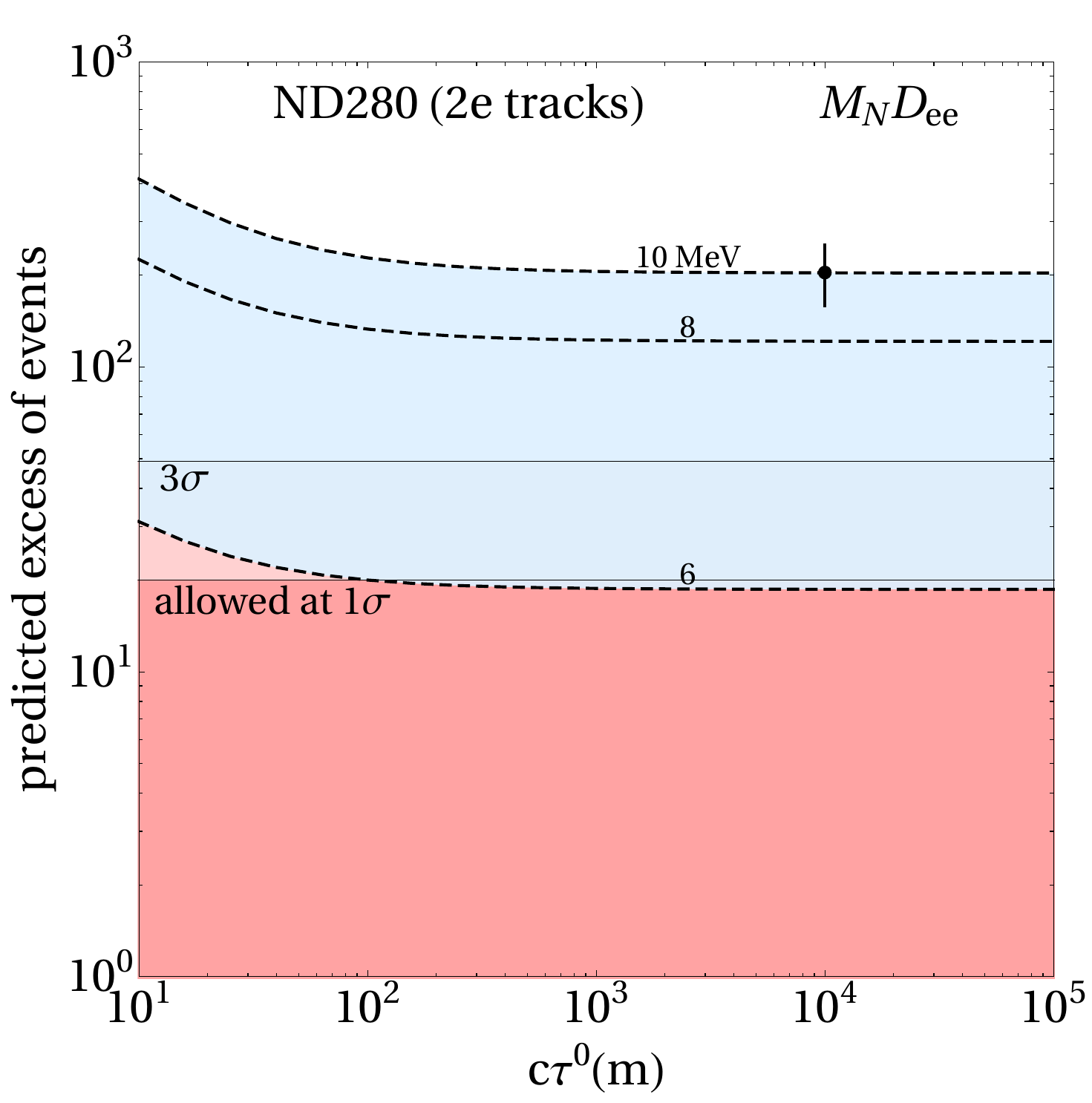}
\includegraphics[width=0.45\textwidth]{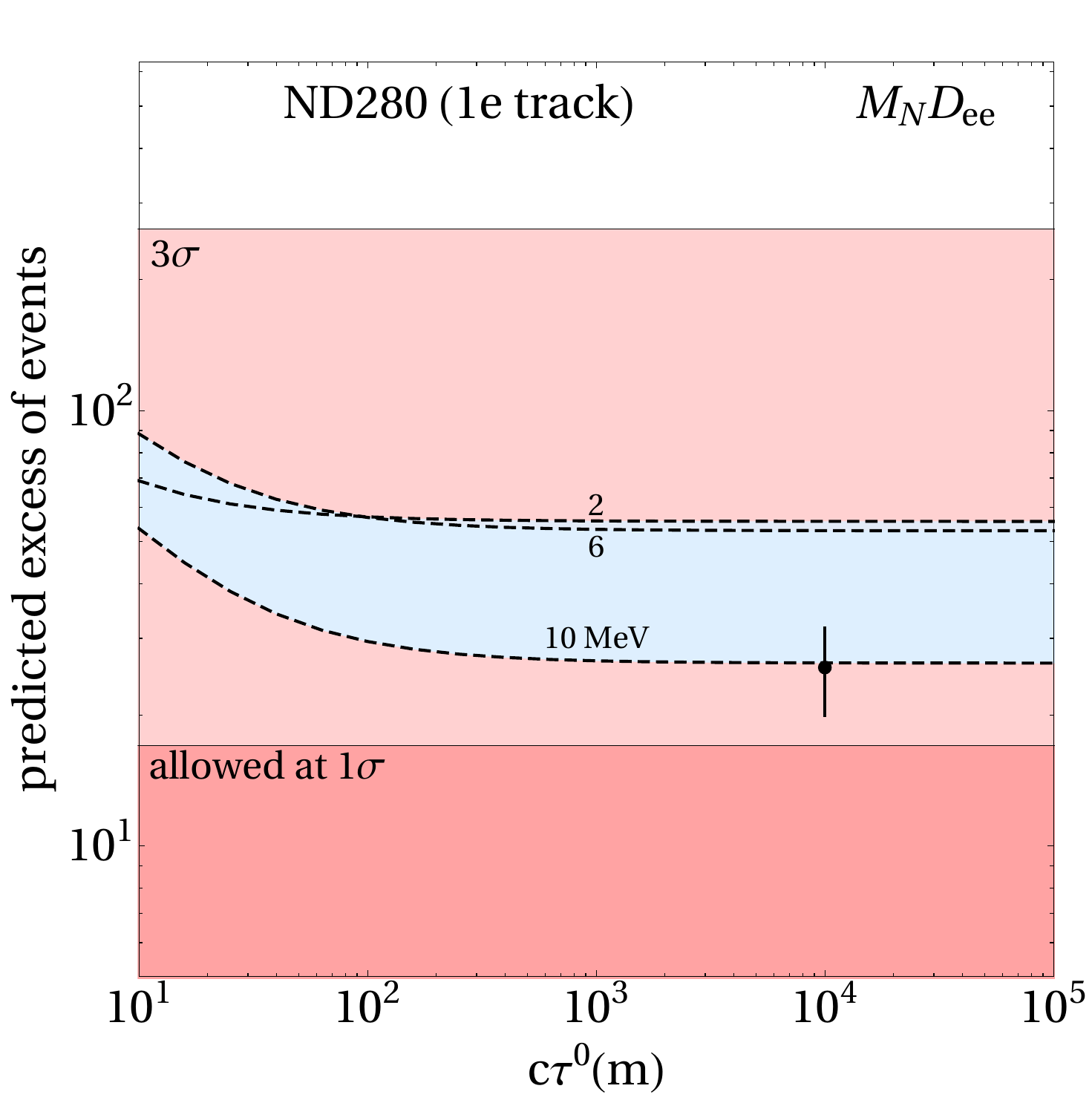}
\caption{
Tests of the Mixing-Decay into $e^+ e^-$ scenario, $M_N D_{ee}$,  at
ND280. {\it Left panel:} Number of expected 2e-shower events produced
by $e^+ e^-$ pair as function
of $c\tau^0$ for different values of $m_N$ (numbers at the curves in
MeV). The point with error bar indicates the uncertainty of the prediction from the MiniBooNE-observed event rate.
Borders of shadowed regions show the $1\sigma$ and $3\sigma$
experimental upper bounds on these numbers.
{\it Right panel:} The same as in the left panel but for
the $1e-$shower events at ND280.
}
\label{fig:T2K1showerEE}
\end{figure*}

\begin{figure*}[h!]
\centering
\includegraphics[width=0.45\textwidth]{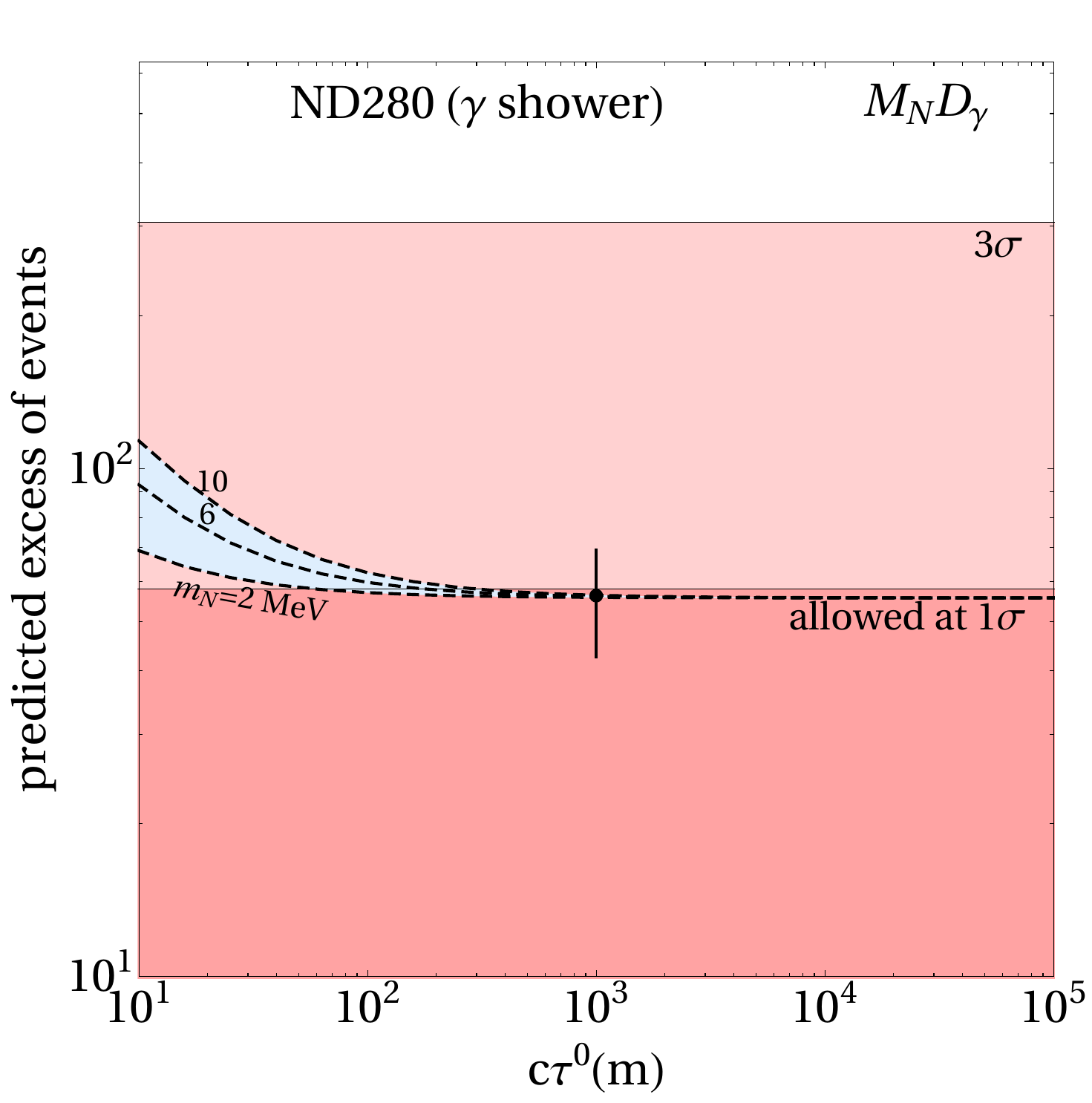}\quad
\includegraphics[width=0.45\textwidth]{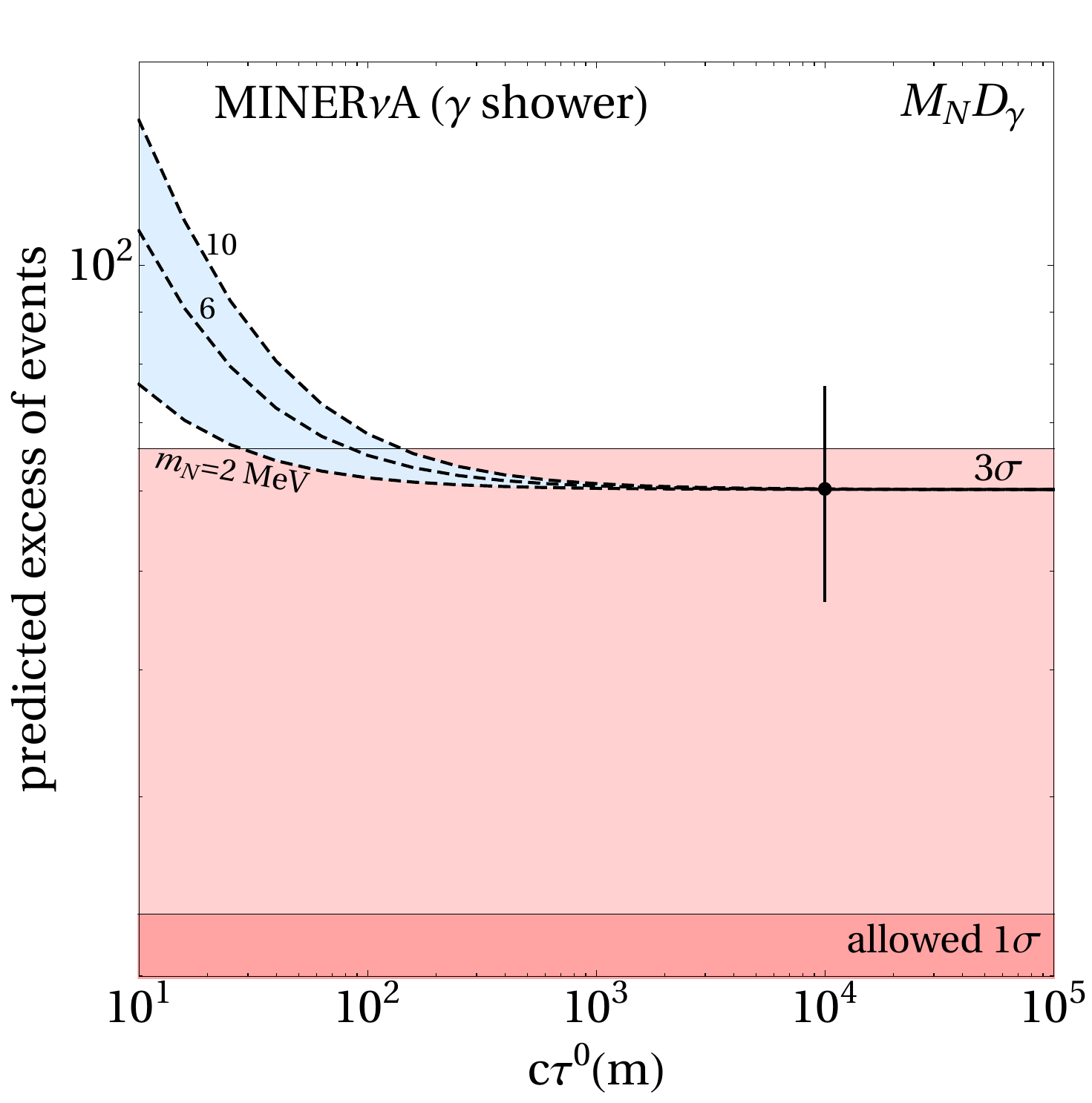}
\caption{
Tests of the Mixing-Decay into $\gamma$ scenario, $M_N D_{\gamma}$. 
The number of expected $\gamma$-shower events is shown
%produced by $\gamma$ 
as a function
of $c\tau^0$ for different values of $m_N$
(numbers at the curves in MeV).
Borders of shadowed regions show the $1\sigma$ and $3\sigma$
experimental upper bounds on these numbers.
The point with error bar indicates the uncertainty of the prediction 
from the MiniBooNE-observed event rate.
{\it Left panel:} ND280, {\it Right panel:} MINER$\nu$A.
}
\label{fig:singlegamma}
\end{figure*}

The typical dependence of the  number of events on $c\tau^0$
(see \cref{subsec:MD}) has the exponential upturn 
% below the upturn point
and  constant asymptotics at $c\tau^0 \rightarrow \infty$  
(see Figs. \ref{fig:T2K1showerEE} and \ref{fig:singlegamma}).
The upturn point is determined by the baseline and typical energy 
of the MiniBooNE experiment \cite{Aguilar-Arevalo:2018gpe}. In our approximation 
of the $E_N-$independent signature factors  such a behavior is the same for all
possible final states $\xi$.

The absolute value of the excess of events 
in a given  experiment is determined by the product (\ref{eq:ndass}).
%%For computations of the number of events we use Eq. (\ref{}).
The final states produced in the $N-$decay are   $\xi = \gamma$
(radiative decay) and $\xi = e^+ e^-$ (three body decay).
Also $2\gamma$ final state can be explored,  but  $\xi = e$ is not
possible.
Let us consider $\xi = e^+ e^-$ and $\xi = \gamma$ in more detail.\\

\emph{1. $\xi = e^+ e^-$: $M_N D_{ee}-scenario$}: 
The $N^{ND}_{ee-2sh}$ result (\ref{eq:2sh-ndws}) provides the direct test, 
and therefore gives the strongest bound.
Bounds from other data rely  on  
the mis-identification of 
$e^+e^- -$showers with  $e-$ or  $\gamma-$showers and 
require small invariant mass of the $e^+e^-$ pair, $W_{ee}$. 
In this scenario an angular 
selection cut of $\cos\theta\geq 0.99$ is well satisfied and therefore the selection  efficiency is close to  100\%.

$(a)$ For the invariant mass of the pair $W_{ee} > W_c = 5 $ MeV,
the electron and positron are resolved in ND280 and therefore 
the bound on $2e-$shower events $N^{ND}_{ee-2sh}$ (\ref{eq:2sh-ndws}) can be used. 
In \cref{fig:T2K1showerEE} (left panel) we show the dependence of
$N^{ND}_{ee - 2sh}$ on $c\tau^0$
for three values of mass, $m_N$,
allowed by timing restriction (see Sec. \ref{subsec:timing} and  \cite{Aguilar-Arevalo:2020nvw}).     
In our computations, we used the expression (\ref{eq:NDevents}) for $N^{ND}_{ee - 2sh}$ 
with parameters of
the experimental setup given in the \cref{tab:numbers} and 
$f_{ee - 2sh}$  found  with eq. (\ref{eq:invariantmass}).
For the $N$ flux at $m_N\lesssim 10$ MeV we use the active neutrino flux 
reduced by the mixing parameter $|U_{\mu N}|^2$ 
as a proxy.

%%%%%ffff7%%%%%%%%%%%%%%%%%%%%%%%%%%%%%%%%%%%%%%%%%%%%%
%%%%%%%%%%%%%%%%%%%%%%%%%%%%%%%%%%%%%%%%%%%%%%%%%%%%%%%%The upturn is at
%$c\tau^0 < 3 \cdot 10^2$ m.
Fig. \ref{fig:T2K1showerEE} shows very strong dependence of the expected number of events on
$m_N$ which comes
mainly from the signature factors. Indeed,
$N^{ND}_{ee - 2sh} \propto f_{ee - 2sh}^{ND}/f_{ee - 1sh}^{MB}$.
In MiniBooNE, with $W_c = 30$ MeV, the $e^+ e^-$ pairs  are not resolved: $W_{ee}< m_N < W_c$,
so that $f_{ee - 1sh}^{MB} = 1$.  In ND280, the values of
mass $m_N$ are close to the
threshold and therefore $f_{ee - 2sh}^{ND}$ increases  strongly with
$m_N$.

According to the figure, the $M_N D_{ee}$ scenario with $m_N > 7$ MeV is excluded.
The bound relaxes with decrease of $m_N$,  
being below $\sim 1\sigma$ for $m_N < 7$ MeV.\\

$(b)$  For $W_{ee} < 5$ MeV, the $e^+e^-$ pairs show up in ND280 as $1sh-$events.
Their number can be restricted by results of studies of the $e-$showers produced by the $\nu_e-$CCQE at ND280,
as well as at PS191, NOMAD and MINER$\nu$A.
Notice that this is an indirect test which relies on  misidentification.

In \cref{fig:T2K1showerEE} (right panel), we show  the expected number
of 1 shower events at ND280 produced by the $e^+ e^-$ pairs.
The dependence of $N_{ee - 1sh}^{ND}$  on $m_N$ is strong but opposite to that
for the  2 shower events:
$N_{ee - 1sh}^{ND}$ decreases with increase of $m_N$, again, due to signature
factor $f_{ee - 1sh}^{ND}$. According to Eq.~(\ref{eq:invariantmass}), for
$m_N$ above the threshold,  $f_{ee - 1sh}^{ND} \propto W_c^2/m_N^2$.
(This reflects the fact that probability of the 3-body $N-$decay
with invariant mass of the pair $W_{ee} < W_c$ decreases.) 
The opposite dependence of number of events on $m_N$ in $1sh-$ and $2sh-$cases can be also inferred 
from the sum rule:  $f_{ee - 2sh}^{ND} = 1 - f_{ee - 1sh}^{ND}$.

We confront the predictions with the bound
(\ref{eq:nd-e}).  According to \cref{fig:T2K1showerEE} (right panel),
the $M_N D_{ee}$ scenario  with  $m_N < 6$ MeV is disfavored at about
$2\sigma$ level in the whole range of
$c\tau^0$. The bound weakens with the increase  of $m_N$.\\

For small $W_{ee}$, the final $e^+ e^-$ state  can also be
mis-identified  with $\gamma-$shower.
In such a case the bounds on $1\gamma-$shower searches  of new physics 
by NOMAD, ND280, PS191, MINER$\nu$A can  be applied 
(see for instance \cref{eq:ngammaND,eq:mv-gamma/ee_LE}). \\

\emph{2. $\xi = \gamma$, $M_N D_\gamma-$scenario}: 
The direct tests of this scenario are provided by the $1\gamma$ shower 
searches of new physics at ND280, MINER$\nu$A 
and NOMAD.  %%(\ref{eq:nom-gamma/ee}), % 
In \cref{fig:singlegamma} we present results  
for ND280 (left) and MINER$\nu$A (right).
NOMAD gives much  weaker bounds than  ND280 and MINER$\nu$A.
In our computations, we used  $f_{1\gamma} = 1$, and the values of $\epsilon$ 
from  the \cref{tab:numbers} (see also \cref{sec:limits}).
According to this figure, the predicted number of $1\gamma$ events
is at the level of $1\sigma$ upper bound from ND280, see  Eq. (\ref{eq:ngammaND}). 
Future ND280 data may improve the bound. MINER$\nu$A gives much stronger restriction, see   
Eq. (\ref{eq:mv-gamma/ee_LE}).
%%% REMOVED diffractive pi0 search
%Eq. (\ref{eq:mv-gamma}\,,\ref{eq:mv-gamma/ee_LE}).  
For $c\tau^0 > 10^2$ m, the prediction
is at $3\sigma$ exclusion and at
$c\tau^0  < 10^2$ m  the bound becomes stronger than $3\sigma$
especially for larger  values of $m_N$.

%%%%%%%%%%%%ffff8%%%%%%%%%%%%%%%%%%%%%%%%%%%%%%%%%%%%%%%%%%%%%%%%%%%%%%%%%%%
%%%%%%%%%%%%%%%%%%%%%%%%%%%%%%%%%%%%%%%%%%%%%%%%%%%%%%%%

The model with $c\tau^0\gtrsim 10^3$ m and $m_N \sim 250$ MeV
which fits this scenario (but with  much larger masses of $N$) was proposed 
in \cite{Fischer:2019fbw}.
It is excluded by timing constraints, and independently disfavored
by our consideration.\\

The bounds obtained here can be applied to the  mixing - double decay scenario
$M_N D_B D_\xi$ considered in \cref{subsec:5th}. In the limits  
$\lambda_B \ll \lambda_N$ and $\lambda_B \gg \lambda_N$,
they can be applied immediately. In the case $\lambda_B  \sim \lambda_N$, 
the predicted number of events should be corrected by factor 
(\ref{eq:dopfac}) which is about 0.4 for ND280.  
For other possibilities we can introduce scaling:  
$\lambda_B = \alpha \lambda_N$ and
$m_B = \beta m_N$, where $\alpha$ and $\beta$ are constants, 
and present results in the same way as for the 2-parameter scenarios, 
namely, as number of events as function of $c\tau_N^0$
for different values of $m_N$.
Model \cite{Abdallah:2020biq} fits this scenario with $\lambda_N \rightarrow 0$
(or $U_N D_\xi$ scenario with $N$ substituted by $B$).

%%%%%%%%%%%%%%%%%%%%%%%%%%%%%%%%%%%%%%%%%%%%%%%%%%%%%%%%%%%%%%%%%%%
\subsection{Upscattering - Decay scenario, $U_N D_\xi$ }
\label{subsec:a}
%%%%%%%%%%%%%%%%%%%%%%%%%%%%%%%%%%%%%%%%%%%%%%%%%%%%%%%%%%%%%%%%%%%

Recall that here $N$ is produced by %QE 
the $\nu_\mu-$upscattering in a detector as well as in matter between
a decay pipe and a detector. In turn,  $N$ decays in the detector (see \cref{subsec:UD}). 
This scenario has final states $\xi$ and signatures similar to those of  $M_N D_\xi$,
since in both cases the final state is produced in the $N-$decay.
%%Also it has the same relevant experiments and  bounds.
The difference is in the geometry of the $N-$production part, and consequently, in the $c\tau^0$ dependence,
as well as in the larger allowed values of $N$ mass: $m_N \gtrsim 100$ MeV. 
Timing constraints are much weaker in this scenario with respect 
to $M_N D_\xi$.

According to \cref{subsec:UD},
the contribution to the number of events from the $\nu_\mu-$upscattering 
in the detector has a smoothed step-like dependence on $c\tau^0$ 
with transition region between the two asymptotics at 
$D^i < c \tau^0  < D^{MB}$, where $D^i \equiv d_i m_N/E_N$ is the reduced size of a detector.
The contribution from  the $\nu_\mu-$upscattering in outer matter 
is negligible at small $c\tau^0$ and it increases, first linearly, 
then reaches its maximum at $D^i < c \tau^0$  followed 
by a decrease  toward a constant value in the asymptotics.
The sum of the two contributions produces a ``bumpy'' form in the transition region (see Figure  
\ref{fig:T2K2showerEE} below).
Substantial difference from the $M_N D_\xi$ scenario in terms of tests and relevance of
experimental bounds is related to the masses of $m_N$,
which affects the signature factors $f$. The latter can suppress or enhance expected numbers of events. 
The final states $\xi$  can be $e^+ e^-$ and $\gamma$ and we  will consider them in order.\\

\emph{1. $\xi = e^+ e^-$ --  $U_ND_{ee}$ scenario}: ND280 data on $e^+ e^-$ pairs 
provide the direct test of this scenario. Due to large mass of $N$, $m_N \gg W_c^{ND} = 5$ MeV, 
the signature factor $f_{ee - 2e sh}$  is close to 1.
We evaluated the efficiency of the angular selection cut $\cos \theta >0.99$ 
for $m_N$ masses of 150, 250 and 350 MeV (indicated in the figures) and gauge boson masses corresponding to the benchmark 
points or Ref. \cite{Bertuzzo:2018itn} (partially coherent) and \cite{Ballett:2018ynz} (incoherent). We found that for incoherent (partially coherent) scattering 
roughly $10\%$ $(40\%)$ of the signal events pass this selection cut.

Furthermore, we found that this angular cut corresponds to the hadronic 
recoil momenta  below the detection threshold 
which is $\sim 400$ MeV in ND280 \cite{Alexander}).
This means that incoherent scattering will not receive further 
efficiency reductions 
from veto on events related to the absence of hadron activity. 

In the left panel of \cref{fig:T2K2showerEE}  we show the predicted number of $2e-$track events,
$N_{ee - 2e tr}^{ND}$,  as function of $c \tau^0$.
The theoretical value $N_{2e-2sh}^{ND}$ has been computed using \cref{eq:NDevents,eq:totnee}.
The $N-$flux at the detector was found using \cite{Abe:2012av}.
The bump in the prediction at $c\tau^0\simeq 0.1$ m is due to the contribution  
from $\nu_\mu-$upscattering in the pion detector (P0D) in addition to scattering in TPC+FGD system, 
and we consider detection of events in the latter only.
The bump is significant, since P0D has larger mass than TPC-FGD.  
The surrounding dirt with length $b = 140$ m has also been taken into account.  
%%given the baseline of 280 meters and $l_p\simeq 100$ m, the dirt extends over more than 100 meters. 

The predicted number of events strongly depends on $m_N$.
This dependence follows from the MB signature factor
$f_{ee - 1sh}^{MB}$  which appears in the expression for $N_{ee - 1sh}^{MB}$
in the denominator of (\ref{eq:NDevents}).
From \cref{eq:invariantmass} 
%%with constant $W_{ee}$  
we have
\begin{equation}
f_{ee - 1e}^{MB} \sim \frac{2(W_c^{MB})^2}{m_N^2},
\label{eq:f-dep}
\end{equation}
while in the numerator $f_{ee - 2etr}^{ND} \approx 1$.  Consequently,
$N_{ee - 2etr}^{ND, obs} \propto m_N^2$.
%%
%%since $f_{1e}^{ND} \ll 1$,  and therefore its dependence on $m_N$ is weak.
%%Indeed, the figure confirms $N_{2e}^{ND, pred} \propto m_N^2$.
Let us underline that this  dependence on $m_N$  comes
from the theoretical number of events at MiniBooNE: with increase of $m_N$, 
the decrease of $f_{ee - 1sh}^{MB}$ (\ref{eq:f-dep}) 
should be compensated by increasing  other factors in
$N_{ee - 1sh}^{MB}$ (e.g., coupling constants) which are also present in the expression for 
$N_{ee -2e sh}^{ND}$.

In \cref{fig:T2K2showerEE} two sets of lines correspond to the partially coherent
$N$ production on nuclei realized for light mediators ($\sim 30$ MeV)
and to the incoherent $N$ production  due to heavy ($> 1$ GeV) mediators 
(see corresponding discussion in \cref{sec:limits}).
The difference between usage of these two types of cross sections 
is not large since the same type of cross section is used in the numerator 
and denominator of \cref{eq:NDevents}. The mild differences appear 
in the intermediate region of $c\tau^0$ where P0D and dirt also contribute.

According to the left panel of \cref{fig:T2K2showerEE}, the experimental bound (\ref{eq:2sh-ndws}) 
excludes the scenario for $c\tau^0 \gtrsim 10^{-2}\,\text{m}$ and 
$m_N > 50$ MeV at more than $3\sigma$ confidence level. 
For smaller values of $c\tau^0$ this exclusion weakens exponentially because $N$ produced in the FGD 
would decay already within FGD and that would be vetoed.
The model in \cite{Ballett:2018ynz} matches this scenario with 
$m_N = 110$ MeV and $c\tau^0\gtrsim 1$ m, where $N$ is produced incoherently,  
since mediator mass for the benchmark point is 1.25 GeV. Such model is  
excluded by the $2e-$tracks  ND280 data.  
(See \cite{Coloma:2019qqj} for the independent test of this model in Icecube). 

%%\textcolor{red}{[[ hadronic veto less stringent than angular selection, discussed above ]]}
%\newtext{\sout{Let us note that in drawing this conclusion we have not treated hadronic-activity 
%cuts that can yield certain suppression on number of expected events 
%for incoherent upscattering scenario; 
%instead, we [[assume]] that the additional tracks induced due to such effects are rare or missed. }
%%The exact treatment requires detector simulation that is beyond the scope of this paper.

%%%%%%%%%%%%ffff7%%%%%%%%%%%%%%%%%%%%%%%%%%%%%%%%%%%%%%
\begin{figure*}[h!]
\centering
\includegraphics[width=0.45\textwidth]{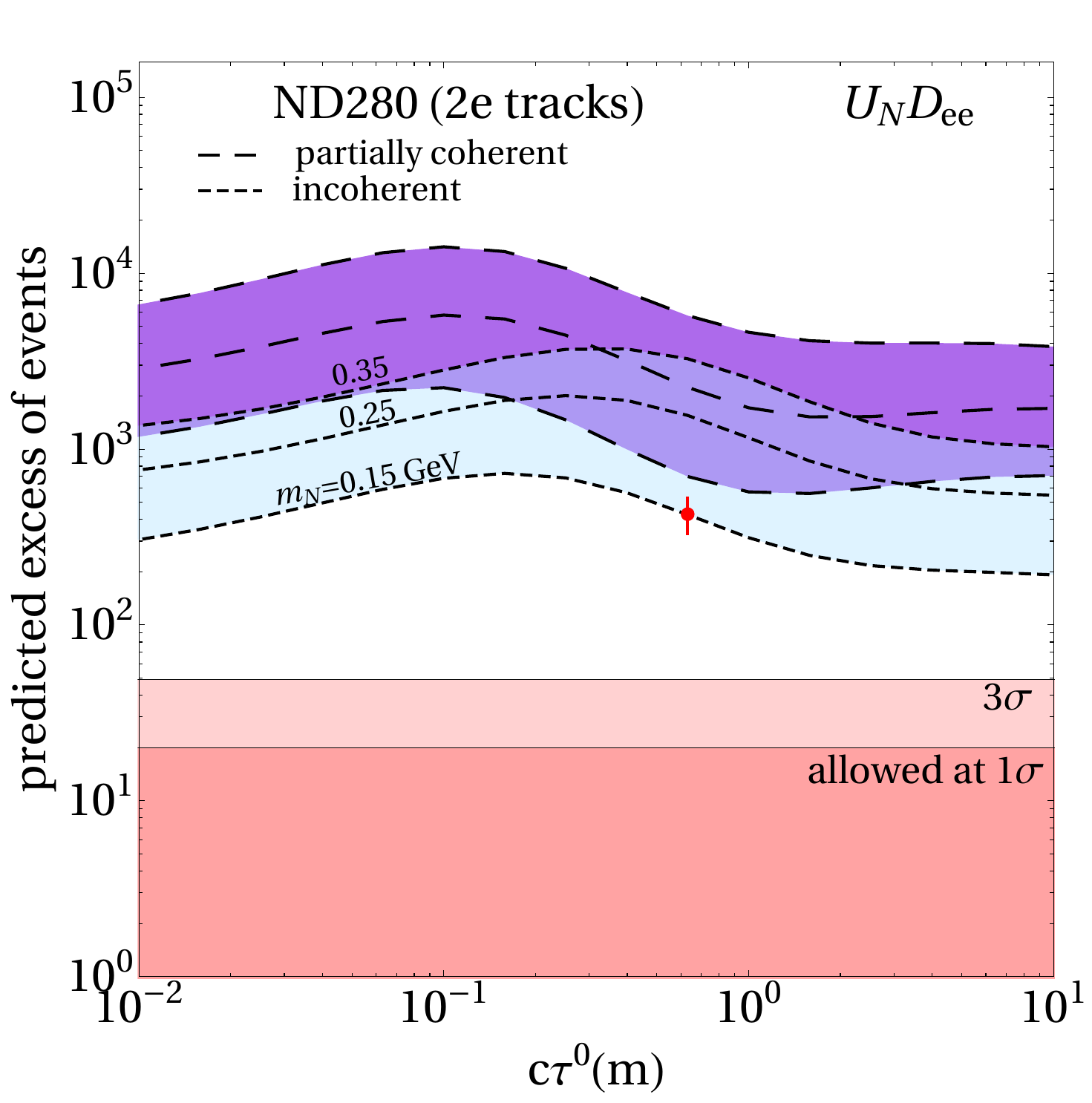}\quad
\includegraphics[width=0.45\textwidth]{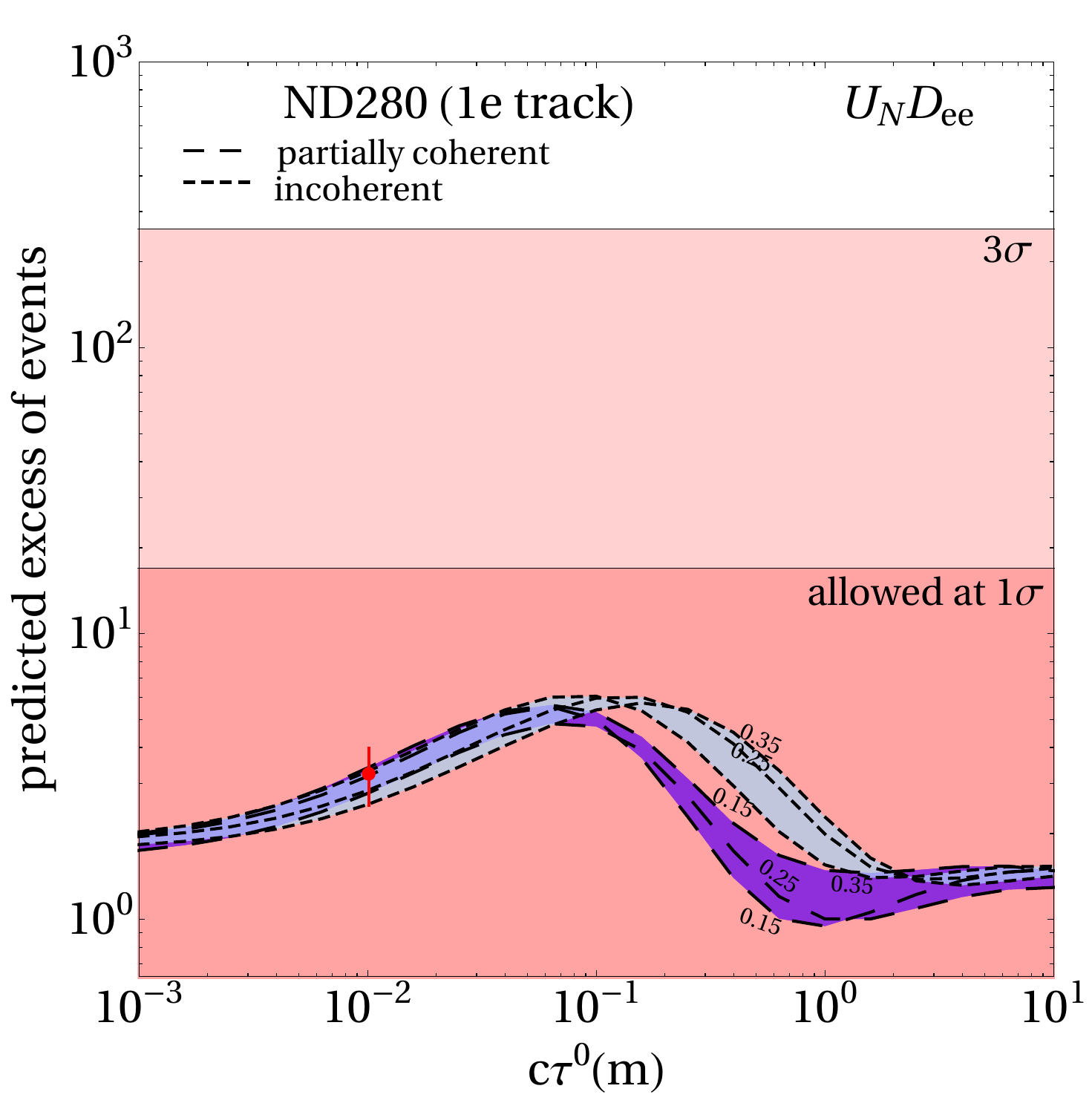}\quad
\caption{
Tests of the Upscatering-Decay into $e^+ e^-$ scenario,
$U_N D_{ee}$ at ND280. {\it Left panel:} The number of expected $2e-$track events
produced by the $e^+ e^-$ pairs as a function
of $c\tau^0$ for different values of $m_N$
(numbers at the curves in GeV).
The point with error bar indicates the uncertainty of the prediction from the MiniBooNE-observed event rate. 
Two sets of lines correspond to contributions computed
with partially coherent and incoherent cross sections.
%%The $N$ production in dirt is taken into account.
The horizontal lines show the $1\sigma$ and $3\sigma$
experimental upper bounds.
{\it Right panel:} the same as in the left panel, but for
the $1e-$track events at ND280.
}
\label{fig:T2K2showerEE}
\end{figure*}
%%%%%%%%%%%%%%%%%%%%%%%%%%%%%%%%%%%%%%%%%%%%%%%%%%%%%%%%

As a representative of indirect test for this scenario 
we use  the $1e-$track events studied at ND280. 
The right panel of \cref{fig:T2K2showerEE} shows the predicted excess 
of 1 track events  induced by the $e^+ e^-$ pairs. These events require very low $W_{ee}$ 
and the $ee - 1sh$ mis-identification.  
The predicted number of excess events
has  dependence on $c\tau^0$ similar to that in the left panel. 
Since the signature  factors for both  ND280 and MiniBooNE 
have the same $1/m_N^2$ dependence,  there is no  signature factor enhancement and 
dependence of predictions on $m_N$ is much weaker than in $\xi = e^+e^-$case. 
The predicted excess of events is below $1\sigma$ limits from \cref{eq:nd-e}.

\begin{figure*}[h!]
\centering
\includegraphics[width=0.4585\textwidth]{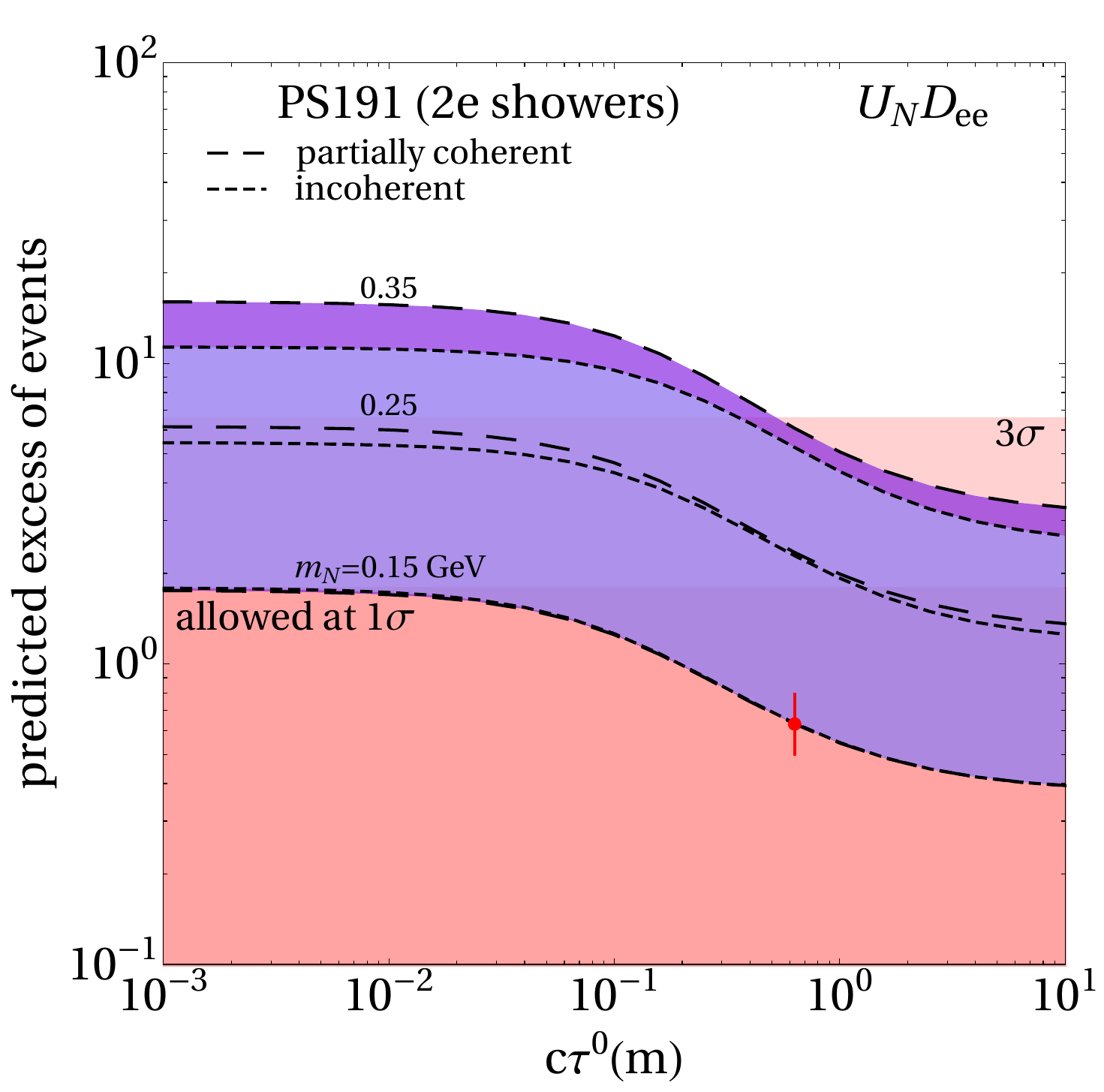}\quad
\includegraphics[width=0.45\textwidth]{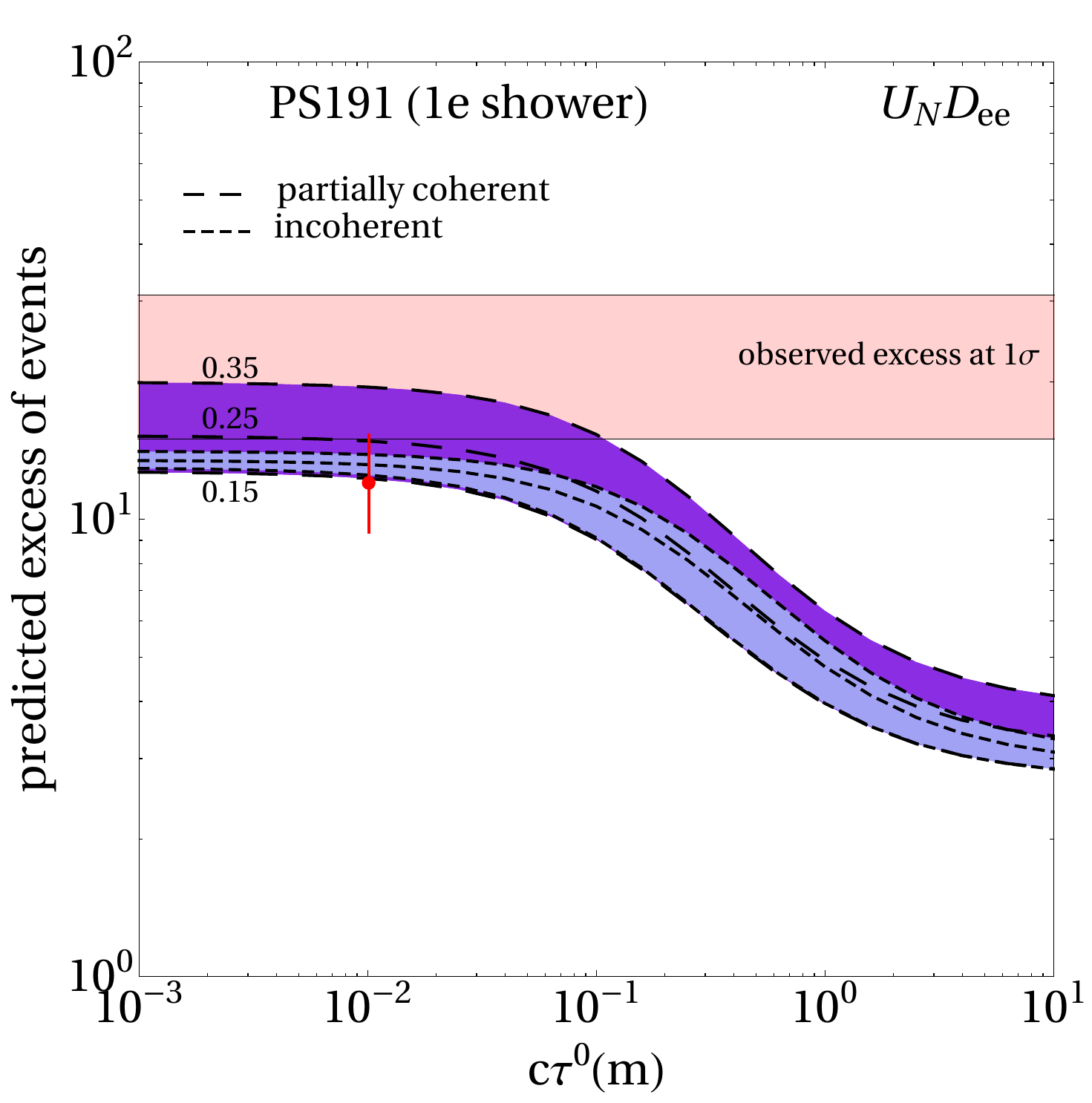}\quad
\caption{
The same as in \cref{fig:T2K2showerEE} but at PS191.
}
\label{fig:PS2showerEE}
\end{figure*}

The direct test of the $U_ND_{ee}$ scenario is given by
the bound on the two track events from PS191 experiment (\ref{ps:vertex}).
In the left panel of \cref{fig:PS2showerEE} we show the dependence of  $N_{ee - 2tr}^{PS}$ 
on $c\tau^0$. For PS191 we did not include the dirt contribution. 
Hence in  both panels one finds the  expected smoothed step form 
of the dependence.
%% between the asymptotics at small and large $c\tau^0$. 
%It has usual dependence of the smooth step form.
The dependence on $m_N$ has the same origin as in \cref{fig:T2K2showerEE}.
The total number of expected events is, however, much smaller than in ND280 
due to low strength $\kappa_\nu$ for PS191, in particular, due to  low number of POT (see \cref{tab:numbers}). 
Strong bound (more than $3\sigma$) on this scenario appears for large values of masses,  
$m_N > 0.25$  GeV,  and short decay lengths: $c\tau^0 < (0.1 - 1)$ m.

In the right panel of \cref{fig:PS2showerEE} we show prediction for the number of $1sh-$events originated from
the $e^+ e^-$ pairs. Mis-identification $e^+ e^-~ - ~1sh$  requires the  low threshold $W_{ee} < W_c^{PS} = 30$ MeV.
According to \cref{fig:PS2showerEE}, the
$U_ND_{ee}$ scenario could explain the observed excess of events at PS191. However, the required
values of parameters  are already excluded at more than $3\sigma$ 
by two track events at ND280 (see \cref{fig:T2K2showerEE}).

%%%%%%%%%%%%ffff10%%%%%%%%%%%%%%%%%%%%%%%%%%%%%%%%%%%%%%
%%%%%%%%%%%%%%%%%%%%%%%%%%%%%%%%%%%%%%%%%%%%%%%%%%%%%%%%

%%%%%%%%%%%%ffff11%%%%%%%%%%%%%%%%%%%%%%%%%%%%%%%%%%%%%%
\begin{figure*}[h!]
\centering
\includegraphics[width=0.45\textwidth]{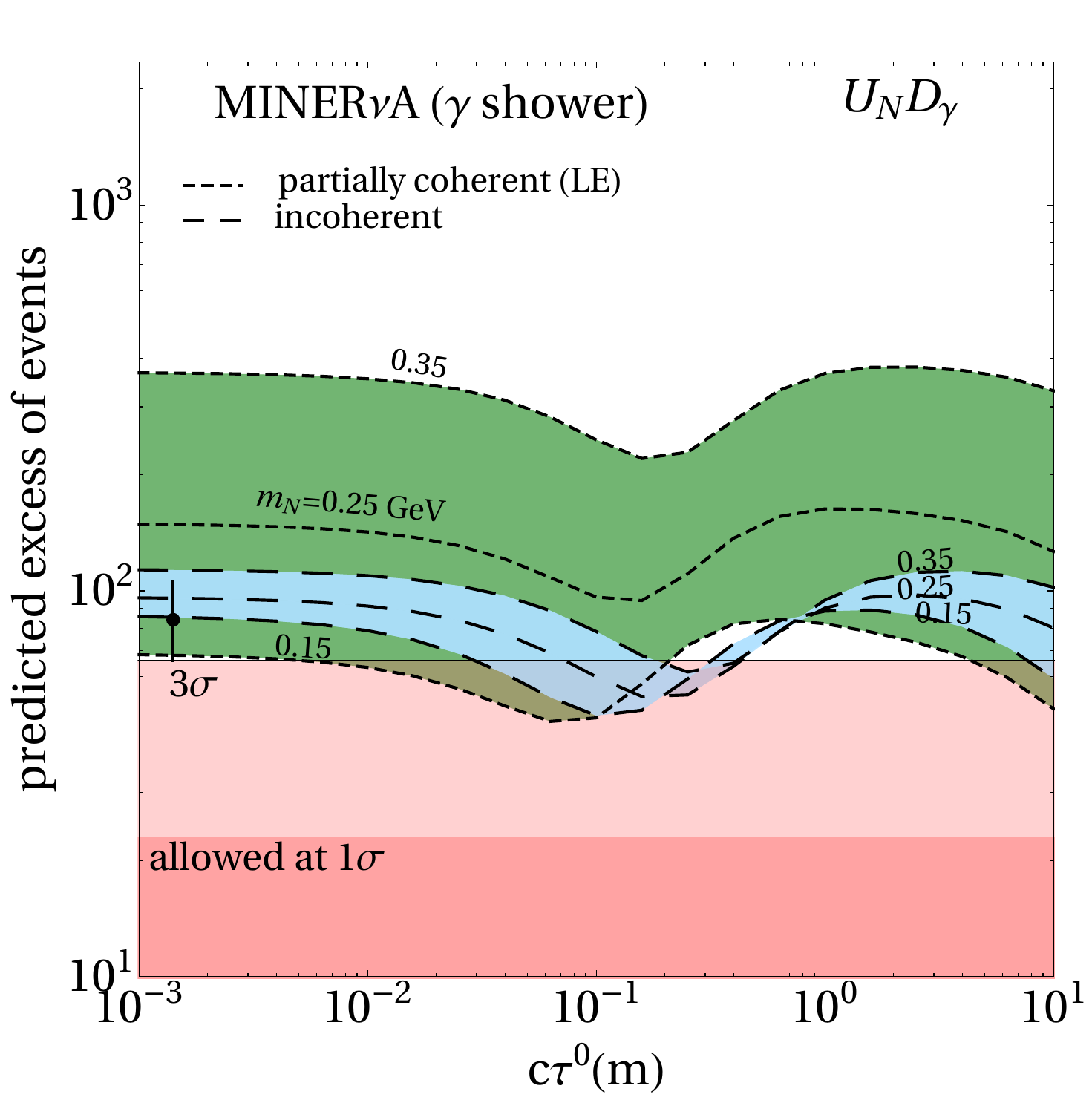}\quad
\includegraphics[width=0.45\textwidth]{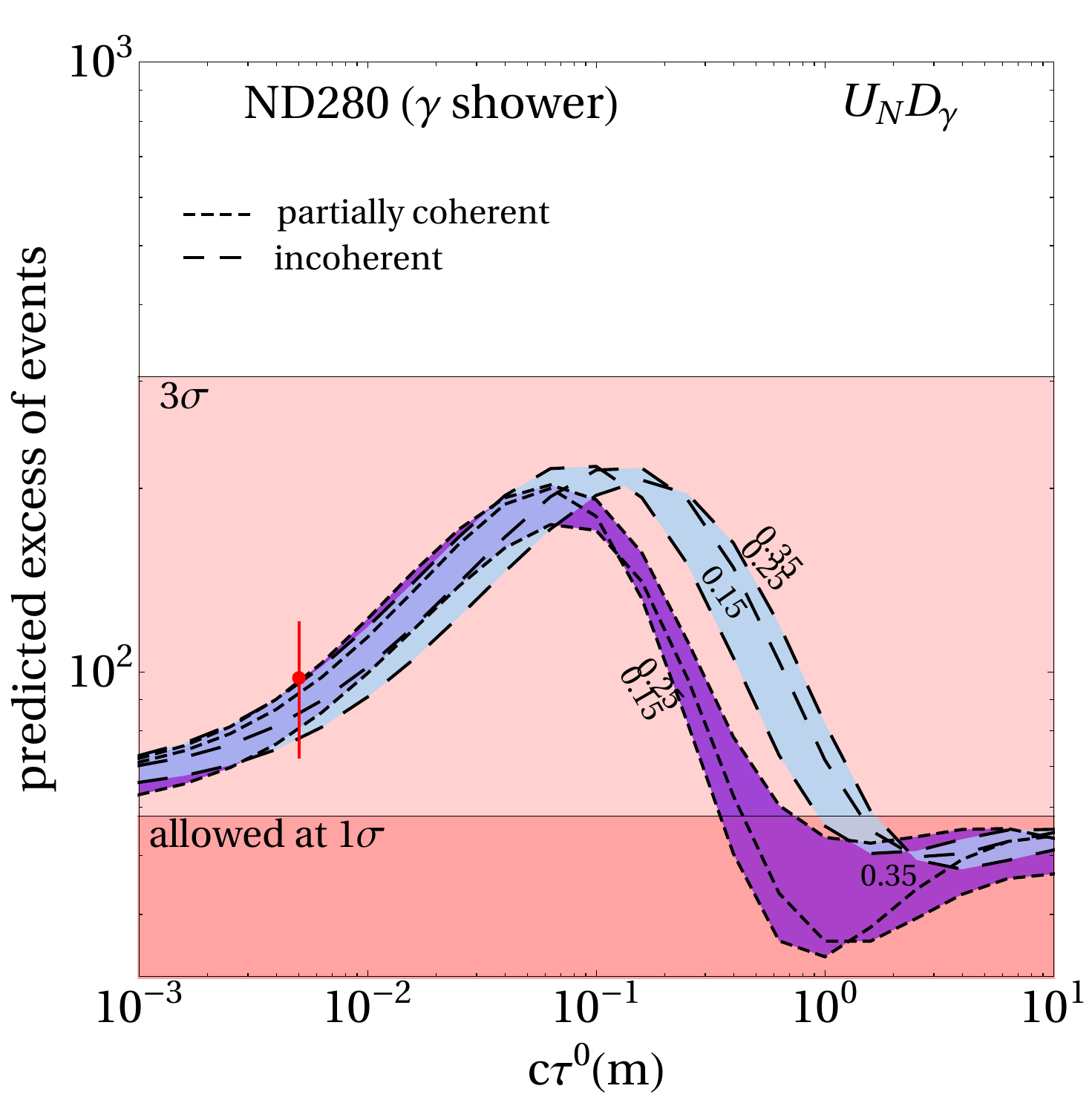}\quad
\caption{
Direct tests of (Bounds on) the Upscatering-Decay into $\gamma-$ scenario,
$U_N D_{\gamma}$ by different experiments. Number of expected
$\gamma$-shower events as function
of $c\tau^0$ for different values of $m_N$
(numbers at the curves in GeV) is shown.
Horizontal lines show the $1\sigma$ and $3\sigma$
experimental upper bounds on these numbers.
The point with error bar indicates the uncertainty of the prediction from the MiniBooNE-observed event rate.
Two sets of lines correspond to contributions computed
with partially coherent and incoherent cross sections.
{\it The  left panel} is for MINER$\nu$A, while {\it the right panel}  corresponds to ND280.
}
\label{fig:ud-gamma}
\end{figure*}
%%%%%%%%%%%%%%%%%%%%%%%%%%%%%%%%%%%%%%%%%%%%%%%%%%%%%%%%

\emph{2. $\xi = \gamma$: $U_ND_\gamma$ scenario}: It can be directly tested at several 
detectors, and in particular,  at MINER$\nu$A and ND280.

%\textcolor{red}{The  MINER$\nu$A bound (\ref{eq:mv-gamma/ee_LE}) is obtained from 
%the $\nu e-$ scattering analysis 
%with  a selection cut $E \theta^2 < $0.0032 GeV [[wrong cut for our study????]]. 
%This corresponds to the selection efficiency $\epsilon^{MV}_\gamma  =  0.1$ [[???]].
%The same reasoning as for the angular cut in ND280 makes us expect this makes hadronic activity negligible.
%[[ the cut of $E_N \theta^2 < $0.0032 GeV corresponds to $\cos\theta_N > 0.9984$ 
%and is much more stringent than at ND280 $\Rightarrow$ hadronic although the threshold 
%for hadron recoil momentum is only 31 MeV  ???? ]] }

In the left panel of \cref{fig:ud-gamma},
the number of isolated $\gamma$ events in MINER$\nu$A $N^{MV}_{\gamma- \gamma sh}$ is shown as a function
of $c\tau^0$. Both contributions from upscattering in the detector 
and in the dirt are included; the latter induces a bump at $c\tau^0 = (1 - 5)$ m depending 
on the value of $m_N$ (if there was no dirt effect included, 
the shape would qualitatively resemble \cref{fig:PS2showerEE}).
Both in MINER$\nu$A and MiniBooNE the signature factors for this channel 
are close to $1$ and the strong dependence of $N^{MV}_{\gamma- \gamma sh}$  on $m_N$ follows 
from coherent cross section: With the increase of $m_N$, 
the cross section for partially coherent scattering 
drops strongly around the typical MiniBooNE energy $E_N^{MB} \sim 0.8$ GeV, while for
MINER$\nu$A with $E_N^{MV} \sim 5$ GeV  the decrease is much weaker
\begin{align}
N^{MV}_{\gamma- \gamma sh} \propto \frac{\sigma^{coh} (E_N^{MV}, m_N)}{\sigma^{coh} (E_N^{MB}, m_N)}\,.
\label{eq:ved}
\end{align}
%%
%For typical MiniBooNE energy $E_N^{MB} \sim 0.8$ GeV the decrease is very strong, while for
%MINER$\nu$A with $E_N^{MV} \sim 5$ GeV  the decrease is much weaker. 
As a result, $N^{MV}_{\gamma- \gamma sh}$
increases with $m_N$.  In the case of incoherent $N-$production, 
the dependence of the cross section on $m_N$ is weak.   
%%being related to integration over the phase space.

According to the left panel of \cref{fig:ud-gamma}, the experimental result (\ref{eq:mv-gamma/ee_LE}) 
excludes the present scenario in the whole range of
$c\tau^0$ and for $m_N > 0.1$ GeV at the $\sim 3 \sigma$ level. 
The model \cite{Gninenko:2009ks} fits  this scenario with  $c\tau^0 = 0.1$ m 
and $m_{N} \sim 0.5$ GeV, and  it is clearly excluded by MINER$\nu$A data. 

In the right panel of \cref{fig:ud-gamma}, we show the excess of 
single $\gamma$ events at ND280. The dependence on $c\tau_0$ has the typical bump
due to contribution from the $N-$ production in P0D.  The dependence of the excess on $m_N$ 
is weak,  since now $E^{MB}_N \approx E^{ND}_N$. 
The scenario is disfavored at the $(1 - 2) \sigma$ level, 
but the bound  can be significantly improved in the future with larger data sets.
%%% OLD TEXT
%The model is disfavored at $(1 - 2) \sigma$ level, 
%but in future with much higher statistics the test can be significantly improved.

\subsection{Upscattering - Double Decay scenario, $U_N D_B D_\xi$}
%%%%%%%%%%%%%%%%%%%%%%%%%%%%%%%%%%%%%%%%%%%%%%%%%%%%%%%%%%%%%%%%%%%%%%%%%%%%

%%%%%%%%%%%%ffff12%%%%%%%%%%%%%%%%%%%%%%%%%%%%%%%%%%%%%%
\begin{figure*}[h!]
\centering
\includegraphics[width=0.45\textwidth]{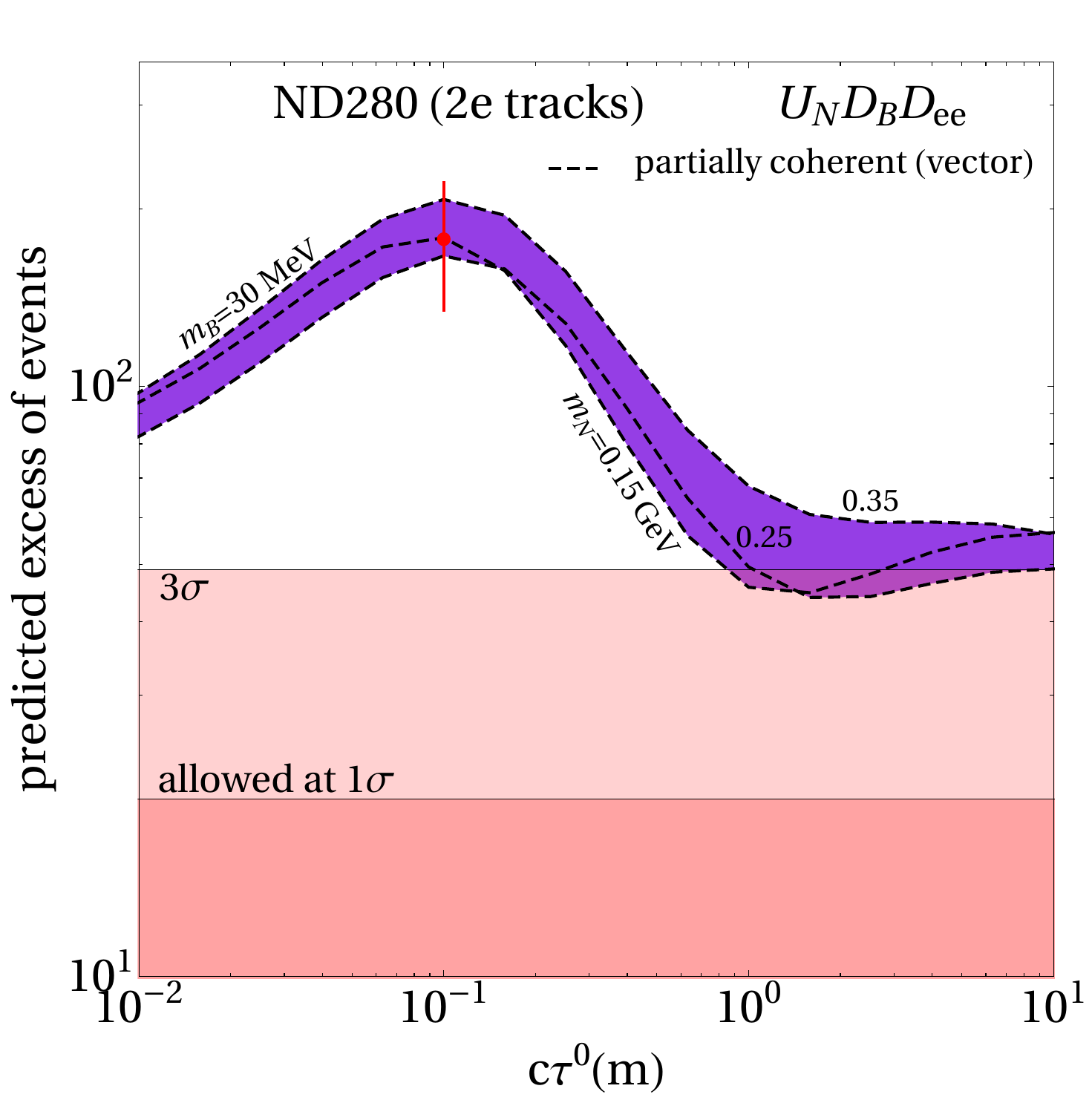}\quad
\includegraphics[width=0.45\textwidth]{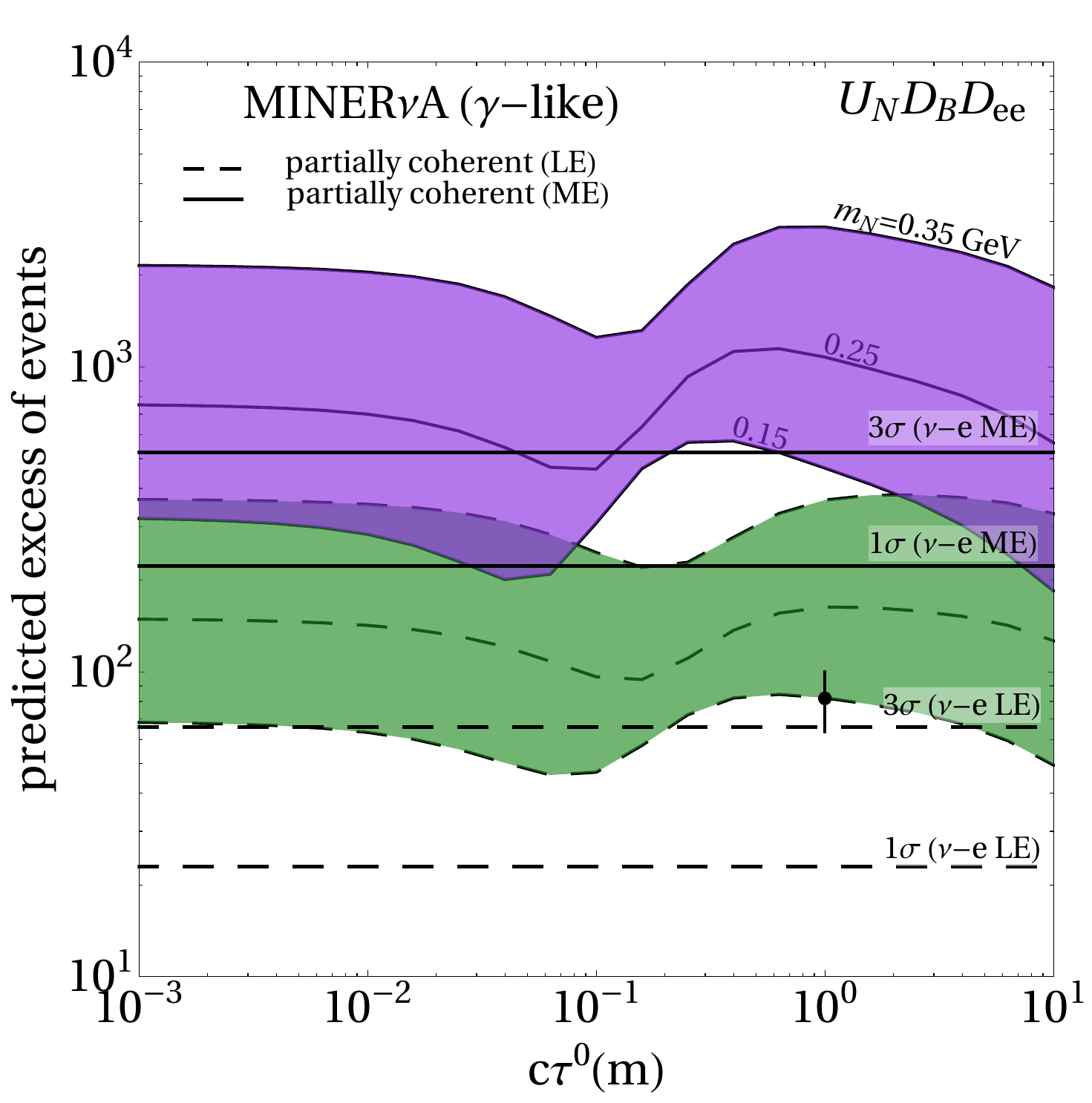}\quad
\caption{
Tests of the Upscattering-Double Decay
into $e^+ e^-$ scenario, $U_N D_B D_{ee}$ at ND280 (left) and
MINER$\nu$A (right).
{\it Left panel:} Number of expected 2e-track events
produced by the $e^+ e^-$ pair at ND280
as a function of $c\tau^0$ for different values of $m_N$
(numbers at the curves in GeV). We take $m_B = 30$  MeV.
The horizontal lines show the $1\sigma$ and $3\sigma$
experimental upper bounds on the 2$e-$track events.
The point with error bar indicates the uncertainty of the prediction from the MiniBooNE-observed event rate.
{\it Right panel:}
Number of expected $\gamma$-like shower events
%produced by the $e^+ e^-$ pair 
at MINER$\nu$A
as a function of $c\tau^0$ for different values of $m_N$
(numbers at the curves in GeV). We take $m_B = 30$  MeV.
Two sets of lines correspond to contribution of 
the ME and LE samples of events.
Partially coherent cross section was used.} 
\label{fig:double}
\end{figure*}
%%%%%%%%%%%%%%%%%%%%%%%%%%%%%%%%%%%%%%%%%%%%%%%%%%%%%%%%

In this scenario (see Fig. \ref{fig:scenE})  $N$ produced via
the $\nu_\mu-$upscattering in a detector and
surrounding materials decays into on-shell boson
$N \rightarrow  B + \nu$,  which in turn decays as
$B \rightarrow e^+ e^-$. Alternatively, $B$ can undergo a radiative decay 
$B \rightarrow B' + \gamma$. $B$ (as well as $B'$)
is new vector or scalar bosons.
In this double decay scenario
there are three vertices with new physics interactions:
$N-$production, $N-$decay and $B-$decay.

If $B$ decays fast, so that the decay length is smaller than (or comparable
to) the size of the detector, effectively the picture of transitions
will be similar to that of $U_N D_\xi$ scenario. Correspondingly,
time evolution, signatures and the most relevant experiments will be similar.
The only difference is that in the $\xi = e^+ e^-$ case
the invariant mass of the pair is fixed by the mass of $B$: 
$W_{ee} = m_B$.
In what follows we will consider the case  $\xi = e^+ e^-$,
that is, the $U_ND_BD_{ee}-$scenario with fast $B$ decay. 

If $m_B > W_c^{ND} = 5$ MeV, ND280 can provide
a direct test of this scenario and therefore give the most stringent bound.
The dependence of number of events, $N_{ee - 2sh}^{ND}$, on
$c \tau^0$ is shown in  the left panel of Fig. \cref{fig:double}. It has the typical 
dependence with two flat asymptotics and
a bump at about $0.1$ m due to $N$ production in the outer P0D detector.
(This is similar to \cref{fig:T2K2showerEE} (left) and \cref{fig:ud-gamma} (right).)
For our computations we use the partially coherent cross section. 
The signature factor enhancement is absent
for $m_B \leq W_c^{MB} = 30$ MeV; MiniBooNE
does not resolve the pair and therefore $f^{MB}_{ee - 1sh} \approx 1$.
On the other hand, for $ m_B \gg W_c^{ND} = 5$ MeV, the ND280 do 
resolve the pair, so that  $f^{ND}_{ee - 2sh} \approx 1$. For larger $m_B$, one would expect
suppression of $f^{MB}_{ee - 1sh}$, and consequently the 
signature factor enhancement of the number of events.
Still, there is a weak dependence of number of events on $m_N$ due to partially
coherent cross section dependence and slightly
higher effective energy of ND280 than that of MiniBooNE.
The reason is the same as for MINER$\nu$A test of $U_ND_\gamma$ scenario described in \cref{subsec:a}.\\

The experimental bounds in \cref{eq:2sh-ndws}
(the same as in \cref{fig:T2K2showerEE} left), disfavor this scenario at more 
than $1 \sigma$ CL in the whole applicable range of $c\tau^0$ ($\gtrsim 10^{-2}$ m) 
and for $m_B > 10$ MeV. In the region $c\tau^0\sim 10^{-1}$ m 
the exclusion of the scenario surpasses $3 \sigma$.
With further decrease of $m_B$ (approaching $W_c^{ND}$)
the number of events is suppressed by the signature factor.
For $m_B < 5$ MeV, the ND280 bound on the $2-$shower events is not applicable,
but one can use various indirect tests. \\

A useful indirect test of the $U_ND_BD_{ee}-$scenario 
is given by the MINER$\nu$A bounds on
$\gamma-$shower events (\ref{eq:mv-gamma/ee_LE}), 
which requires $ee - \gamma$ shower mis-identification.
In \cref{fig:double} (right panel), we show predictions for the number of
$\gamma-$shower events  at MINER$\nu$A.
The dependence of $N^{MV}_{ee-\gamma}$ on $c\tau^0$ has
typical smooth step form with the bump due to $N-$ production in  dirt.
The bump is at larger  decay length than in other
experiments, $c\tau^0 =  (0.5 - 3)$ m due to larger distance between 
the detector and outer material.
The purple and green regions correspond to ME and LE  datasets.
The strong dependence of the number of events on $m_N$ is due to
the coherent cross section enhancement, as explained around 
\cref{eq:ved}. 
Much stronger dependence of  $m_N$ in the right panel  
compared to the one in the left panel is related to higher
neutrino energies at MINER$\nu$A
and therefore weaker suppression of the cross section with
increase of $m_N$, than at ND280 and MiniBooNE.
Also for this reason, the prediction for the ME sample is higher than for the LE
sample (in addition, the ME dataset comes with $\sim 3$ more POT). 
The signature factor enhancement is absent here.
%\textcolor{red}{This does not take into account the selection efficiencies.}

The predictions are at the level of $3\sigma$ upper bounds
on $\gamma-$shower events from \cref{eq:mv-gamma/ee_LE,eq:mv-gamma/ee}.
%We find that the LE sample provides stronger bound than ME one.
%from the  $\nu_e - e$ scattering
%The scenario is excluded at more than $3 \sigma$ level
%in  the whole range of $c\tau^0$ and
%$m_N  > 0.1$ MeV. 

Our prediction is in rough agreement with \cite{Arguelles:2018mtc}, 
apart from the fact that we find stronger exclusion from the LE dataset than from the ME dataset. 
This could stem from the fact that we made simplifying assumptions 
on the experimental efficiencies, where a simulation was performed in ref.~\cite{Arguelles:2018mtc}.

The model \cite{Bertuzzo:2018itn}
matches this scenario for  $c\tau^0 = \mathcal{O}(10^{-9})$ cm, $m_B = 30$ MeV
and $m_N \sim 0.25$ GeV, and therefore is disfavored  by MINER$\nu$A.

However, such a parameter point is 
not excluded by ND280 because of the very small $c\tau^0$. Any 
realization of \cite{Bertuzzo:2018itn} with  $c\tau^0 \gtrsim 10^{-2}$ m is, however,  
tested at least at the level of $3\sigma$ in accord with the left panel of \cref{fig:double}.

The models with scalar mediator \cite{Datta:2020auq,Dutta:2020scq} are not affected by the constraint from MINER$\nu$A due to suppressed upscattering cross section. ND280 can still test this class of models through the search for 2$e$ tracks, analogously to \cite{Bertuzzo:2018itn}. Finally, there could be additional 
tests involving particle misidentification.

\subsection{Mixing - Decay into $\nu_e$ scenario, $M_N D_\nu U_e$} 
%%%%%%%%%%%%%%%%%%%%%%%%%%%%%%%%%%%%%%%%%%%%%%%%%%%%%%%%%%%%%%%%%%%%%%%%%%%%%%%%%%%%%%%%%%%%%%%%%%%%

In this scenario (see Fig. \ref{fig:scenF}), $N$ is produced via mixing in $\nu_\mu$, then $N$
decays along the beamline into $\nu_e$, $N \rightarrow \nu_e  + B$, and in
turn, $\nu_e$ upscatters  in a detector producing the $e-$like events
in the low energy range (if $B$ has large enough mass). In this way
an additional $\nu_e$ flux is generated.

\begin{figure*}[h!]
\centering
\includegraphics[width=0.6\textwidth]{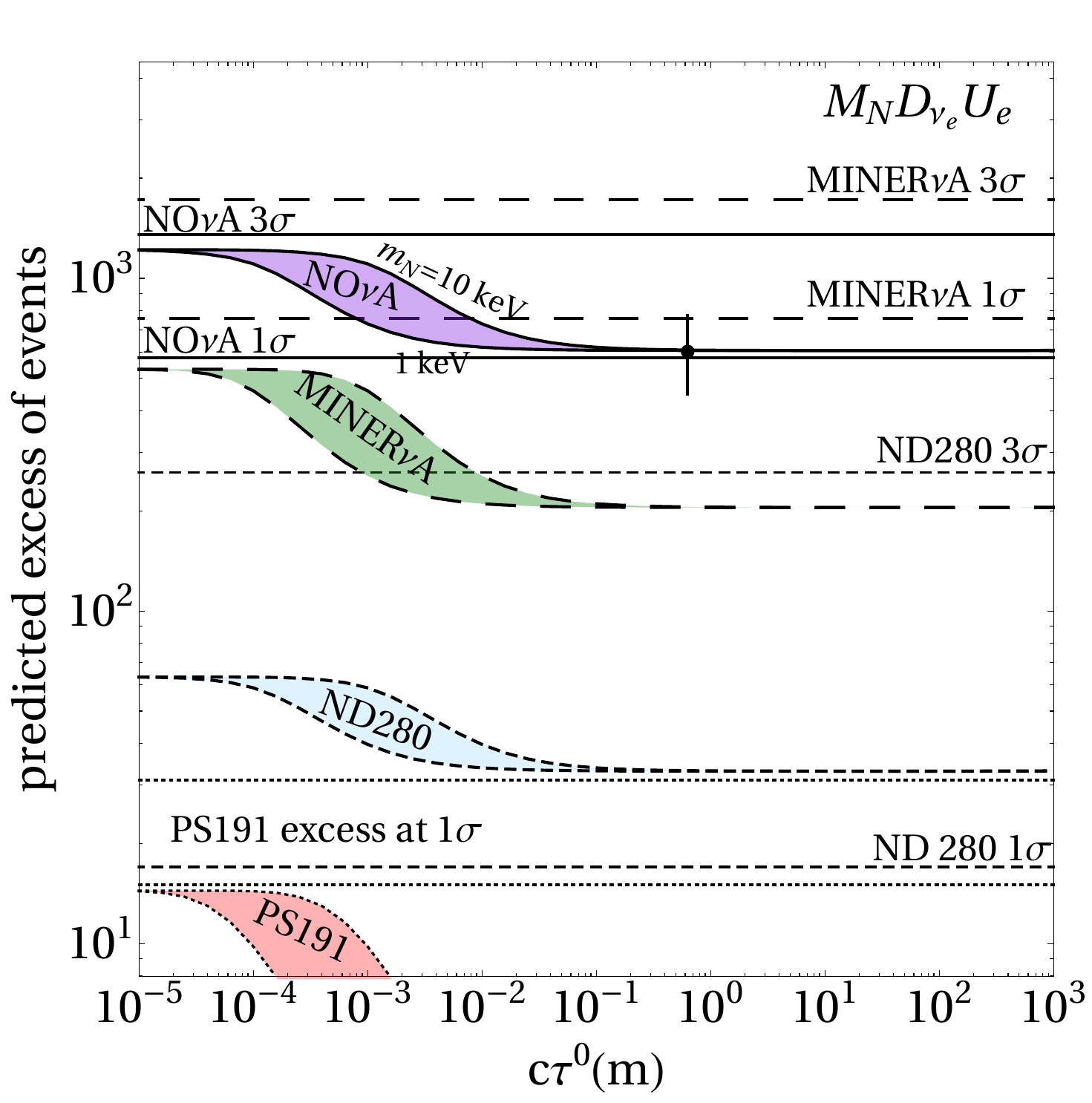}\quad
\caption{
Direct tests of the Mixing - Decay into $\nu_e$ scenario, $M_N D_\nu$.
Number of expected events as a function
of $c\tau^0$ for different values of $m_N$ (numbers at the curves in
keV)
are shown for ND280, MINER$\nu$A, PS191 and NO$\nu$A.
Horizontal lines correspond to  the $1\sigma$ and $3\sigma$
experimental upper bounds for each of these experiments.
The point with error bar indicates the uncertainty of the prediction from the MiniBooNE-observed event rate. 
} 
\label{fig:Kopp}
\end{figure*}

The direct tests of this scenario are provided by studies of
the $e-$like events at ND280, MINER$\nu$A, PS191 and NOvA (\cref{fig:Kopp}).
The number of events due to $M_N D_\nu U_e$ scenario in these experiments,
$N^i_{e - esh}$, has been computed using \cref{eq:decfact3,eq:gbbb}.
According to the analysis in \cref{subsec:jk}, $N^i_{e-esh}$,   
as functions of $c\tau^0$, has smooth step-like form
with constant asymptotics at $c\tau^0 \rightarrow 0$
and $c\tau^0 \rightarrow \infty$ (see \cref{eq:xx}),
and with transition region at
\begin{equation}
{c\tau^0}^i \sim l^i \frac{m_N}{E^i}\,.
\end{equation}
Here, $l^i$ is the baseline.
The asymptotics do not depend on $m_N$, and the transition region
shifts with $m_N$, proportionally to $m_N$.

%%\cref{fig:Kopp}, we confront here the remaining among
%%the scenarios elaborated in \cref{sec3}, $MD_\nu$, with the
%%experimental
%%data. We have calculated the predicted number of events for  ND280,
%%MINER$\nu$A, PS191 and NO$\nu$A.

The limits for single $e-$shower events are given in
\cref{eq:nd-e,eq:mv-e,ps:1shower}.
%%
%%The prediction for the number of additional showers flattens
%%in $c\tau^0\to 0$ and $c\tau^0\to \infty$ limits,
%%according to \cref{eq:xx}.
%%
For MINER$\nu$A,
the predicted number of events is  well below the
$1\sigma$ limit. The prediction for ND280 is
slightly above $1\sigma$, while, interestingly, the calculated
event number for PS191 is almost consistent with the 
observed excess (\cref{sec:limits}).
NO$\nu$A  disfavors this scenario at the level
of $1\sigma$ at large $c \tau^0$  and  above
$2 \sigma$ at small $c \tau^0$.
Notice  that NO$\nu$A has already collected much more data with
respect to the analysis presented in \cite{Adamson:2016tbq} on which
our limits are based. Therefore
an updated analysis can further improve the bounds.

The models \cite{Dentler:2019dhz,deGouvea:2019qre} realize
this scenario with $c \tau^0 \sim 10^{-3}$ cm and $m_M = (1 - 10)$ keV.
Therefore, with present data the best fit point of MiniBooNE 
is disfavored at about $2\sigma$.

%%%%%%%%%%%%ffff13%%%%%%%%%%%%%%%%%%%%%%%%%%%%%%%%%%%%%%
%%%%%%%%%%%%%%%%%%%%%%%%%%%%%%%%%%%%%%%%%%%%%%%%%%%%%%%%

%%%%%%%%%%%%%%%%%%%%%%%%%%%%%%%%%%%%%%%%%%%%%%%%%%%%%%%%%%%%%%%%%%%%%%%%%%
\section{summary and conclusions}
\label{sec:sum}
%%%%%%%%%%%%%%%%%%%%%%%%%%%%%%%%%%%%%%%%%%%%%%%%%%%%%%%%%%%%%%%%%%%%%%%%%%%%

\noindent
  We performed a model independent study of the non-oscillatory
explanations
of the MiniBooNE excess in terms of the phenomenological scenarios.
Here the scenarios are  series of transitions
and  processes which connect the initial interactions
of the accelerated protons with target and the appearance
of single shower ($e$-like) events  in the MiniBooNE detector.
The processes include the production of new particles 
their propagation,  decays, as well as interactions with a medium.
We parametrized scenarios by masses and decay rates of new particles
as well as by cross sections. \\
%%Actually, the latter, to a large extent, does not play
%%a relevant role in our consideration.\\

We carried out a systematic search of the  simplest scenarios
which can be classified by the number of new interaction points
(vertices). We have  found 2 scenarios with 2 vertices,
4 scenarios with 3 vertices, {\it etc.} More possibilities are related to
the nature of new  propagating particles (fermions or bosons)
as well as to the type of particle(s) in the final state which produce
single shower events in MiniBooNE.
We show that these scenarios are reduced to few 
qualitatively different configurations.\\

For these configurations, general  formulas have been derived
for the numbers of events due to new physics. Dependence of these numbers of events 
on parameters of the scenarios were considered.
In particular, we find three
qualitatively different dependences on the decay length $c \tau^0$:
(i) flat dependence with upturn at
small $c \tau^0$ (scenarios with mixing), (ii) smoothed step-like dependence
(scenarios with upscattering in detector), (iii) bump followed by constant 
asymptotics at large $c \tau^0$ (scenarios with upscattering in dirt).
In a sense, we developed the effective theory of new physics at 
low energy accelerator experiments.\\

We described tests of the scenarios employing neutrino experiments
which have setups similar to MiniBooNE:
experiments at near detectors of NO$\nu$A and T2K ND280 as well as
at PS191 and MINER$\nu$A.
While reproducing the MiniBooNE excess, the scenarios lead
to additional events in these experiments.
In other words, scenarios allow to directly connect
the observed MiniBooNE excess of events
to expected excesses in other experiments.
In practice, we normalize the expected number of events
in a given experiment to the MiniBooNE excess, and in this way various
parameters and uncertainties cancel out.

For each experiment under consideration we obtained
the upper bounds on possible numbers of events due to new physics.
We confronted these bounds with expected number of events
related to MiniBooNE excess.

We find that in spite of the large strength of MiniBooNE (mass, POT)
other experiments produce substantial bounds due to better particle ID,
higher neutrino energies, specific dependence of cross section on mass of produced
particle, {\it etc.}
In particular, we find the  signature factor enhancement
and the coherent cross section enhancement. \\

%We find that all scenarios are either excluded or
%disfavored by the data from at least one considered experiment.
Each of the studied scenarios can be tested, with certain
part of parameter space excluded, using available neutrino data.
In particular, $U_N D_{ee}$ and $U_N D_B U_{ee}$ scenarios are restricted  
by the $2e$-tracks data from ND280, while  $U_ND_\gamma$ is
%%disfavored by more than $3\sigma$ using constraints on
excluded by data on isolated photons from MINER$\nu$A.
As far as the MD scenarios with $m_N<10$ MeV are concerned, they are disfavored by the ND280
$2e-$tracks (higher masses are already excluded by MiniBooNE timing
data). According to the $M_N D_\nu U_e$ scenario 
significant excess of events should already be seen at NO$\nu$A 
with the present tension at the $2-3\sigma$ level.

%{\color{blue}

%In terms of previous proposals, \emph{i.e.} specific models,
%we find that \cite{Gninenko:2009ks} is disfavored at $\gtrsim 3 \sigma%$ by MINER$\nu$A data. The model discussed in \cite{Fischer:2019fbw} has already been excluded by MiniBooNE itself (recent timing analysis); hence we studied the realization of such proposal with smaller right-handed neutrino masses ($\mathcal{O}(10)$ MeV) and found that both ND280 and MINER$\nu$A can test this model at the $3\sigma$ level. The proposals in \cite{Ballett:2018ynz,Ballett:2019pyw} are disfavored by $2e$ track search at ND280. Here, it should be pointed out that the benchmark point in these studies corresponds to $c \tau^0>10$ cm. The same authors have 
%come up with a proposal \cite{Abdullahi:2020nyr} in which right-handed neutrino decays promptly. Such case is unconstrained by ND280 because the tracks inside FGD detector are vetoed. In the similar way, proposals in \cite{Datta:2020auq,Dutta:2020scq,Abdallah:2020vgg} also evade limits from ND280 and are further unconstrained by MINER$\nu$A data which is, in contrast, disfavoring \cite{Bertuzzo:2018itn}. Finally, we found that class of models with light scalars and sterile neutrinos 
%\cite{deGouvea:2019qre,Dentler:2019dhz,Bai:2015ztj,PalomaresRuiz:2005vf} would induce $\mathcal{O}(10^3)$ events in NO$\nu$A detector, statistically corresponding to roughly $2\sigma$ limit.
%}

{\color{black}
Concerning specific models, we find that \cite{Gninenko:2009ks} is disfavored at more than $3\sigma$ by MINER$\nu$A data. The model \cite{Fischer:2019fbw} has already been excluded by
the  MiniBooNE timing analysis; hence we studied
the realization of such a model with smaller right handed
neutrino masses ($\mathcal{O}(10)$ MeV); both ND280 and MINER$\nu$A can exclude the models at the $3\sigma$ level.
The proposals in \cite{Ballett:2018ynz,Ballett:2019pyw} with the benchmark point $c\tau^0 > 10$  cm are disfavored by the $2e-$track searches at ND280. The same model with prompt RH neutrino decay \cite{Abdullahi:2020nyr}
is unconstrained by ND280 because the tracks inside FGD detector
are vetoed. In the similar way, proposals in \cite{Datta:2020auq,Dutta:2020scq,Abdallah:2020vgg,Abdallah:2020biq} also evade
limits from ND280 and are also unconstrained by
MINER$\nu$A data. In contrast, the latter disfavors model \cite{Bertuzzo:2018itn}.
We found that in class of models with light scalars and sterile
neutrinos \cite{deGouvea:2019qre,Dentler:2019dhz,Bai:2015ztj,PalomaresRuiz:2005vf} about  $O(10^3)$  additional events is expected
in NO$\nu$A detector, which  corresponds to the  $2\sigma$ upper limit.
New experimental results  will further strengthen these bounds.
Hence, using $3\sigma$ as a criterion, we found that benchmark points in \cite{Gninenko:2009ks,Fischer:2019fbw,Ballett:2018ynz,Ballett:2019pyw,Bertuzzo:2018itn} are ruled out, \cite{deGouvea:2019qre,Dentler:2019dhz,Bai:2015ztj,PalomaresRuiz:2005vf} are in tension while \cite{Abdullahi:2020nyr,Datta:2020auq,Dutta:2020scq,Abdallah:2020vgg,Abdallah:2020biq} are still allowed.

The scenarios and bounds we have elaborated can
be useful for construction of new models which are aimed at
explanations of the  MiniBooNE excess.
}
Our consideration can also be applied to new physics search without reference (connection) 
to the MiniBooNE excess. In this case, the expected number of events at MiniBooNE can be smaller or much smaller than the observed excess. 
Hence, given our general approach, this work can also be regarded as the effective theory of new physics at accelerator based neutrino experiments, being relevant for future projects such as DUNE.

%%%%%%%%%%%%%%%%%%%%%%%%%%%%%%%%%%%%%%%%%%%%%%%%%%%%%%%%%%%%%%%%%%%%%%%%%%%%%%%%%%%%%%%%%%%%%%%%%

\section*{Acknowledgements}
\noindent
OF would like to thank Bill Louis for fruitful email correspondence. VB would like to thank 
Sumit Ghosh and Joachim Kopp for useful discussions. 
Additionally, we would like to thank Laura Fields, Matheus Hostert and the authors of 
\cite{Dentler:2019dhz} for a fruitful email exchange.

%=============================================================================
\bibliographystyle{JHEP}
\bibliography{refs}
%=============================================================================

\end{document}